\documentclass[a4paper,fleqn,usenatbib]{mnras}

\usepackage{newtxtext,newtxmath}

\usepackage[T1]{fontenc}
\usepackage{ae,aecompl}

\usepackage[export]{adjustbox}
\usepackage{amsmath}	
\usepackage{amssymb}	
\usepackage{pdflscape}
\usepackage{psfrag}
\usepackage{pstricks}
\usepackage{fancybox}
\usepackage{mathrsfs}
\usepackage{pstool}
\usepackage{stackengine}

\newcommand{\kms}{\mathrm{km}\,\mathrm{s}^{-1}}
\newcommand{\nhat}{\hat{n}}
\newcommand{\vhat}{\hat{v}}

\newcommand{\vv}{\vec{V}}
\newcommand{\rr}{\vec{r}}
\newcommand{\gradv}{\left[\frac{dv}{dr}\right]}
\newcommand{\tdyn}{t_\mathrm{dyn}}
\newcommand{\HCOp}{\mathrm{HCO}^{+}}

\newcommand{\Tex}{T_\mathrm{ex}}
\newcommand{\e}{\mathrm{e}}
\newcommand{\K}{\mathrm{K}}
\newcommand{\Tkin}{T_\mathrm{kin}}

\newcommand{\mmol}{m_\mathrm{mol}}
\newcommand{\Rt}{R_t}
\newcommand{\rt}{r_t}
\newcommand{\Tt}{T_t}
\newcommand{\gammat}{\gamma_t}
\newcommand{\vexp}{V_\mathrm{exp}}
\newcommand{\rhohat}{\hat{\rho}}
\newcommand{\Xmol}{X_\mathrm{mol}}

\DeclareMathOperator{\sign}{sgn}

\newcommand{\circleBox}[1]{
  \pscirclebox[linewidth=.5pt,framesep=.4pt]{#1}
}

\title[ALMA imaging of IRAS 15103-5754]{ALMA imaging of the nascent planetary nebula IRAS 15103-5754}

\author[G\'omez et al.]{
Jos\'e F. G\'omez,$^{1}$\thanks{On sabbatical leave at Joint ALMA Observatory, Chile}\thanks{E-mail: jfg@iaa.es}
Gilles Niccolini,$^{2}$
Olga Su\'arez,$^{2}$
Luis F. Miranda,$^{1}$
\newauthor
J. Ricardo Rizzo,$^{3}$
Lucero Uscanga,$^{4}$
James A. Green,$^{5}$
Itziar de Gregorio-Monsalvo$^{6}$
\\
$^{1}$Instituto de Astrof\'{\i}sica de Andaluc\'{\i}a, CSIC, Glorieta de la Astronom\'{\i}a s/n, E-18008 Granada, Spain\\
$^{2}$Universit\'e C\^ote d'Azur, Observatoire de la C\^ote d'Azur, CNRS, Laboratoire Lagrange, France\\
$^{3}$Centro de Astrobiolog\'{\i}a (INTA-CSIC), Ctra. M-108, km. 4, E-28850 Torrej\'on de Ardoz, Spain\\
$^{4}$Departamento de Astronom\'{\i}a, Universidad de Guanajuato, A.P. 144, 36000 Guanajuato, Gto., Mexico\\
$^{5}$CSIRO Astronomy and Space Science, Australia Telescope National Facility, PO Box 76, Epping, NSW 2121, Australia\\
$^{6}$Joint ALMA Observatory, Alonso de C\'ordova 3107, Vitacura 763-0355, Santiago, Chile
}

\date{Accepted XXX. Received YYY; in original form ZZZ}

\pubyear{2018}

\begin{document}
\label{firstpage}
\pagerange{\pageref{firstpage}--\pageref{lastpage}}
\maketitle

\begin{abstract}
We present continuum and molecular line { (CO, C$^{18}$O, HCO$^+$)} observations carried out with the Atacama Large Millimeter/submillimeter Array toward the  ``water fountain'' star IRAS 15103-5754, { an object that could be the youngest PN known}. We detect two continuum sources, separated by $0.39\pm 0.03$ arcsec. The emission from the brighter source seems to arise mainly from ionized gas, { thus} confirming the PN nature of the object. The molecular line emission is dominated by a circumstellar torus with a diameter of $\simeq 0.6$ arcsec (2000 au) and expanding at $\simeq 23$ km s$^{-1}$. We see at least two gas outflows. The highest-velocity outflow (deprojected velocities up to 250 km s$^{-1}$), { traced by the CO lines,} shows a biconical morphology, whose axis is misaligned $\simeq 14^\circ$ with respect to the symmetry axis of the torus, and with a different central velocity (by $\simeq 8$ km s$^{-1}$).  An additional high-density outflow { (traced by HCO$^+$)} is oriented nearly perpendicular to the torus.  We speculate that IRAS 15103-5754 was a triple stellar system that went through a common envelope phase, and one of the components was ejected in this process. A subsequent low-collimation wind from the remaining binary stripped out gas from the torus, creating the conical outflow. The high velocity of the outflow suggests that the momentum transfer from the wind is extremely efficient, or that we are witnessing a very energetic mass-loss event.
\end{abstract}




\section{Introduction}

Planetary nebulae (PNe) represent one of the last stages of evolution of stars with main-sequence masses $\simeq 0.8-8$ M$_\odot$.
The PN phase  starts when the effective temperature of the central star reaches $\simeq 25\,000$ K, and the circumstellar envelopes ejected during the asymptotic giant branch (AGB) are then photoionized. PNe display a huge variety of morphologies, in contrast with the spherical symmetry expected from stars in previous phases. Different processes can break this spherical symmetry. For instance, jets launched during the post-AGB phase \citep{sah98,buj01} can open cavities in the circumstellar envelopes. When the PN phase starts, the photoionization front and low-density winds from the central star will proceed preferentially along those cavities. Alternatively, the presence of a binary companion will also alter the morphology of the circumstellar envelope, by capturing part of it and/or creating a common envelope when the material overfills the Roche lobes of the secondary \citep{iva13}.

The processes that take place at the beginning of photoionization, while the ionization front advances through the envelope, will determine the physical characteristics and the morphology in later evolutionary stages. This phase of growth of the ionized nebula only lasts for a $\la 100$ years \citep{bob98}, so it is difficult to catch a source at that exact instant. There is evidence that the object IRAS 15103$-$5754  is just starting the photoionization phase, making it the youngest PN known.  Our Herschel and VLT observations prove that it is a PN \citep{gom15}. However, it shows water maser emission over a velocity range of $\simeq 75$ km s$^{-1}$. Only four other PNe are known to harbour water masers \citep{mir01,deg04,gom08,usc14}, and all these are supposed to be extremely young PNe, as this emission is expected to last for only $\simeq 100$ yr after the end of the AGB \citep{lew89,gom90}. However, IRAS 15103$-$5754 has several key characteristics only observed in this source, all indicating that it is now at the very onset of photoionization \citep{gom15,sua15}.

\begin{itemize}
\item It is the only known ``water fountain'' (WF) that is already a PN. WFs \citep{ima07,des12} are evolved stars with high-velocity water masers (velocity spreads $>75$ km s$^{-1}$), tracing jets of short dynamical ages ($< 100$ yr). They might be one of the first manifestations of collimated mass loss in evolved objects. This type of high-velocity maser jet is also present in IRAS 15103$-$5754. However, all other known WFs are in the post-AGB or late-AGB phase. The small group of water-maser-emitting PNe do not show high velocities. This indicates that IRAS 15103$-$5754 is just leaving the post-AGB phase.
\item The velocity of the water masers increases linearly at longer distances from the central star. This is a sign of an explosive or ballistic event. While this linear velocity trend is present in the ionized and molecular gas of post-AGB stars and PNe, they are not seen in the energetic maser emission of other WFs. 
\item It is the only known PN with non-thermal radio continuum emission. Although the physical conditions in PNe are appropriate for the existence of synchrotron radiation \citep{cas07}, this emission is usually overwhelmed by free-free radiation from electrons in the photoionized regions. The existence of this type of non-thermal emission in IRAS 15103$-$5754 would mean that the photoionized region is still small, and does not veil synchrotron emission.
\item The spectrum of non-thermal emission is changing rapidly. There has been a progressive flattening of the spectral index of radio continuum emission, mainly driven by a decrease of flux density at lower frequencies, over the past 25 years. In particular, the spectral index changed from $\simeq -0.54$ in 2010-2011 to $-0.28$ in 2012. We interpreted this change as due to the progressive growth of electron density in a nascent photoionized region.
\end{itemize}

Given the outstanding characteristics of IRAS 15103$-$5754, and the rapid evolution this source is undergoing in time-scales of a few years, its study represents a unique opportunity to witness the morphological and kinematical changes occurring at the birth of a PN, because the probability of finding any other source at this particular phase is extremely small. 
In this paper, we present ALMA observations of the (sub)mm continuum emission and spectral lines. Our aim is to study the morphology and kinematics of the circumstellar structure, in order to determine the physical processes taking place at the beginning of the PN phase.

\section{Observations and data processing} \label{sec:obs}

Observations were carried out during cycle 3 of ALMA, at bands 6 (1.3 mm) and 7 (0.85 mm)  in different sessions between 2016 March 14 and 2016 August 11. { Depending on the session, the array comprised between 38 and 49 antennas.} All observations were carried out in dual linear polarization, with a spectral set-up comprising four spectral windows. Multiple spectral lines were detected in the data. For the continuum, we combined the line-free channels of all spectral windows at each frequency band. In this paper, we are presenting data from a subset of the observed spectral lines (the brightest ones in our data), specified in Table \ref{tab:obslines}, using data from two spectral windows in each frequency band. The data from the rest of detected lines will be presented in a subsequent paper. Doppler tracking with a central velocity of $-23$ km s$^{-1}$ with respect to the kinematical definition of Local Standard of Rest (LSRK) was applied. { Sources J1427-4206 and J1524-5903 were used to calibrate the spectral bandpass and complex gains, respectively. The absolute flux density scale was determined with sources J1617-5848 and Titan. Mean precipitable water vapour in our observations were 1.1 and 0.7 mm at band 6 and 7, respectively.}

{ All celestial coordinates quoted in the text and figures are in the International Celestial Reference System (ICRS).}
The phase centre of the observations was set as { R.A.(ICRS)} $=15^h14^m17.94^s$, { Dec(ICRS)} $=-58^\circ05' 21.9''$, 
the position reported by \cite{gom15} for the radio continuum emission in IRAS 15103-5754 from centimetre observations using the Australia Telescope Compact Array (ATCA). However, this position was determined with a relatively compact ATCA configuration (10 arcsec angular resolution), and the absolute astrometry seems to be in error by $\simeq 4$ arcsec. 
More recent ATCA observations carried out with higher angular resolution (G\'omez et al. in preparation) provides an updated position for the emission at cm wavelengths that is within 0.5 arcsec of the position of the emission detected with ALMA (section \ref{sec:continuum}), and of the infrared counterpart of the source (2MASS J15141845-5805203). The astrometric accuracy for our ALMA data is $\simeq 50$ mas.

\begin{table}
	\centering
	\caption{Set-up of processed spectral lines}
	\label{tab:obslines}
	\begin{tabular}{lllll} %
		\hline
Line$^a$ 		& $\nu_{\rm rest}$$^b$ & Bandwidth$^c$  & $d\nu$$^d$  & $\nu_{\rm center}$$^e$\\
			& (GHz) 		& (MHz)		 &	(kHz)  & (GHz)	 \\
		\hline
C$^{18}$O(2-1)	&	219.560  & 1875.0	 & 976.562 & 219.000\\
CO(2-1)		&	230.538	 & 234.375	 & 122.070 & 230.538\\
CO(3-2) 	&	345.796	 & 1875.0	 & 976.562  & 345.796\\
HCO$^+$(4-3)&	356.734	 & 1875.0	 &	976.562  & 356.000\\
\hline
\end{tabular}

$^a$ Spectral transition.\\
$^b$ Rest frequency of the transition.\\
$^c$ Bandwidth of the spectral window containing the transition.\\
$^d$ Spectral resolution within the window.\\
$^e$ Central frequency of the spectral window.
\end{table}

Data were calibrated using the ALMA pipeline, { and with versions 4.5.1 and 4.5.3 of the Common Astronomy Software Applications (CASA) package}. Further processing and imaging were carried out with { version 4.6.0 of CASA}. For each frequency band, data from different sessions were combined into a single measurement set. Images were obtained with a Briggs' weighting and a robust parameter of 0.5 (as defined in task {\sc clean} of {\sc casa}), and deconvolved with the {\sc clean} algorithm. Self-calibration was carried out on the continuum data in both phase and amplitude, and the corresponding gains were interpolated and applied to the line data. The continuum images were obtained with multifrequency synthesis. { Continuum emission was subtracted from the visibilities before imaging the line emission.} The line data were averaged on the fly during imaging, to obtain final data cubes with a velocity resolution of 1 km s$^{-1}$, { except for the C$^{18}$O line, which was observed with a coarser resolution of 1.34 km s$^{-1}$}.  Velocities are given with respect to the LSRK. The synthesized beams of all images is $\simeq 0.2-0.3$ arcsec, although their exact values are given in the corresponding figures. All images presented in this paper have been corrected by the primary beam response.

\section{Results}

\subsection{Continuum emission}
\label{sec:continuum}

\begin{figure*}
	\includegraphics*[width=0.9\textwidth]{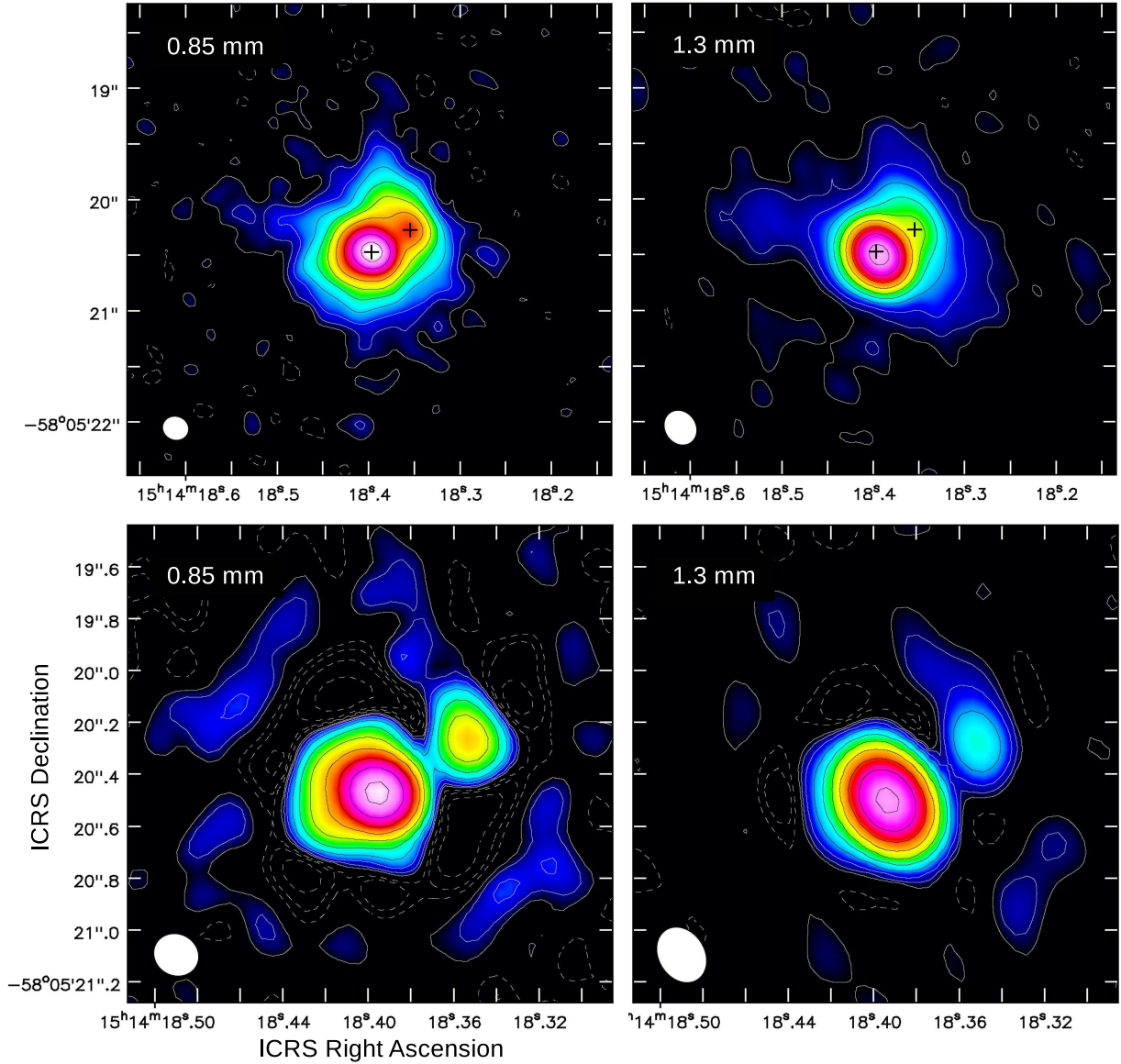}
	\caption{{ Top panels:} Continuum emission at 0.85 mm (left) and 1.3 mm (right). Crosses mark the position of the two point sources obtained from the 0.85 mm image (Table \ref{tab:cont_fits}). { Contour levels are $-2^n\sigma$ (dashed contours) and $2^n\sigma$ (solid contours), starting at $n=1$ and with increment step $n=1$), where $\sigma$ is the rms of each map} (50 and 35 $\mu$Jy/beam at 0.85 and 1.3 mm, respectively). Synthesized beams are $0.23\times 0.21$ arcsec, p.a. $=62^\circ$ (0.85 mm) and $0.32\times 0.27$ arcsec, p.a. $=28^\circ$ (1.3 mm). { Bottom panels: Same as the top panels, but using only baselines $>300$ kilolambda. Note that the images in the top and bottom panels have different spatial scales.}}
	\label{fig:continuum}
\end{figure*}

{ The top two panels in Fig. \ref{fig:continuum} show} the continuum emission at 0.85 and 1.3 mm. 
The total flux densities are $161.9\pm 1.1$ and $71.3\pm 0.5$ mJy, respectively.
The higher-resolution image at 0.85 mm shows two distinct sources (which we name A and B, with A being the brighter source), separated by $\simeq 0.39\pm 0.03$ arcsec. Source A also dominates the emission at 1.3 mm, while B  is evident as an extension to the NW. There is also some weak (but significant) extended emission engulfing these two central sources. Moreover, an extension to the east is seen in both images at a $\simeq 4\sigma$ level. 
{ Images created  by filtering the extended emission (using only projected baselines $>300$ kilowavelengths, two bottom panels in Fig. \ref{fig:continuum}) show sources A and B more clearly. This indicates that these two sources are compact, and that they probably represent two distinct stars.
We tried to fit elliptical Gaussians to the central part of the image (including the two point sources and the extended emission around it), but we only obtained reliable fitting parameters at 0.85 mm, as shown in Table \ref{tab:cont_fits}). The parameters of the sources at 1.3 mm in this table were obtained from the image with  projected baselines $>300$ kilowavelengths. } The extended emission at 0.85 mm (from the map using all visibilities, Fig. \ref{fig:continuum}) has a deconvolved size of $580\pm 90 \times 420\pm 70$ mas and a position angle (measured from north to east) of $-40^\circ\pm 20^\circ$.  Assuming a distance of 3.38 kpc \citep{vic15}, the linear size of the extended component would be $\simeq 2000\times 1400$ au. { Note that throughout our paper we use this distance estimate obtained by \citet{vic15}, as it is the only one available in the literature. This distance was obtained by fitting the spectral energy distribution of the source and assuming a certain value of total luminosity. 
Although it is probably a reasonable estimate, it is unlikely to be accurate to three significant figures. }

\begin{table*}
	\centering
	\caption{Parameters of the continuum sources}
	\label{tab:cont_fits}
	\begin{tabular}{llllll} %
		\hline
Source & R.A. (ICRS) & Dec. (ICRS)  & $S_\nu$(0.85 mm) & $S_\nu$(1.3 mm) & Spectral index\\
		&  			&  				&	(mJy)		   & (mJy) \\
		\hline
A & 15:14:18.396 $\pm 0.006$ & $-58$:05:20.47 $\pm 0.05$ &  $77.8\pm 1.2$ & $50.46\pm 0.16$ & $0.97\pm 0.04$\\
B & 15:14:18.353 $\pm 0.006$ & $-58$:05:20.26 $\pm 0.05$ &  $7.7\pm 1.2$ & $1.23 \pm 0.15$ & $4.1\pm 0.4$\\
\hline
\end{tabular}
\end{table*}

For the whole emission, we estimated a spectral index $\alpha$ (defined as $S_\nu \propto \nu^\alpha$) of $1.835\pm 0.021$. 
This spectral index is close to, but significantly lower than two. In particular, we note that the spectral index of source A is $0.97\pm 0.04$ (Table \ref{tab:cont_fits}). 
A spectral index $\alpha<2$
is not compatible with dust only, so at least part of the emission must arise from other physical processes. In Fig. \ref{fig:spindex}, we show a map of the spectral index, obtained from a ratio between the emission at 0.85 and 1.3 mm, after convolving both images to a common angular resolution of 0.33 arcsec. The white contour represents $\alpha=2$, so the dark area inside this contour cannot arise from dust alone. This area is associated with source A. The minimum spectral index we obtained in that area is $\alpha \simeq 0.6$, which is the expected one for free-free emission from an ionized region with a radial  dependence of electron density of $n_e\propto r^{-2}$ \citep[see, e.g.,][]{oln75,pan75}. 

However, the outer regions in the images (including the location of source B) do have spectral indices compatible with dust ($\alpha>2$ in Fig. \ref{fig:spindex}). The emission from source B has a spectral index of $4.1\pm 0.4$, which indicates that it arises from optically thin dust emission with an emissivity index $\beta= \alpha -2 \simeq 2$ (whith $\kappa_\nu\propto \nu^\beta$, where $\kappa_\nu$ is the mass absorption coefficient per unit dust mass). This value of $\beta$ is consistent with the value expected at mm wavelengths for dust particle sizes $\ll1$ mm \citep*{dra84,bec00}. Assuming $\kappa_\nu = 0.294$ cm$^2$ g$^{-1}$ at 1 { mm} \citep{dra03}, and a dust temperature of 100 K, { we estimate a dust mass $M_d = 2.5\times 10^{-3}$ M$_\odot$ for source B. 
Taking this result at face value, and assuming a gas-to-dust ratio of 100, it would imply a total mass of 0.25 M$_\odot$ within a radius $\la 500$ au (as the emission is unresolved with beam sizes $\simeq 0.3$ arcsec) for source B. If the continuum emission from source B were due to dust in an AGB envelope expanding at $\simeq 15$ km s$^{-1}$, the mass-loss rate would be $\simeq 1.5\times 10^{-3}$ M$_\odot$ yr$^{-1}$. This value would be too high, by at least one order of magnitude, for an AGB star \citep{blo95}. However, we note that our estimates of mass and mass-loss rate are very uncertain, as they depend on several parameters that are not well constrained (e.g., distance, dust temperature, gas-to-dust ratio, mass absorption coefficient). A possible alternative is that source B has accreted material from the envelope of source A, rather than having expelled its own envelope. Moreover, it is also possible that the continuum source B is not really tracing a star, but it is a dense clump in the circumstellar material around IRAS 15103-5754. However, { if it were only a density accumulation, a local maximum} should also be seen in the molecular line data, especially in the optically thin { C$^{18}$O line}, but this is not the case (see next section). { Thus,  continuum source B should then be tracing denser and warmer material than the torus. If dust emission becomes optically thick at that position, the higher temperature at this location would make it appear as a compact continuum source, but not necessarily as a local maximum in the gas tracers. The most likely explanation for a local density and temperature enhancement is the existence of an internal energy source. Therefore, we suggest that this continuum source is actually tracing warm dust around a separate star}.

In summary, it seems that a significant fraction of the observed emission in IRAS 15103-5754 arises from free-free processes in ionized gas. Higher-resolution observations covering a wider wavelength range would be necessary to disentangle the different components (dust and free-free radiation) of the emission.

\begin{figure}
	\includegraphics*[width=0.45\textwidth]{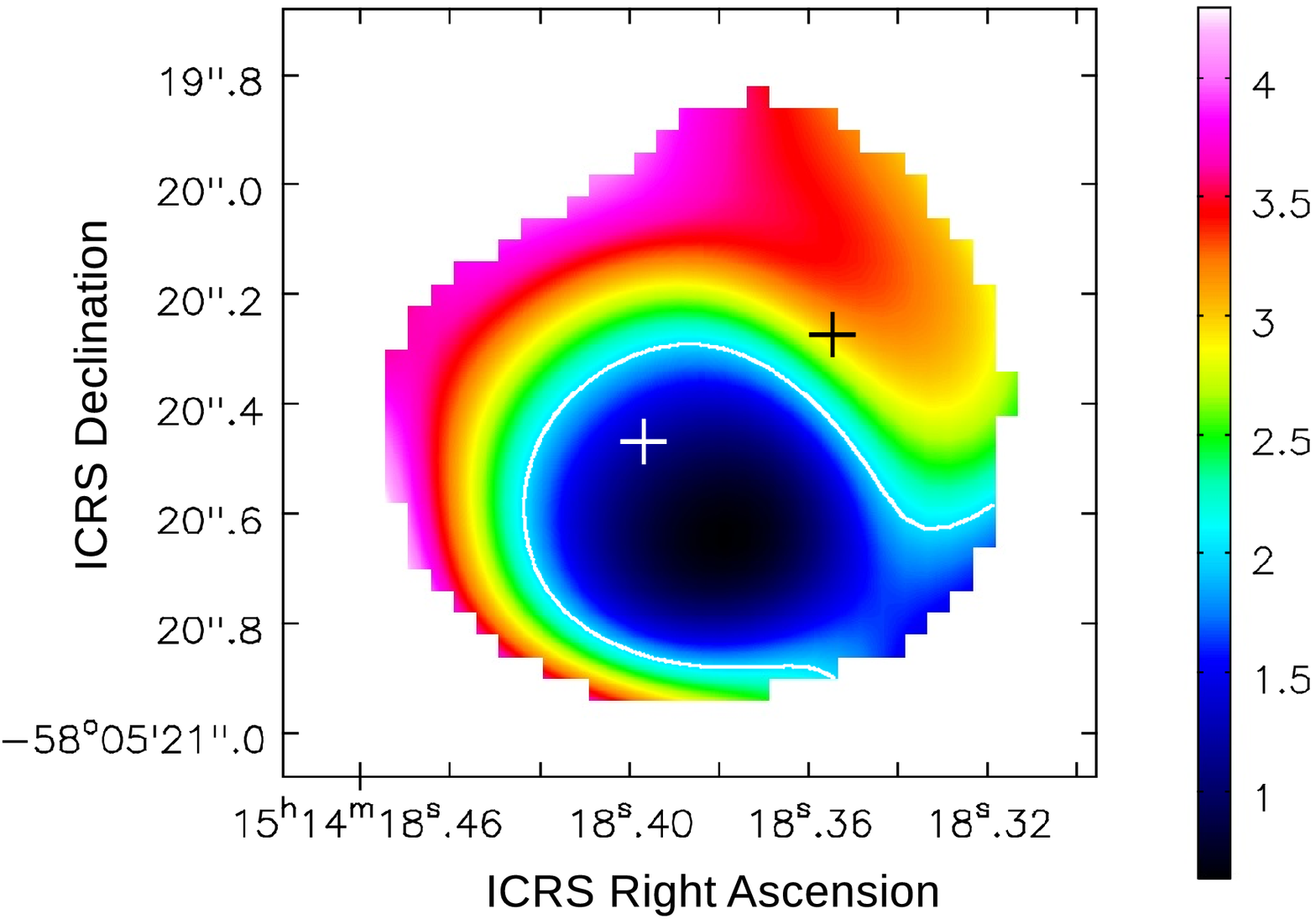}
	\caption{Spectral index map between 0.85 and 1.3 mm, for areas with a signal-to-noise ratio higher than 20 at both wavelengths. Crosses mark the positions of sources A and B. The white contour represents $\alpha=2$.}
	\label{fig:spindex}
\end{figure}

\subsection{Line emission: an expanding torus}

\label{sec:toroid}
The integrated intensity map of the C$^{18}$O(2-1) line traces a nearly edge-on toroidal structure centred on source A (Fig. \ref{fig:torus_mom0}). 
The toroidal structure also dominates the emission at the central velocities of CO and HCO$^+$, at levels $> 128\sigma$ (Fig. \ref{fig:torus_mom0}), although additional weaker structures also appear. In particular, the lower levels in the CO maps trace a butterfly-like structure that might be related to the outflow motions described in the following sections. The major axis of the projected torus is at p.a. $\simeq -36^\circ$. This is close to the p.a. estimated for the extended continuum component ($-40^\circ$, Sec. \ref{sec:continuum}).

\begin{figure*}
	\includegraphics*[width=0.45\textwidth]{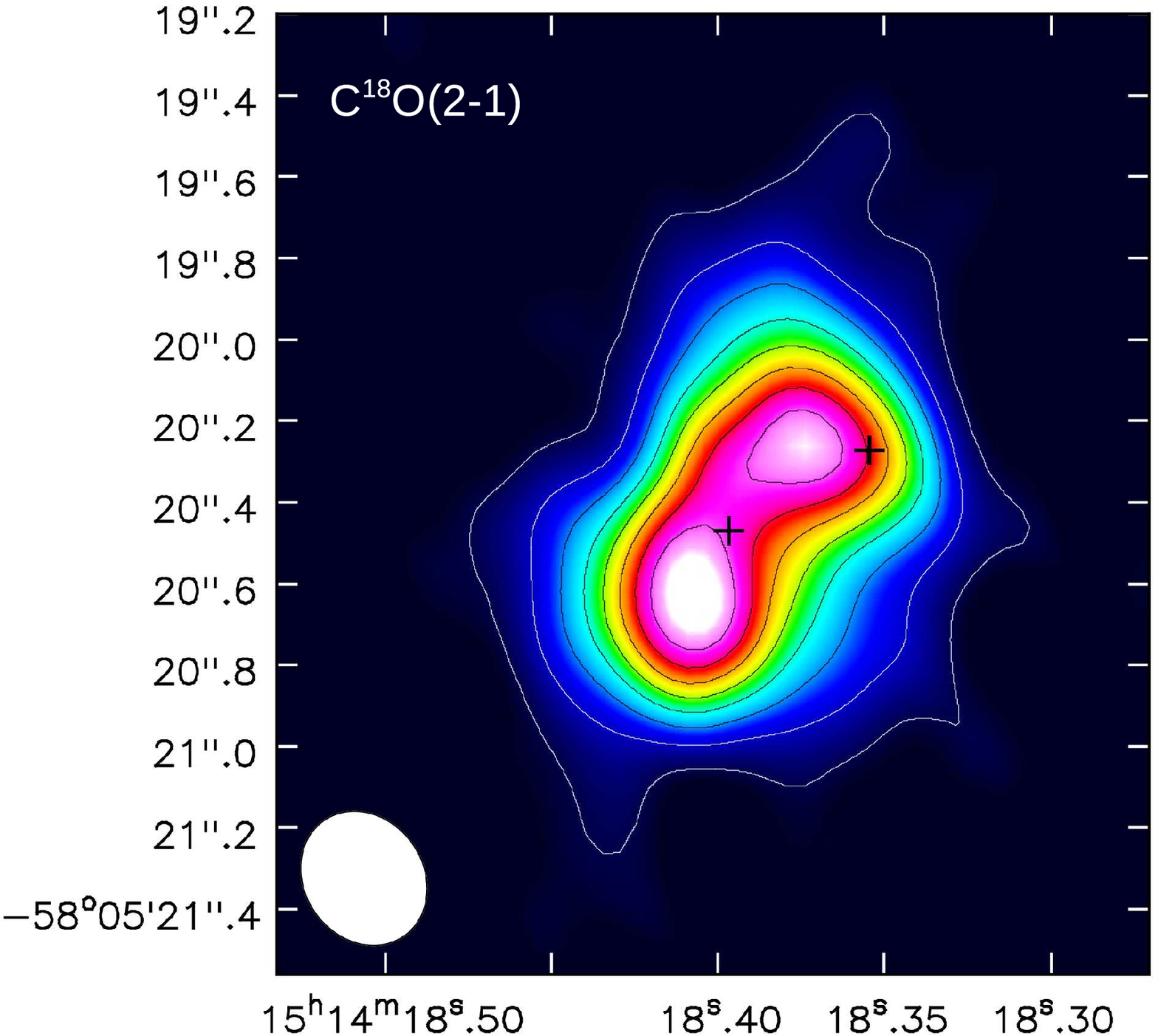}
	\includegraphics*[width=0.45\textwidth]{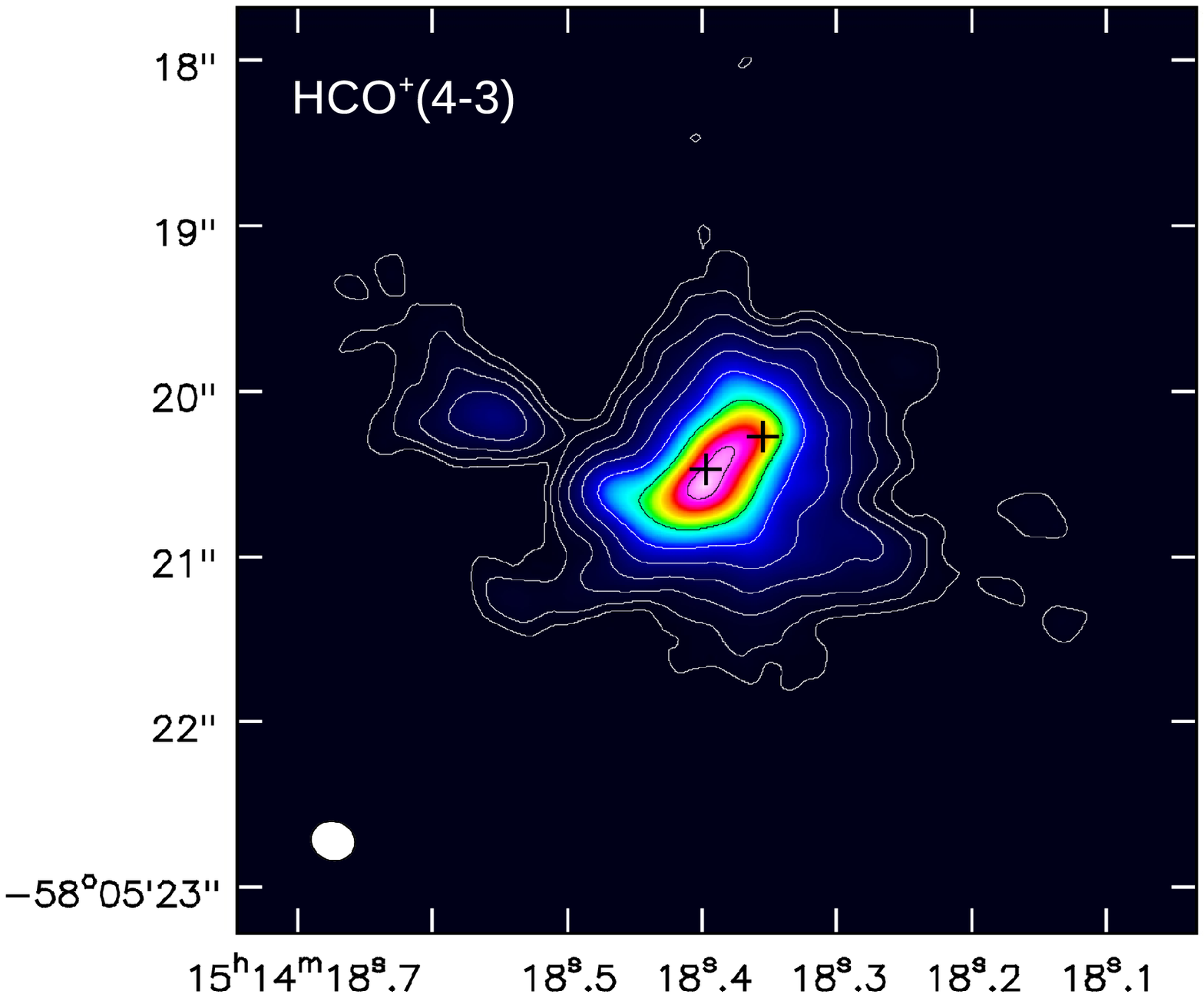}
	
	\vspace*{0.5cm}
	\includegraphics*[width=0.45\textwidth]{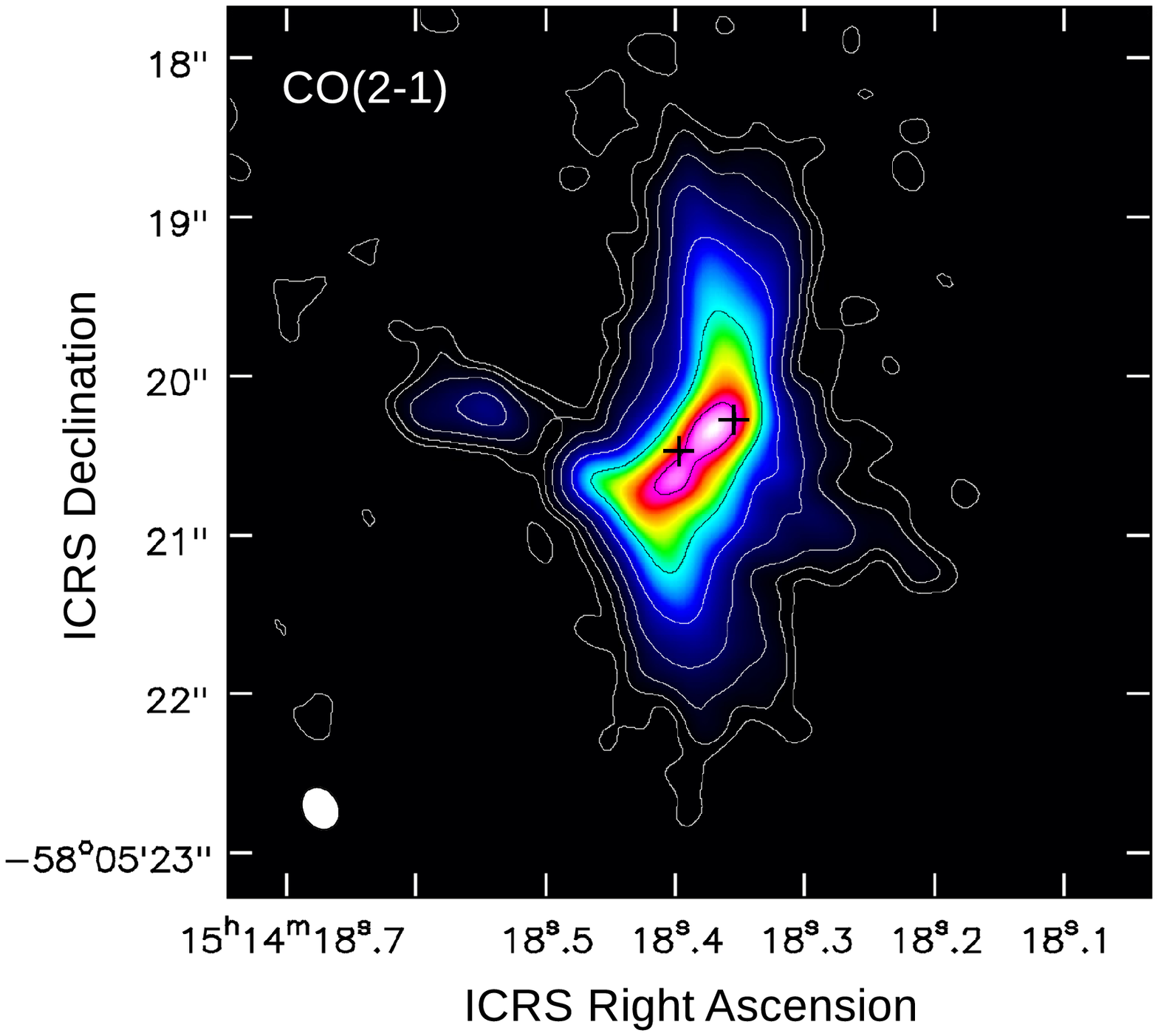}
	\includegraphics*[width=0.45\textwidth]{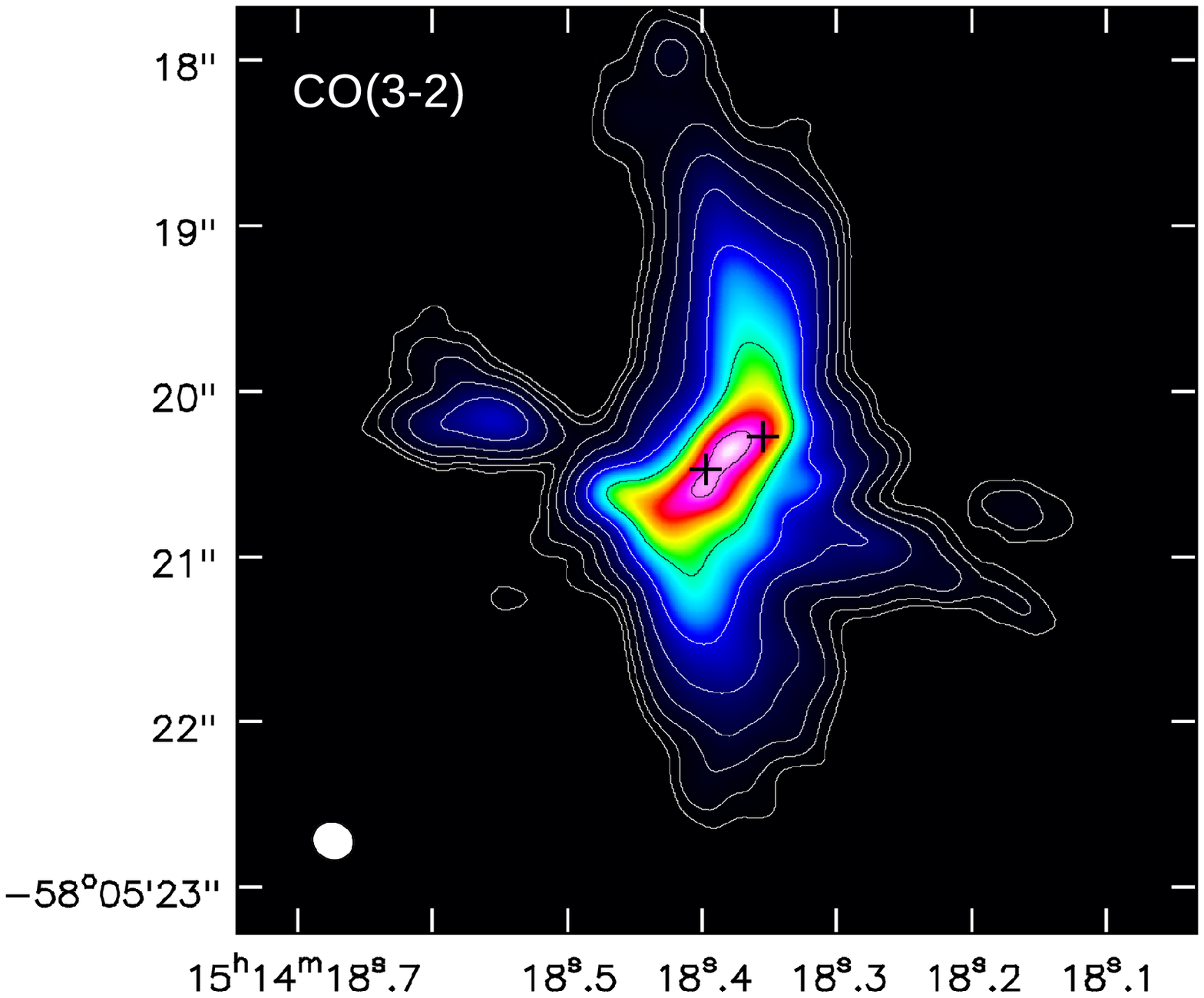}
	\caption{Top left: integrated intensity of C$^{18}$O(2-1) between $-66$ and 6 km s$^{-1}$ (the whole velocity extent of the emission). The lowest contour and the increment step are 3 times 18 mJy km s$^{-1}$ (the rms of the map). The synthesized beam is $0.34\times 0.29$ arcsec, p.a. $=30^\circ$. Top right: integrated intensity of HCO$^+$(4-3) between $-63$ and $-3$ km s$^{-1}$. Contour levels are $2^n\sigma$ (starting at $n=1$ and with increment step $n=1$), where $\sigma$ is the rms of the map (60 mJy km s$^{-1}$). The synthesized beam is $0.26\times 0.24$ arcsec, p.a. $=68^\circ$. Bottom left: integrated intensity of CO(2-1) between $-63$ and $-3$ km s$^{-1}$. Contour levels are $2^n\sigma$, starting with $n=1$ and with increment step $n=1$ ($\sigma = 50$ mJy km s$^{-1}$). The synthesized beam is $0.28\times 0.22$ arcsec, p.a. $=25^\circ$.
	Bottom right: integrated intensity of CO(3-2) between $-63$ and $-3$ km s$^{-1}$. Contour levels are $2^n\sigma$, starting with $n=1$ and with increment step $n=1$ ($\sigma = 70$ mJy km s$^{-1}$). The synthesized beam is $0.25\times 0.23$ arcsec, p.a. $=62^\circ$. In all panels, crosses mark the positions of the two sources resolved at 0.85 mm (Table \ref{tab:cont_fits}).}
	\label{fig:torus_mom0}
\end{figure*}

Position-velocity (PV) diagrams along the projected major and minor axes (Fig. \ref{fig:PV}) of the torus (p.a. $=-36^\circ$ and $-126^\circ$, respectively) clearly show the kinematical pattern of an expanding torus in C$^{18}$O and HCO$^+$. We can rule out contracting motions, given the observed asymmetry in the optically thicker lines (HCO$^+$ and CO transitions). Optically thick lines trace the temperature at the surface of the torus. In that case the brighter SW half is farther away from us, and we see the inner surface, closer to the star (and thus hotter). Because this brigther (farther away) half is redshifted, it necessarily indicates that the torus is expanding. In the case of the C$^{18}$O line, which is expected to be optically thin, the asymmetries in the PV diagrams reflect variations of optical depths. Thus, the column density seems to be higher in the NE half of the torus.

We note that the HCO$^+$ and CO PV diagrams along the minor axis also show contribution from high-velocity gas (sec. \ref{sec:outflows}). This high-velocity gas is more prominent in CO images, and therefore, the kinematical pattern of the torus is not seen so clearly in them.

\begin{figure*}
	\includegraphics*[width=0.9\textwidth]{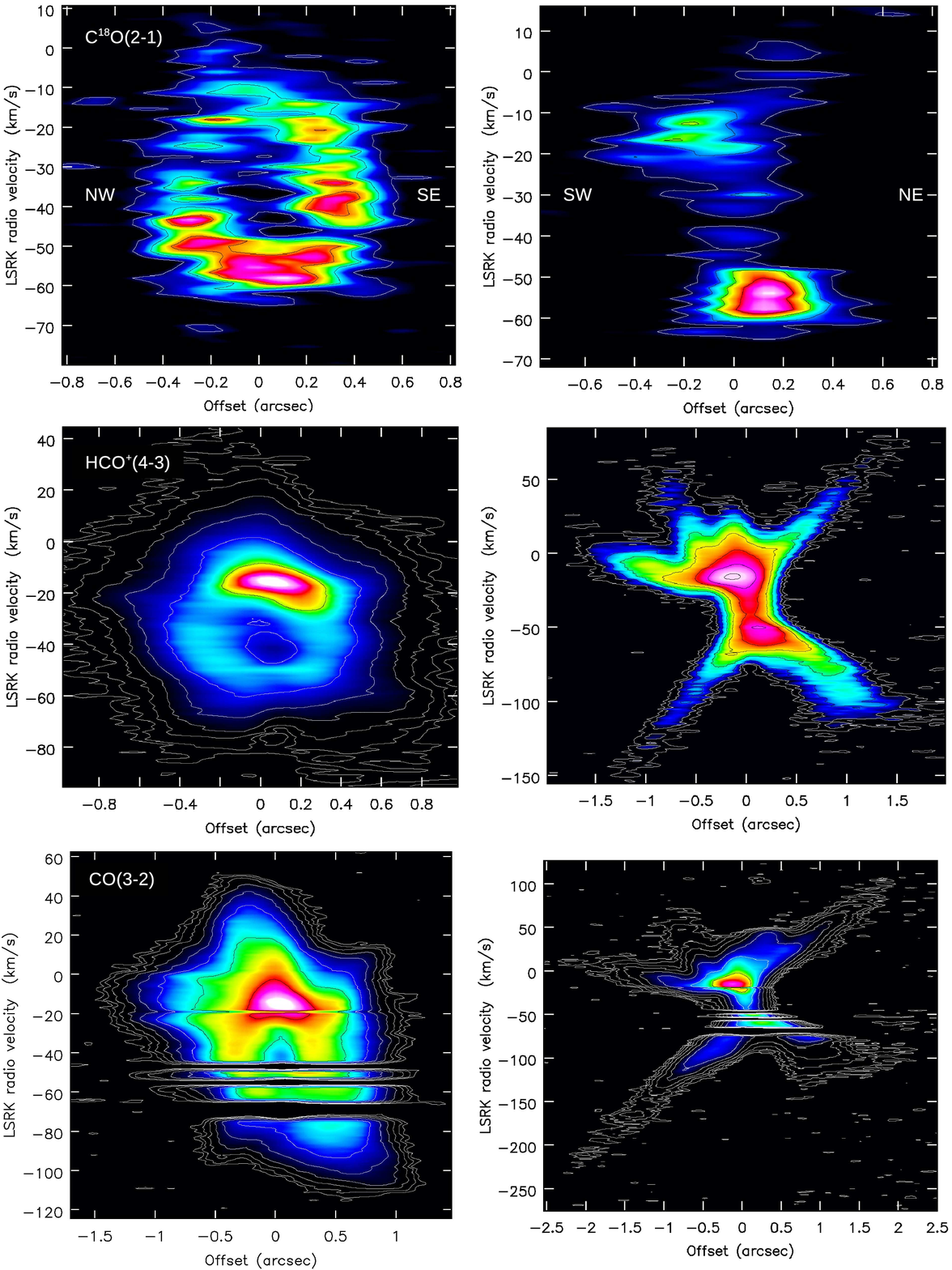}
	\caption{PV diagrams of C$^{18}$O(2-1), HCO$^+$(4-3), and CO(3-2) (top, middle, and bottom panels, respectively) along the projected major (left panels, p.a $=-36^\circ$) and minor (right panels, p.a $=-126^\circ$) axes of the toroidal structure, and averaging over a spatial width of 0.28 arcsec. In all cases, the offset increases in the same sense as right ascension (negative offsets are regions of lower right ascension). The geometrical locations of the positive and negative positional offsets of each diagram are also labelled for clarity in the top panels. Contour levels in the C$^{18}$O(2-1) panels are 3, 6, 9, and 12$\sigma$. In the HCO$^+$(4-3) and CO(3-2) panels, they are $2^n\sigma$, starting with $n=1$ and with increment step $n=1$ . The 1$\sigma$ rms values are 1.1, 2.5, and 1.6 mJy beam$^{-1}$ for C$^{18}$O(2-1), HCO$^+$(4-3), and CO(3-2), respectively. The horizontal dark stripes in the CO emission correspond to channels in which the atmospheric effects could not be properly calibrated.}
	\label{fig:PV}
\end{figure*}

Because the C$^{18}$O(2-1) line is optically thinner, it can provide a better estimate of the mass of the torus. The estimated mass of the structure is { 0.4-1.0} M$_\odot$ assuming { local thermodynamic equilibrium (LTE) with a temperature of  60-180 K (estimated from the HCO$^+$ lines, see below)}, an abundance of { $^{12}$C$^{16}$O} with respect to H$_2$ of $3\times 10^{-4}$ \citep[standard value of O-rich evolved stars, see, e.g.,][]{tey06} and a $^{16}$O/$^{18}$O ratio of $\simeq 330$, corresponding to that of the interstellar medium at a galactocentric distance of $\simeq 6.8$ pc \citep{wil99}.

To estimate other physical properties of the torus, we have used the HCO$^+$ images and PV diagrams, as they have a higher S/N ratio than the C$^{18}$O images, and the torus emission is less contaminated from high-velocity, low-density gas than in the case of the CO transitions. Moreover, the HCO$^+$ data do not suffer from any problems with atmospheric calibration that plagues some of the channels in the CO data sets (horizontal stripes in Fig. \ref{fig:PV}). 
In any case, { the obtained geometric and kinematic} parameters are consistent, within the errors, for all observed lines. We obtain a torus mean diameter  of $\simeq 0.6$ arcsec { (2000 au at 3.38 kpc)}, a deconvolved diameter of its transversal section of $\simeq 0.43$ arcsec (1400 au), and an angle of $=26^\circ$ between the symmetry axis of the torus and the plane of the sky. Regarding kinematics, we obtain a deprojected expansion velocity of 23 km s$^{-1}$ and a central LSRK velocity of $-33$ km s$^{-1}$. With the diameter and expansion velocity, we estimate a dynamical age of 200 yr for the expanding torus. The motions in this torus are largely unbound, as the escape velocity at a distance of 1100 au from a central star of $\la 1$ M$_\odot$ \citep[as expected after the AGB phase, see e.g.,][]{ren10} is only $\la 1.3$ km s$^{-1}$. This provides further support for our suggestion that the torus is expanding, rather than contracting.

We have modelled the HCO$^+$ emission from an expanding torus, as described in Appendix \ref{app:toroid_model}. In this model, we fixed the { geometric} parameters mentioned in the previous paragraph. The remaining free parameters are the abundance of HCO$^+$ with respect to H$_2$, the turbulence velocity, and the radial dependence of temperature and density. 
Despite the simplicity of our model, the resulting position-velocity diagrams (Fig. \ref{fig:PVmodel}) reasonably resemble the observed ones (Fig. \ref{fig:PV}), for a torus with a maximum (inner) temperature of 180 K, and a minimum (outer) temperature of 60 K. { We note that the brightness temperatures of the CO lines in the torus are $\simeq 30-40$\% higher than those of the HCO$^+$ line. This could mean that the HCO$^+$(4-3) transition is not fully thermalized in the torus, or that the CO emission is contaminated by hotter, less dense gas (below the critical density of HCO$^+$) outside the torus. Thus, we consider $\simeq 30-40$\% as the typical uncertainty in our temperature determination.}
Because the HCO$^+$ line becomes optically thick very quickly, our results do not depend on the particular choice of density or abundance (for the latter we assumed $\simeq 10^{-7}$, see also Sec. \ref{sec:HCOoutflow}). 
It is interesting to note that we need to include a value of turbulence velocity of $\ga 2$ km s$^{-1}$ to reproduce to observed intensity asymmetry in the HCO$^+$ emission (the SW half being brighter than the the NE half). Lower values of turbulence produce a symmetric brightness, even in the optically thick regime. This is a reasonable value for turbulence velocity in PNe \citep{gue98,sab08}.

Our model of  HCO$^+$ emission is only intended as an illustration of the general geometric characteristics of the torus, as it is too simple for a close reproduction of the observations.  Therefore we did not attempt any detailed fit to the data, other than a first-order approximation to the temperatures in the torus.
A noticeable difference between the images and our model is that the intensity decrease is { steeper} in the data. This indicates that the temperature variation departs from the dependence assumed  in our model. Observations with a higher angular resolution of both optically thick and thin transitions would provide more accurate information about the temperature and density gradients to aid the modelling.

\begin{figure*}
	\includegraphics*[width=0.45\textwidth]{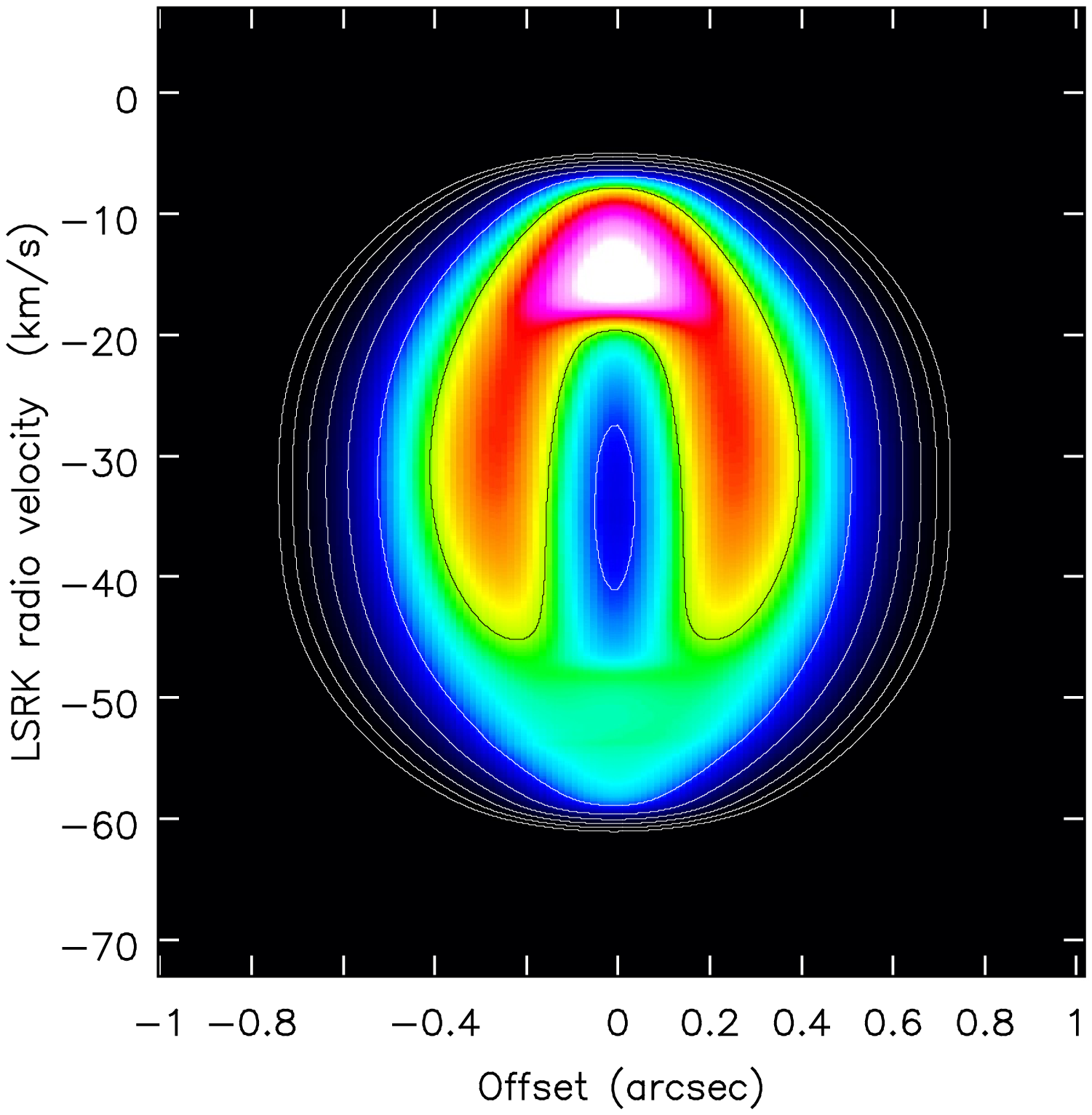}
	\includegraphics*[width=0.45\textwidth]{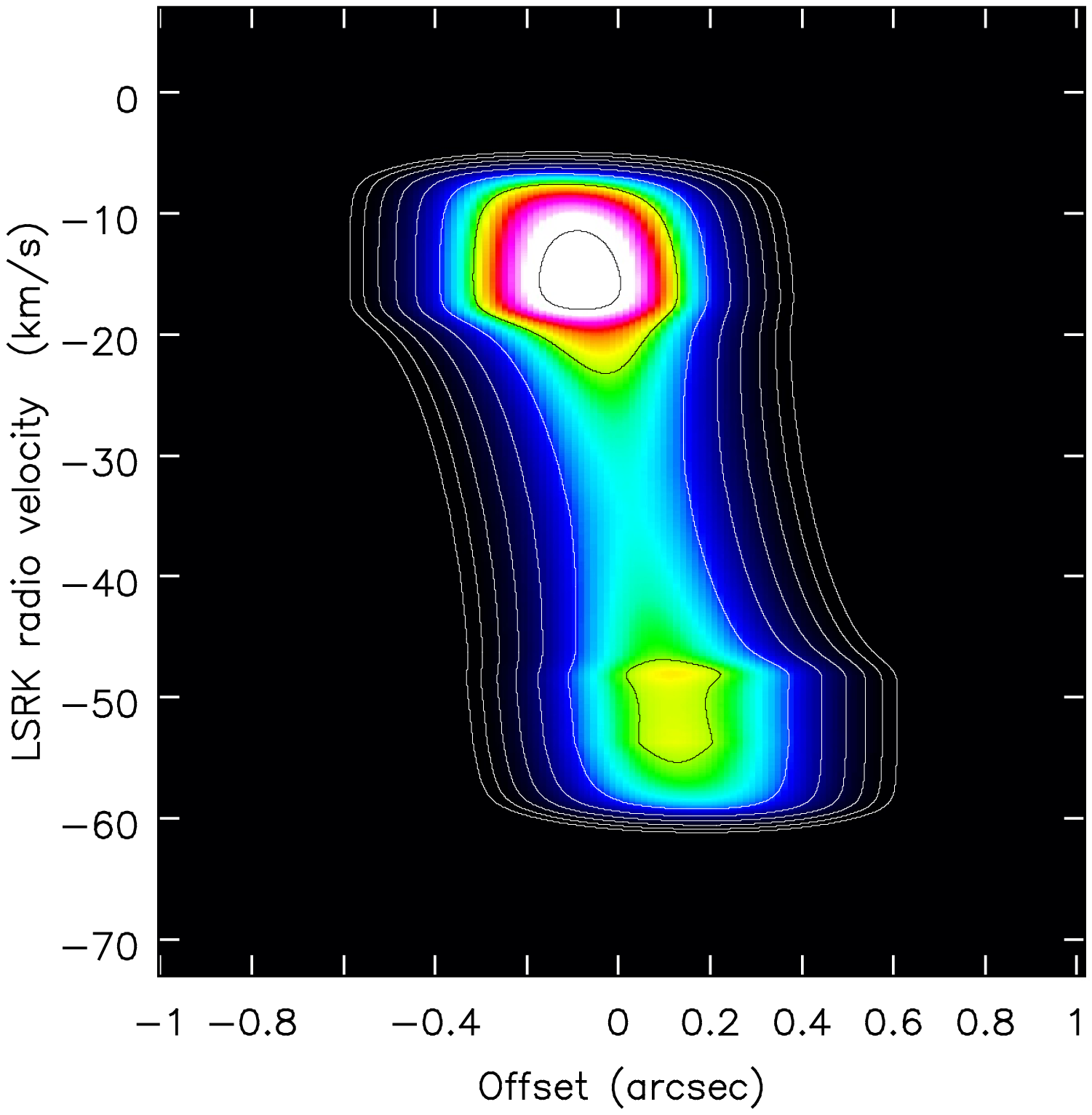}
	\caption{PV diagrams of the torus model for HCO$^+$, presented in Appendix \ref{app:toroid_model}. The axes, averaging width, contour levels, { and units} are the same as the HCO$^+$ diagrams in { Fig. \ref{fig:PV}, although the spatial and velocity scales are smaller in the present figure, in order to show the details more clearly.}}
	\label{fig:PVmodel}
\end{figure*}

\subsection{Line emission: high-velocity outflows}
\label{sec:outflows}

High-velocity gas (with projected velocities up to $\simeq 200$  km s$^{-1}$ with respect to the central velocity of the toroidal structure) are seen in both CO transitions, and in the HCO$^+$ emission. 
The position velocity diagrams along the minor axis of the projected torus of the HCO$^+$ and CO line emission (Fig. \ref{fig:PV})  clearly shows the presence of outflows with multiple lobes. It is to be expected that outflowing gas would show a bipolar pattern perpendicular to the torus, so redshifted gas would be to the NE, and blueshifted gas to the SW. This is opposite to the trend seen with the water maser jet \citep{gom15}, which shows a prominent blueshifted lobe to the NE. However, the PV diagrams show a complex pattern with both redshifted and blueshifted gas on each side of the source. A bipolar outflow with a large opening angle and with an axis close to the plane of the sky can explain part of this structure of overlapping blueshifted and redshifted gas, so that the front side of each lobe would have a velocity sign opposite to that of the back side. However, a single outflow cannot completely explain the complexity of the high-velocity gas. Possible separate outflows are described below.

\subsubsection{A biconical outflow}

\label{sec:co_outf}
Fig. \ref{fig:CO_outflow} shows the map of high-velocity gas ($> 30$ km s$^{-1}$ from the central velocity) traced by the two observed CO transitions. The high-velocity gas shows a butterfly structure, with both redshifted and blueshifted gas on each side of the star, which might represent a single outflow close to the plane of the sky, or a mixture of several outflows. The general axis of the outflow is not exactly perpendicular to the torus, but it has a position angle $\simeq 36^\circ$, so there is a deviation of $\simeq 10^\circ$ with respect to the torus symmetry axis, which is small, but significant. The outflow structure is most clearly seen in the CO(3-2) transition.

{ Assuming LTE and optically thin emission in both CO transitions for the high-velocity gas, we obtain a total mass of $0.03-0.10$ M$_\odot$ for the outflow, for a range of excitation temperatures $60-180$ K. Considering that the brightness temperature of the high-velocity CO emission is $\le 40$ K, the optically thin assumption seems reasonable. If the actual excitation temperatures were lower, then the emission would be partially thick, and our calculations would underestimate the outflow mass. We could not obtain consistent values of excitation temperatures from the ratio CO(3-2)/CO(2-1) at different velocities, which probably means that these transitions depart from LTE.}


\begin{figure}
	\includegraphics*[width=0.45\textwidth]{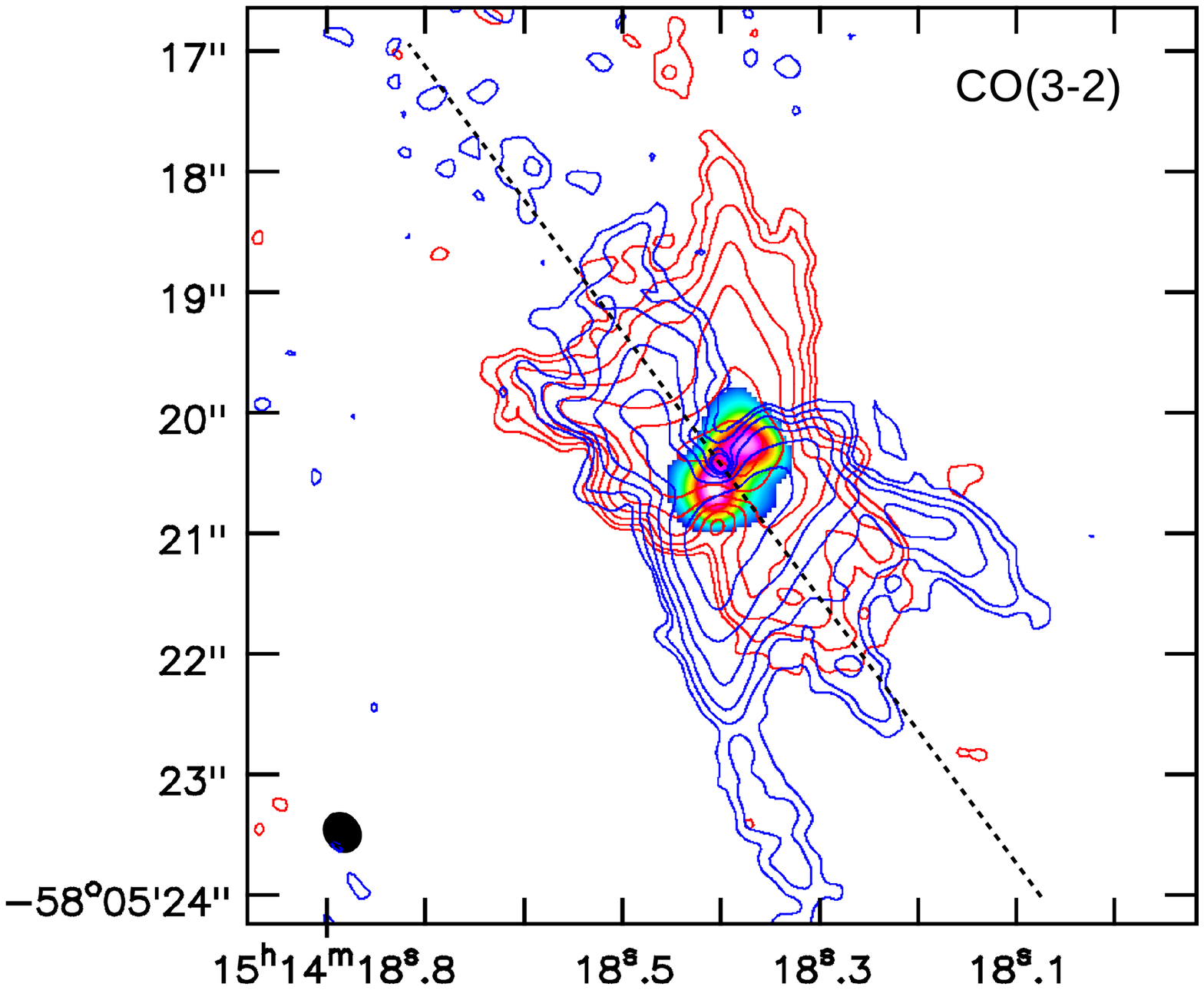}
	\includegraphics*[width=0.45\textwidth]{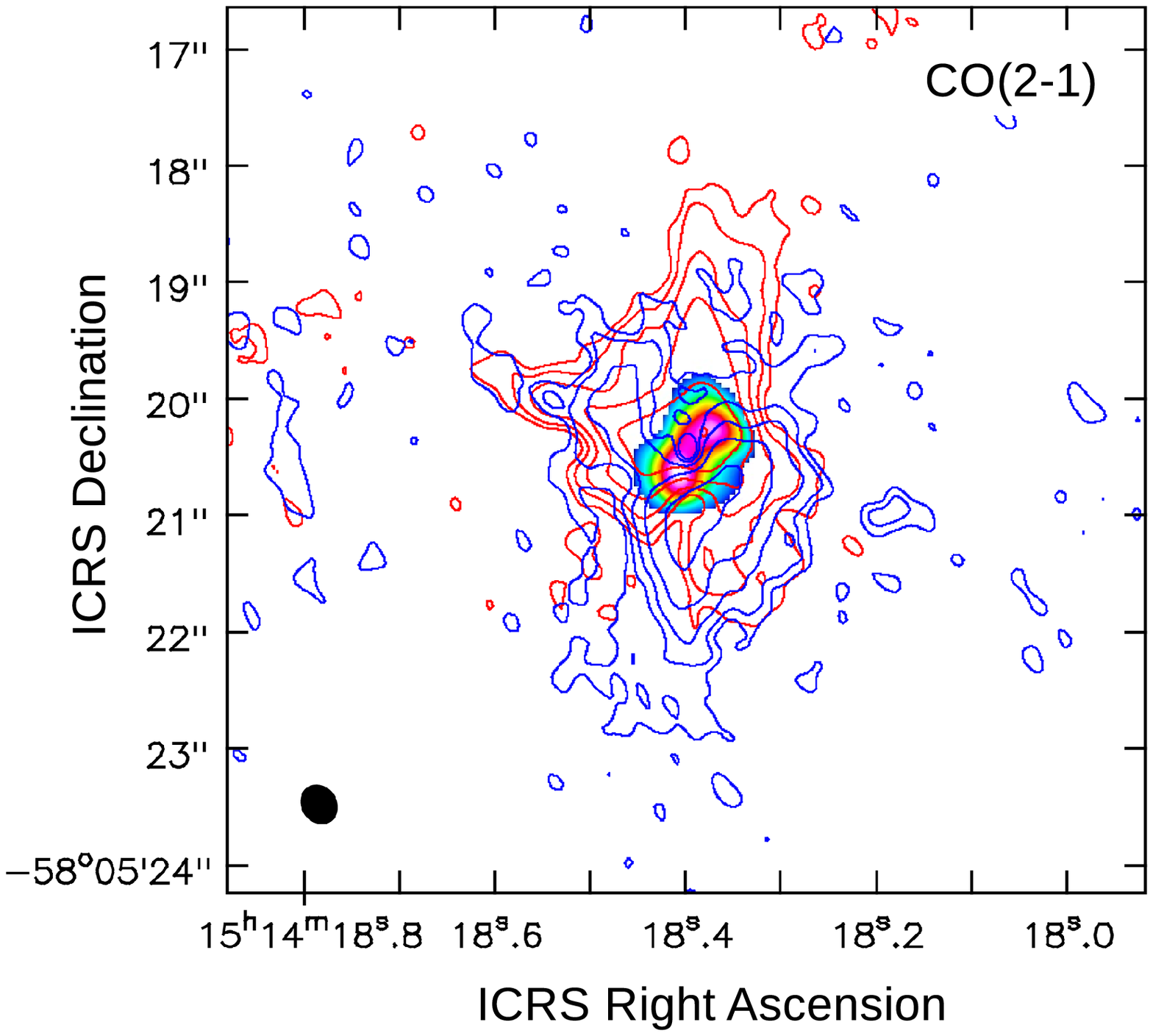}
	\caption{Top: map of high-velocity traced by CO(3-2) (redshifted and blueshifted gas, moving $>30$ km s$^{-1}$ from the central velocity of $-33$ km s$^{-1}$). Contour levels are $2^n\sigma$, starting with $n=1$ and increment step $n=1$. The dashed line traces the direction of the outflow axis. Bottom: the same as the top panel, for CO(2-1) emission. The $1\sigma$ rms is 60 mJy beam$^{-1}$ km s$^{-1}$ in both maps. The colour image in both panels is the integrated emission map of C$^{18}$O(2-1), tracing the torus.}
	\label{fig:CO_outflow}
\end{figure}

A position velocity diagram along this main outflow axis, p.a. $=36^\circ$ (Fig. \ref{fig:PVCOoutflow}), clearly shows a very-high-velocity outflow, extending up to at least $\simeq 200$ km s$^{-1}$ from the central velocity, with the most extreme blueshifted and redshifted gas to the southwest and northeast, respectively. The velocity of the outflow increases linearly with the spatial offset. A similar linear trend is seen in an opposite outflow (blueshifted and redshifted size to the  northeast and southwest, respectively), but with lower velocities. The total projected extent of the outflow is $\simeq 6$ arcsec ($\simeq 20000$ au at 3.38 kpc). It is interesting to note that the main linear trends (blueshifted gas to the southwest and redshifted gas to the northeast) are symmetrical with respect to a $V_{\rm LSRK}$ velocity $\simeq -41$ km s$^{-1}$, which does not { exactly} coincide with the central velocity of the torus ($\simeq -33$ km s$^{-1}$). Different PV diagrams, with a p.a. range $16^\circ-56^\circ$ also show mean outflow { velocities blueshifted ($V_{\rm LSRK}-40$ to $-46$ km s$^{-1}$) with respect to the central velocity of the torus. Although these estimates of velocity are uncertain, the central velocity of the torus in all observed transitions (including the optically thin C$^{18}$O line)
are systematically offset from that of the outflow along different position angles, suggesting that this velocity difference is real.}

\begin{figure}
	\includegraphics*[width=0.45\textwidth]{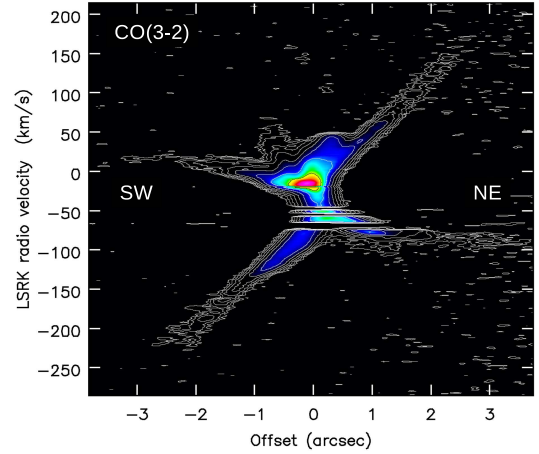}
	\includegraphics*[width=0.45\textwidth]{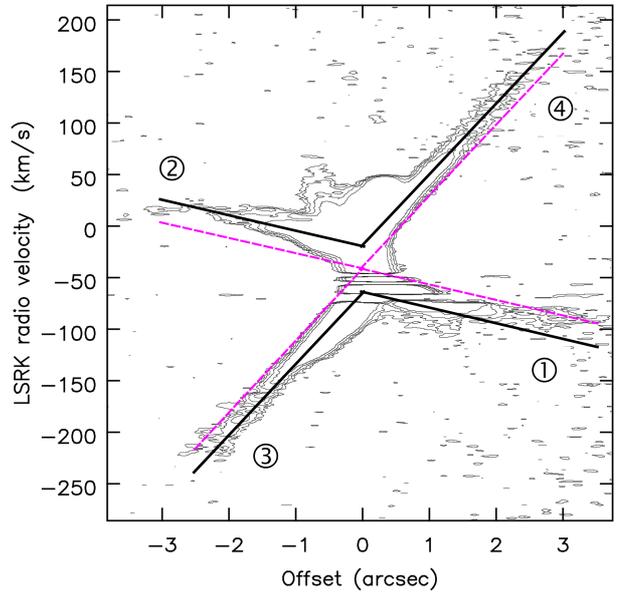}
	\caption{Top: PV diagram of CO(3-2)  along p.a. $=36^\circ$, the main axis of the outflow of Fig. \ref{fig:CO_outflow} (and marked with the dashed line in that figure). The offset increases in the same sense as right ascension (negative offsets are regions of lower right ascension). Contour levels are $2^n\sigma$, starting with $n=1$ and increment step $n=1$. The $1\sigma$ rms is 1.6 mJy beam$^{-1}$. Bottom: Position velocity diagram following the model presented in Appendix \ref{app:cone_model}, superimposed on the lowest contours of the PV diagram. The final adopted model is represented by
     black solid  lines. The magenta dashed lines
     line represent the position velocity relation in the case of $v_0 = 0$,
     as described by Eq.~(\ref{eq:vdotn}). The parts of the diagram corresponding to the the branches of the cones defined in Fig. \ref{fig:coneGeometry} are
     marked by circled numbers. }
	\label{fig:PVCOoutflow}
\end{figure}

Maps of CO(3-2) integrated intensity (Fig. \ref{fig:CO_ranges} in Appendix \ref{app:range_maps}) over different velocity ranges can shed some light on the underlying outflow structure. At high velocities, the emission seems to form arcs (blueshifted to the SW and redshifted to the NW, as expected from the toroidal orientation), which are farther away from the source at increasingly higher velocities. At lower velocities, there are also gas streams inside the opening of the arcs, but with opposite velocity sign. We can explain most of these trends with an outflow along the walls of two cones, with the velocity increasing linearly with distance to the central star.

We have modelled the { outflow geometry and kinematics} as arising from the walls of a biconical structure, in order to reproduce the position-velocity diagram in Fig. \ref{fig:PVCOoutflow}. 
Our model is explained in Appendix \ref{app:cone_model}. This model is able to reproduce the PV diagram with an opening angle of the cone ($2\times\theta_0$) of $\simeq 56^\circ$, and an inclination between the cone axis and the plane of the sky of $\simeq 16^\circ$ -- which is also offset by $\simeq 10^\circ$ from the inclination angle of the torus axis (see Section \ref{sec:toroid}), so the total 3-D misalignment is $\simeq 14^\circ$  -- with the axis of the SW cone closer to us. With this geometry, we estimate a maximum deprojected outflow velocity of $\simeq 250$ km s$^{-1}$. The velocity gradient along the conical surface is 73 km s$^{-1}$ arcsec$^{-1}$, equivalent to $\simeq 21.6$ m s$^{-1}$ au$^{-1}$ at a distance of 3.38 kpc. This gradient indicates a dynamical age of $\simeq 220$ yr for the outflow. Two additional characteristics of the biconical outflow model are of particular interest.  One is that the model confirms the systematic difference between the central velocities of the high-velocity outflow and the torus, as a velocity for the former of $\simeq -41$ km s$^{-1}$  is required to reproduce the observed diagrams. The second is that the straight lines in the PV diagram (Fig. \ref{fig:PVCOoutflow}) do not point to a single (central) velocity at spatial offset zero. The blueshifted and redshifted branches in the diagram intersect at different velocities. An additional equatorial velocity component $V_\circ=23$ km s$^{-1}$, perpendicular to the outflow axis, is necessary on both outflow sides to explain this feature in the PV diagram. 
This equatorial velocity coincides with the expansion velocity of the torus, and these could have the same origin, as explained in Section 4.

\subsubsection{A high-density outflow}

\label{sec:HCOoutflow}

The map of high-velocity HCO$^+$ emission ($> 30$ km s$^{-1}$ with respect to the central velocity)  in Fig.
 \ref{fig:HCO_outflow} also shows mixed redshifted and blueshifted gas on both sides of the source. However, 
 the general structure of the outflow traced by this molecule is different from that of the CO transitions, 
 with a more compact extension, and a different orientation (the HCO$^+$ outflow is nearly perpendicular to 
 the central torus). The HCO$^+$(4-3) has a very high critical density \citep[$1.8\times 10^6$ cm$^{-3}$,][]{jan95}, 
 which is several orders of magnitude higher than that of the CO transitions 
 ($2.7\times 10^3$ and $8.4\times 10^3$ cm$^{-3}$ for the CO(2-1) and (3-2) transitions, respectively). This 
 means that the HCO$^+$ transition specifically traces higher-density gas and that a significant fraction of the CO outflow is below the critical density of HCO$^+$.

\begin{figure}
	\includegraphics*[width=0.45\textwidth]{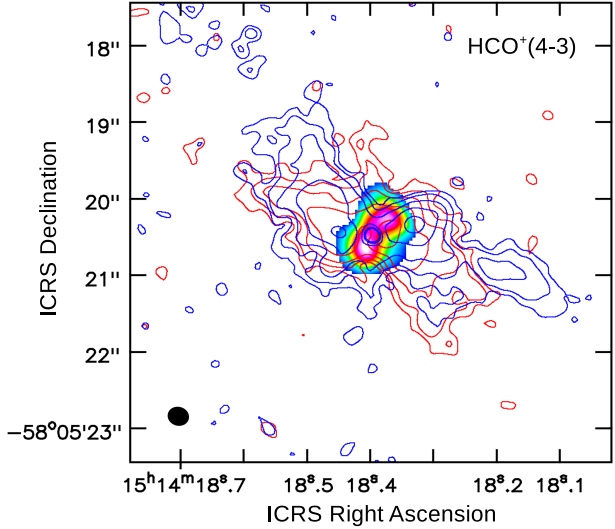}
	\caption{Map of high-velocity traced by HCO$^+$(4-3) (redshifted and blueshifted gas, moving $>30$ km s$^{-1}$ from the central velocity). Contour levels are $2^n\sigma$, starting with $n=1$ and increment step $n=1$. The $1\sigma$ rms is 65 mJy beam$^{-1}$ km s$^{-1}$. The colour image is the integrated emission map of C$^{18}$O(2-1), tracing the torus.}
	\label{fig:HCO_outflow}
\end{figure}

In the maps of integrated intensity over different velocity ranges (Fig. \ref{fig:HCO_ranges} in Appendix \ref{app:range_maps}), we can identify some of the trends of the CO outflow. However, the high-velocity arcs are not so evident. The outflow is dominated at lower velocity (30-70 km s$^{-1}$) by long filaments (blue to the NE and red to the SW), which are similar to the ones seen in the CO outflow at the same velocities. At higher velocities ($> 80$ km s$^{-1}$), the outflow is oriented nearly perpendicular to the torus, and it seems to trace an outflow different from the biconical outflow described above. Assuming the axis of this high-density outflow coincides with the symmetry axis of the torus, we estimate a dynamical age $\ga 200$ yr, thus compatible with the estimated age of the torus and the biconical CO outflow.

Because this HCO$^+$ transition specifically traces dense gas, it would miss the lower-density areas of the outflow traced by CO. { Thus, we would expect the mass of the outflow traced by HCO$^+$ to be lower than that of CO (0.03-0.10 M$_\odot$, sec. \ref{sec:co_outf}). Our results are consistent with this expectation, although the mass estimates are not very accurate, due to the uncertainties in the molecular abundances and excitation temperatures. We obtain an estimate of 
0.02-0.04 M$_\odot$ for the HCO$^+$ outflow, assuming excitation temperatures 60-180 K and a molecular abundance of HCO$^+$ with respect to H$_2$ of $\simeq 10^{-7}$, which is the main value obtained by \cite{sch16} in a sample of PNe. However, they obtained abundances with a large dispersion, with values up to $\simeq 7.4\times 10^{-7}$. This higher molecular abundance would give an HCO$^+$ outflow mass $\simeq 0.003-0.005$ M$_\odot$.}

\subsection{Comparison of molecular structures with the infrared nebula}

Fig. \ref{fig:neii} shows the comparison between a mid-infrared image of the source  (in the [Ne{\sc ii}] filter centred at 12.2 $\mu$m, \citealt{lag11})
and the three different molecular structures presented above. Because there is no precise astrometrical information in the infrared image, we have aligned it with our mm/submm data assuming that the centre of the torus coincides with the waist at the centre of the infrared nebula.

\begin{figure}
	\includegraphics*[width=0.45\textwidth]{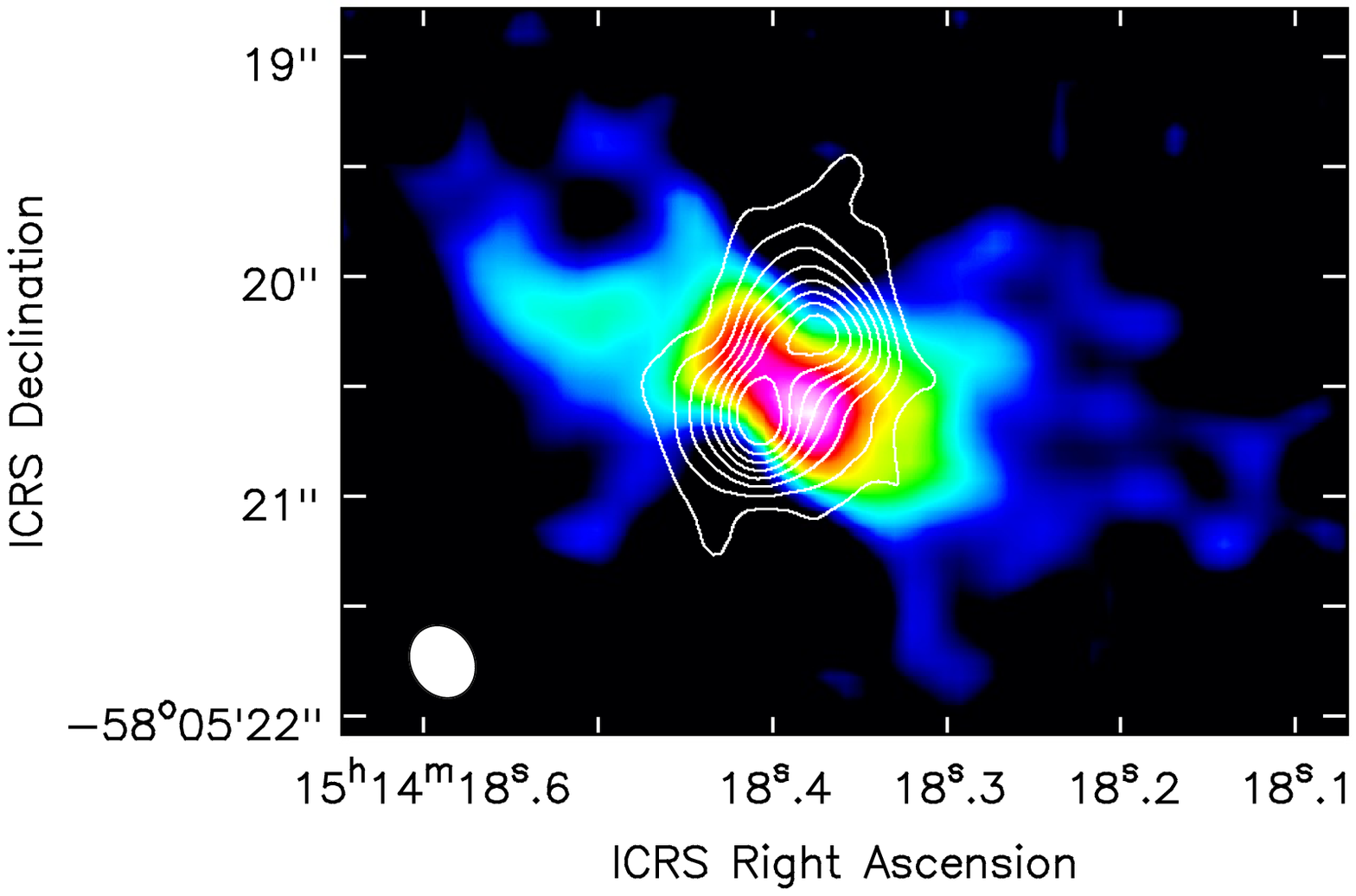}\\
	
	\includegraphics*[width=0.45\textwidth]{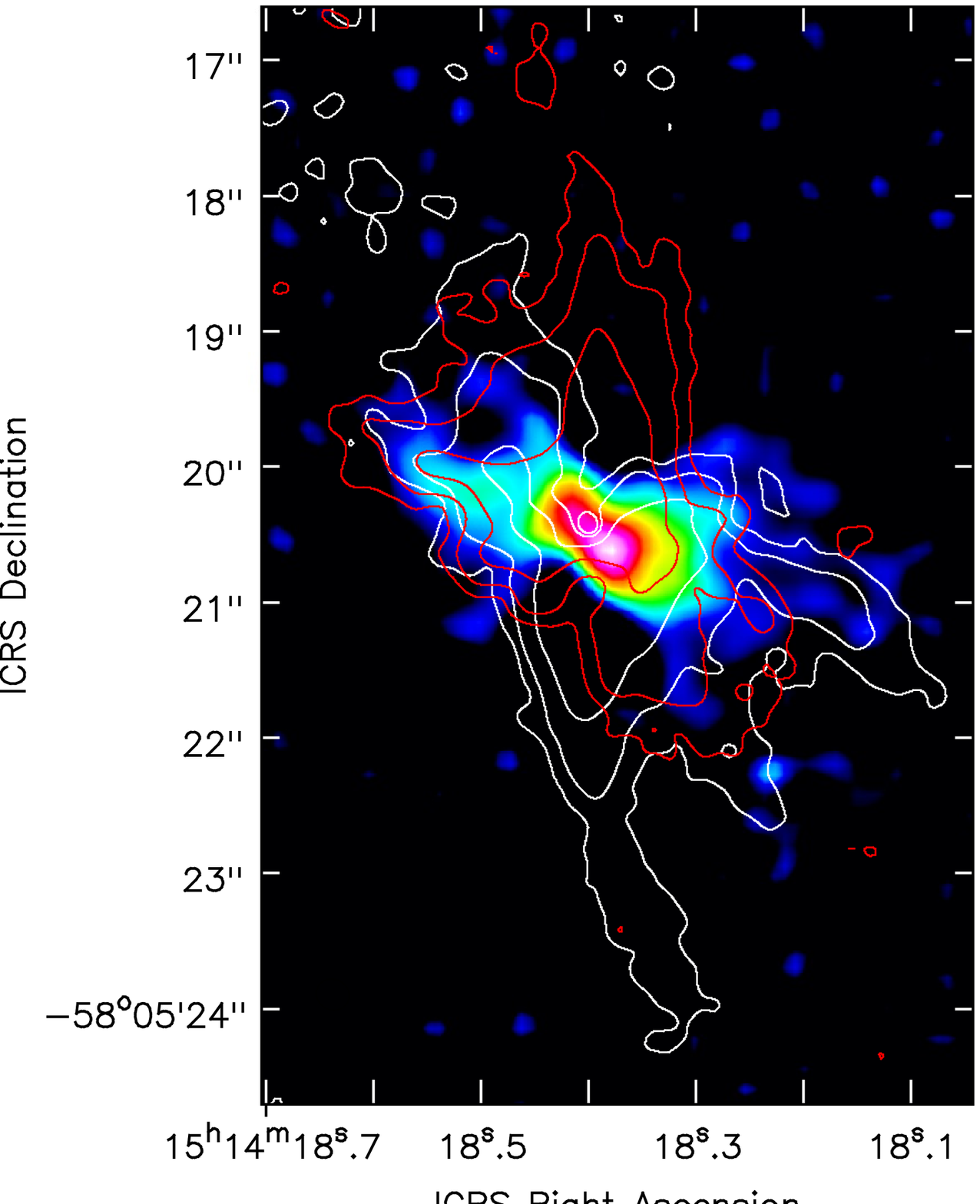}\\

	\includegraphics*[width=0.45\textwidth]{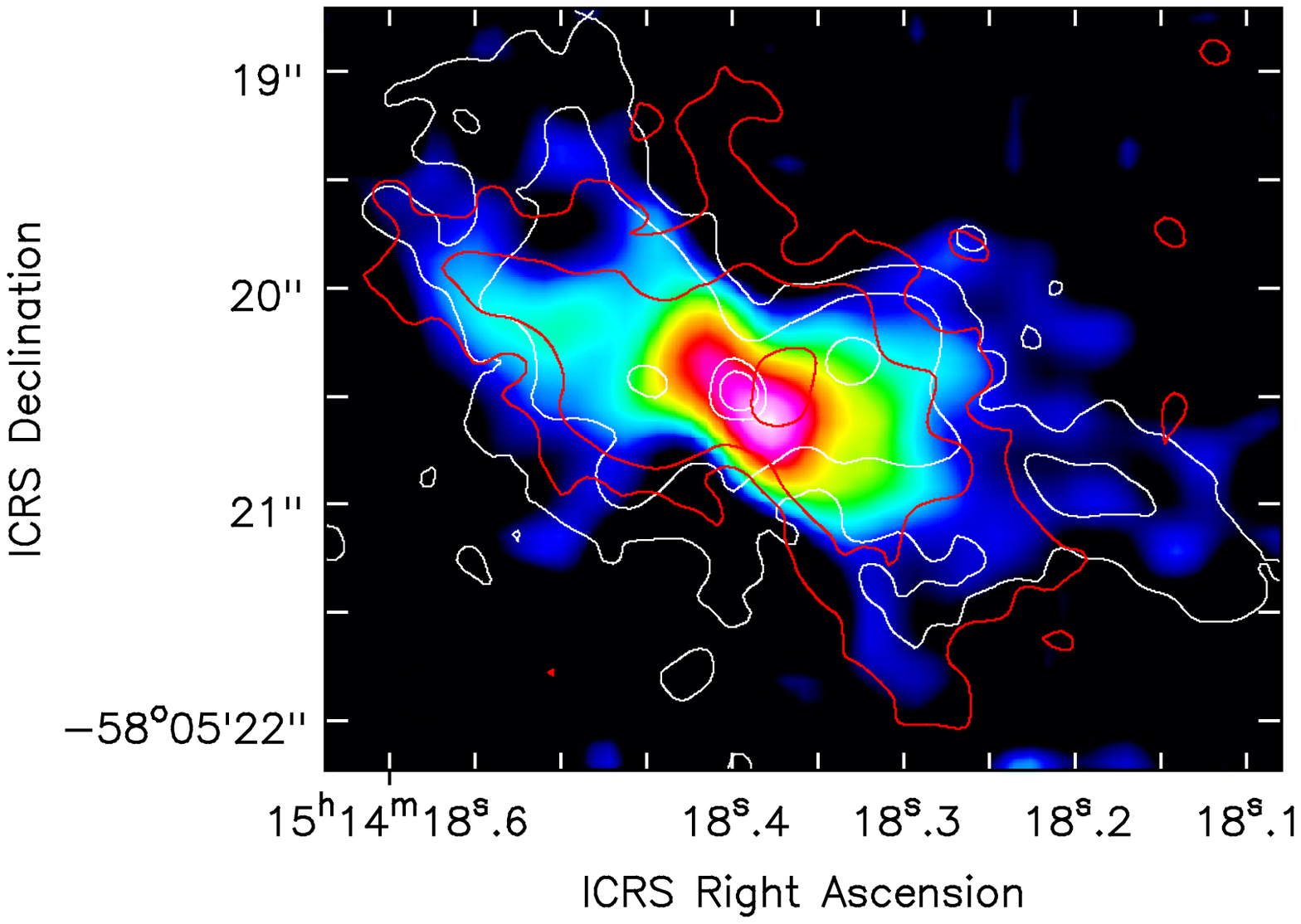}
	\caption{Overlay of the mid-infrared image at 12.2 $\mu$m \citep{lag11} with different molecular structures. Top: Contour map of the C$^{18}$O integrated emission. Contour levels are the same as in Fig. \ref{fig:torus_mom0}. Middle: Contour map of integrated emission of CO(3-2), as in Fig. \ref{fig:CO_outflow}, but with contour levels are 2, 8 and 32 times 60 mJy beam$^{-1}$ km s$^{-1}$). Bottom: Contour map of integrated emission of HCO$^+$(4-3),  as in Fig. \ref{fig:HCO_outflow}, with contour levels 2, 8 and 32 times 65 mJy beam$^{-1}$ km s$^{-1}$. In the middle and bottom panels, red and white contours represent redshifted and blueshifted gas, respectively.}
	\label{fig:neii}
\end{figure}


The expanding torus (top panel in Fig. \ref{fig:neii}) is perpendicular to the main axis of the brightest, inner part of the infrared nebula. The high-density outflow traced by the HCO$^+$ emission is well correlated with the mid-infrared emission (bottom panel in Fig. \ref{fig:neii}), with a similar extension and orientation. We note that the distribution of water masers observed in 2011 \citep{gom15} also had a similar alignment (p.a. $=56^\circ$), although it has significantly changed in the recent years 
\citep[p.a. $\simeq 180^\circ$ in 2015-2016,][]{gom18}. This can be interpreted as a possible change in the radio continuum background, which is being amplified in the maser process.

Near-infrared images of IRAS 15103-5754 \citep{ram12} show an extended structure of $8.4\times 6.3$ arcsec, with a p.a. $\simeq 50^\circ$. The orientation of this extended nebula is consistent with the torus axes and the HCO$^+$ outflow. 
The relationship of high-velocity CO outflow with the infrared nebula (middle panel in Fig. \ref{fig:neii}) is not that obvious, as it is misaligned with the mid- and near-infrared images, although the outflow extent is contained within the latter. Thus, the conical CO outflow represents a mass-loss episode that has not yet influenced the shape of the nebula at infrared wavelengths.

\section{Discussion}

IRAS 15103-5754 is a very young PN that shows a complex system of mass-loss processes. Our ALMA observations, with high sensitivity and angular resolution can shed some light on the physical processes taking place in this object. Fig. \ref{fig:toy} shows a schematic diagram of some of the main features we have observed.

\begin{figure}
	\includegraphics*[width=0.48\textwidth]{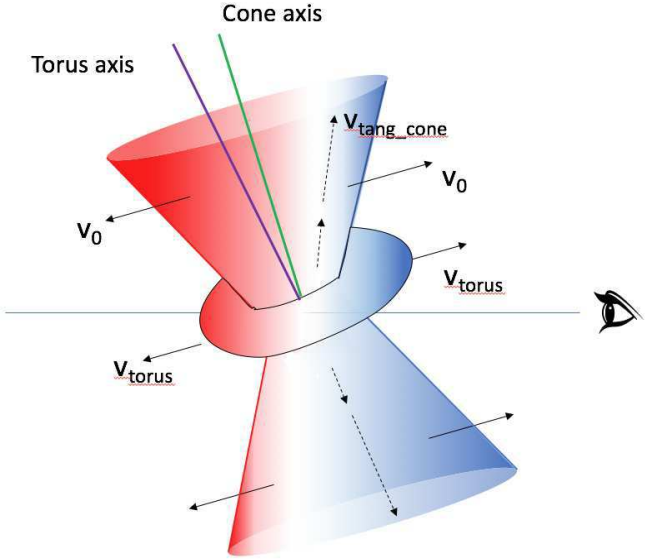}\\
	\caption{Representation of the proposed geometry in IRAS 15103-5754. The drawing represents a torus expanding with a velocity $V_{\rm torus}$, and a biconical outflow, with a velocity along the walls ($V_{\rm tang\mbox{\_} cone}$) linearly increasing with distance from the centre, and with an additional equatorial component ($V_\circ$). We note that ($V_\circ$) is not the equatorial component of $V_{\rm tang\mbox{\_} cone}$, but an additional constant component that would result in a separation of the walls of the cone. The axis of the torus and the conical outflow are misaligned. The high-density HCO$^+$ outflow (not drawn) would be aligned with the torus axis.}
	\label{fig:toy}
\end{figure}

The molecular content of the nebula is dominated by a torus of $\simeq 0.7-1.8$ M$_\odot$ and 1000 au radius, expanding at a velocity ($V_{\rm torus}$ in Fig. \ref{fig:toy}) of $\simeq 23$ km s$^{-1}$. The fact that most of the mass around the PN is distributed in a torus, is the expected outcome of a binary system that went through a common envelope phase \citep{iva13}, when the expanding envelope from the primary overfills the Roche lobe of both stars. In this situation, the envelope tends to be detached along the orbital plane \citep[e.g.,][]{san98,mor06,nor06,che17} and the binary system would lie at the centre of an expanding torus. We note that the two continuum sources we detected (Fig. \ref{fig:continuum}) cannot constitute the binary system that was the origin of  the toroidal mass loss, as source B is outside this torus, at a projected angular distance of $\simeq 0.39$ arcsec ($\ga 1300$ au) from its centre. Thus, source A is near the location of the binary that created the torus. It is possible that this continuum source is actually an unresolved close binary, or the secondary was engulfed by the primary (creating a single star) or threaded by tidal forces (forming a circumstellar disc around the primary). 

In this scenario, source B would be a third star. Thus, IRAS 15103-5754 is (now or at some point in the past) at least a triple system. Source B could either have originally been in a wide stable orbit of $\ga 1300$ au radius or be an escapee from a closer triple system, which was ejected during the common envelope evolution. The initial distribution of the triple system does have an impact on the subsequent evolution. For instance, \cite{sok16} discusses different possible evolutionary scenarios depending on the characteristics of the triple system, leading to different properties in the resulting planetary nebulae. 

We propose that a possible scenario fitting our observations of IRAS 15103-5754 is that of a triple system in which one of the components is ejected. Under this scenario, there is initially a tight binary system orbiting the star that evolves into the AGB. When the envelope reaches the tight binary, it breaks up, and the lower-mass component is ejected.  IRAS 15103-5754 meets several expectations from such a scenario, as presented by \cite{sok16}. In particular, we note the two slightly different symmetry axes. One axis is defined by the torus (Fig. \ref{fig:torus_mom0}), the inner part of the infrared nebula \citep[][and Fig. \ref{fig:neii}]{lag11}, and the high-density HCO$^+$ outflow (Fig. \ref{fig:HCO_outflow}). The other axis is the biconical CO outflow (Figs. \ref{fig:CO_outflow} and \ref{fig:CO_ranges}).  Moreover, there is a velocity shift between the central velocity of the torus and that of the CO outflow. This points to an initial mass loss from a central triple system and a second mass loss after one of the components is ejected. This explains the different axes and central velocities, since the main orbit and the barycentre of the central stellar system will change after ejection of a star. The final result will be a nebula that departs from point symmetry, as seen in the outer parts of the infrared nebula.

An interesting question is the physical process that is the origin of each outflow episode. As mentioned above, the expanding torus seems to be a natural outcome of a common envelope phase, which could originate when the envelope of the AGB star engulfed a tight binary system orbiting around it. The expanding torus represents an equatorial density enhancement, and any subsequent mass loss will proceed more easily in the direction along the poles of the torus. Thus, an isotopic wind from the central star(s) would show up as low-collimation bipolar flow oriented along this direction. The biconical CO outflow shows a low degree of  collimation, but it is not oriented exactly along the poles of the torus. So, a process with a certain degree of collimation must be at work. It is possible that, after ejection of the tertiary, the secondary creates a magnetic dynamo that produces an explosive, low-collimation outflow \citep{nor06}, perpendicular to the orbital axis of the remaining binary. This wind could rip off material from the torus, giving rise to a biconical outflow with an equatorial velocity component, as determined in our PV diagrams.

Considering that this source is a WF, but one that seems to be entering the PN phase, it is interesting to  compare our results with those for other WFs and young PNe. The expansion velocity  of the torus ($\simeq 23$ km s$^{-1}$) is similar to the ones derived for other WFs \citep{riz13}, while its relatively high mass ($\simeq 1$ M$_\odot$) is consistent with the mass needed to explain the spectral energy distribution of this type of source \citep{dur14}. However, the torus masses in water-maser-emitting PNe are significantly lower \citep[$<0.15$ M$_\odot$,][Uscanga et al. in preparation]{san12}. 
ALMA observations of the post-AGB WF  IRAS 16342-3814 \citep{sah17} show a central torus whose diameter and expansion velocity (1300 au and 20 km s$^{-1}$) are very similar to the ones found for IRAS 15103-5754. The outflow, however, has a more collimated, jet-like morphology in IRAS 16342-3814. This outflow could be similar to the HCO$^+$ outflow we see in IRAS 15103-5754. Although there is an extreme high-velocity outflow ($\simeq 310$ km s$^{-1}$) in IRAS 16342-3814 with an orientation different from the main outflow, there is no indication of a low-collimation outflow  as we detected with CO in IRAS 15103-5754. 

This low-collimation outflow in IRAS 15103-5754 has the kinematical signature of a biconical outflow with linearly increasing velocities away from the central star.  Conical structures and outflows can be seen in PNe \citep[e.g.,][]{kwo08,sah99} and post-AGB stars \citep{hug04,ima05,kon11,buj13}. However, the velocities of the molecular gas do not usually reach values as high as in IRAS 15103-5754 ($\simeq 250$ km s$^{-1}$), although the underlying winds can reach values of thousands of km s$^{-1}$ \citep{bor01,san01}. Actually, values that high in the molecular gas around post-AGB stars have only been observed in rare cases \citep[e.g.,][]{alc96,sah17}.
Here we are proposing that the conical outflow in IRAS 15103-5754 consists of molecular gas from the torus, stripped out by a wind, which can have much higher velocities. The velocity of the molecular gas in IRAS 15103-5754 is also higher than other cases of possible gas outflows from tori, as in the PN NGC 6302 \citep{san17} or the post-AGB star IRAS 08544-4431 \citep{buj18}. This indicates that the momentum transfer from the wind to the disc material must be extremely efficient in IRAS 15103-5754 or that we are witnessing an especially energetic outburst.

\section{Conclusions}

We presented continuum and molecular-line observations of the young PN IRAS 15103-5754 carried out with ALMA at 0.85 and 1.3 mm. Our main conclusions are as follows.
\begin{itemize}
\item We detect two continuum sources, separated by $\simeq 0.39\pm 0.03$ arcsec. A significant fraction of the emission has a spectral index $<2$, with a minimum of 0.6. This is not compatible with dust only, and suggests a significant contribution from free-free emission from ionized gas.
\item The line emission is dominated by a central torus of { 0.6 arcsec (2000 au at 3.38 kpc)} diameter and a mass { 0.4-1} M$_\odot$, expanding at $\simeq 23$ km s$^{-1}$.
\item There are at least two high-velocity outflows. The first one, mainly traced by CO, has the kinematical signature of a biconical outflow, with an opening angle of 56$^\circ$, and linearly increasing velocities with distance from the central star, reaching deprojected velocities up to 250 km s$^{-1}$ and a total extent of { 6 arcsec (20000 au)}. The axis of the outflow is misaligned by $\simeq 14^\circ$ with respect to the axis of the torus. Moreover, the central velocity of the outflow is shifted by $\simeq 8$ km s$^{-1}$ from the central velocity of the torus. 
\item A second, high-density outflow traced by HCO$^+$ is nearly perpendicular to the torus, and aligns well with the infrared nebula.
\item We propose that the characteristics of IRAS 15103-5754 could be due to a central triple stellar system, which went through a common envelope phase, leading to the formation of the circumstellar torus. One of the components was ejected in this process, changing the central velocity and orbital plane. A subsequent low-collimation wind stripped the gas from the torus, generating the conical outflow.
\item The extremely high velocity of the conical outflow indicates that the momentum transfer between the wind and the molecular gas was extremely efficient, or that the mass loss was very energetic.
\end{itemize}

\section*{Acknowledgements}

JFG and LFM are supported by  Ministerio de Econom\'{\i}a, Industria y Competitividad (MINECO, Spain) grants
AYA2014-57369-C3-3 and AYA2017-84390-C2-1-R
(co-funded by FEDER). JFG also acknowledges the support of Ministerio de Educaci\'on, Cultura y Deporte
(Spain), under the mobility program for senior scientists at
foreign universities and research centres, as well as the hospitality of Universit\'e C\^ote d'Azur/Laboratoire Lagrange during the preparation of this paper. J.R.R. is supported by grant ESP2015-65597-C4-1-R (MINECO/FEDER). LU acknowledges support from  the University of Guanajuato (Mexico) grant ID CIIC 17/2018, and from PRODEP (Mexico). 
This paper makes use of the following ALMA data: ADS/JAO.ALMA\#2015.1.00777.S. ALMA is a partnership of ESO (representing its member states), NSF (USA) and NINS (Japan), together with NRC (Canada) and NSC and ASIAA (Taiwan) and KASI (Republic of Korea), in cooperation with the Republic of Chile. The Joint ALMA Observatory is operated by ESO, AUI/NRAO and NAOJ. 





\begin{thebibliography}{99}
\bibitem[\protect\citeauthoryear{Alcolea, Bujarrabal, \& Sanchez Contreras}{1996}]{alc96} Alcolea J., Bujarrabal V., Sanchez Contreras C., 1996, A\&A, 312, 560 
\bibitem[\protect\citeauthoryear{Beckwith, Henning, \& Nakagawa}{Beckwith et al.}{2000}]{bec00} Beckwith, S.~V.~W., Henning, T., \& Nakagawa, Y.\ 2000,  in Mannings V., Boss
A. P., Russell S. S., eds, Protostars and Planets IV. Univ. Arizona Press,
Tucson, AZ, p. 533
\bibitem[\protect\citeauthoryear{Berestetskii, Lifshitz, \& Pitaevskii}{Berestetskii et al.}{1982}]{landau}Berestetskii V.~B., Lifshitz E.~M., Pitaevskii L.~P., 1982, in Quantum Electrodynamics. Pergamon Press, Oxford
\bibitem[\protect\citeauthoryear{Bl\"ocker}{1995}]{blo95} Bl\"ocker T., 1995, A\&A, 299, 755 
\bibitem[\protect\citeauthoryear{Bobrowsky et al.}{1998}]{bob98} Bobrowsky M., Sahu K.~C., Parthasarathy M., Garc{\'{\i}}a-Lario P., 1998, Natur, 392, 469 
\bibitem[\protect\citeauthoryear{Borkowski \& Harrington}{2001}]{bor01} Borkowski K.~J., Harrington J.~P., 2001, ApJ, 550, 778 
\bibitem[\protect\citeauthoryear{Bujarrabal et al.}{2001}]{buj01} Bujarrabal V., Castro-Carrizo A., Alcolea J., S{\'a}nchez Contreras C., 2001, A\&A, 377, 868 
\bibitem[\protect\citeauthoryear{Bujarrabal et al.}{2013}]{buj13} Bujarrabal V., Castro-Carrizo A., Alcolea J., Van Winckel H., S{\'a}nchez Contreras C., Santander-Garc{\'{\i}}a M., Neri R., Lucas R., 2013, A\&A, 557, L11 
\bibitem[\protect\citeauthoryear{Bujarrabal et al.}{2018}]{buj18} Bujarrabal V., Castro-Carrizo A., Van Winckel H., Alcolea J., Sanchez Contreras C., Santander-Garcia M., Hillen M., 2018, A\&A, 614, A58
\bibitem[\protect\citeauthoryear{Casassus et al.}{2007}]{cas07} Casassus S., Nyman L.-{\AA}., Dickinson C., Pearson T.~J., 2007, MNRAS, 382, 1607 
\bibitem[\protect\citeauthoryear{Chen et al.}{2017}]{che17} Chen Z., Frank A., Blackman E.~G., Nordhaus J., Carroll-Nellenback J., 2017, MNRAS, 468, 4465 
\bibitem[\protect\citeauthoryear{de Gregorio-Monsalvo et al.}{2004}]{deg04} de Gregorio-Monsalvo I., G{\'o}mez Y., Anglada G., Cesaroni R., Miranda L.~F., G{\'o}mez J.~F., Torrelles J.~M., 2004, ApJ, 601, 921 
\bibitem[\protect\citeauthoryear{Desmurs}{2012}]{des12} Desmurs, J.-F.\ 2012, in Booth R.~S.,  Humphreys E.~M.~L., \& Vlemmings W.~H.~T., eds, 
Proc. IAU Symp. 287, Cosmic Masers from OH to H0.  
Cambridge Univ. Press, Cambridge, p. 217
\bibitem[\protect\citeauthoryear{Draine}{2003}]{dra03} Draine B.~T., 2003, ARA\&A, 41, 241 
\bibitem[\protect\citeauthoryear{Draine \& Lee}{1984}]{dra84} Draine B.~T., Lee H.~M., 1984, ApJ, 285, 89 
\bibitem[\protect\citeauthoryear{Duran-Rojas et al.}{2014}]{dur14} Duran-Rojas M., et al., 2014, 
in Morisset C., Delgado-Inglada G., Torres-Peimbert S., eds, Asymmetrical Planetary Nebulae VI, 19
\bibitem[\protect\citeauthoryear{G{\'o}mez et al.}{2008}]{gom08} G{\'o}mez J.~F., Su{\'a}rez O., G{\'o}mez Y., Miranda L.~F., Torrelles J.~M., Anglada G., Morata {\'O}., 2008, AJ, 135, 2074 
\bibitem[\protect\citeauthoryear{G{\'o}mez et al.}{2015}]{gom15} G{\'o}mez J.~F., et al., 2015, ApJ, 799, 186 
\bibitem[\protect\citeauthoryear{G{\'o}mez et al.}{2018}]{gom18} G{\'o}mez J.~F., Miranda L.~F., Uscanga L., Su{\'a}rez O., 2018, in Tarchi A., Reid M.~J., Castangia P., eds, Proc. IAU Symp. 336, Astrophysical Masers: Unlocking the Mysteries of the Universe. Cambridge University Press, p. 377
\bibitem[\protect\citeauthoryear{G{\'o}mez, Moran, \& Rodr{\'{\i}}guez}{1990}]{gom90} G{\'o}mez Y., Moran J.~M., Rodr{\'{\i}}guez L.~F., 1990, RMxAA, 20, 55 
\bibitem[\protect\citeauthoryear{Guerrero, Villaver, \& Manchado}{1998}]{gue98} Guerrero M.~A., Villaver E., Manchado A., 1998, ApJ, 507, 889 
\bibitem[\protect\citeauthoryear{Huggins et al.}{2004}]{hug04} Huggins P.~J., Muthu C., Bachiller R., Forveille T., Cox P., 2004, A\&A, 414, 581 
\bibitem[\protect\citeauthoryear{Imai}{2007}]{ima07} Imai, H.\ 2007,
in Chapman J.~M., Baan W.~A., eds, Proc. IAU Symp. 242,
Astrophysical Masers \& Their Environments. Cambridge Univ. Press,
Cambridge, p. 279
\bibitem[\protect\citeauthoryear{Imai et al.}{2005}]{ima05} Imai H., Nakashima J.-i., Diamond P.~J., Miyazaki A., Deguchi S., 2005, ApJ, 622, L125 
\bibitem[\protect\citeauthoryear{Ivanova et al.}{2013}]{iva13} Ivanova N., et al., 2013, A\&ARv, 21, 59 
\bibitem[\protect\citeauthoryear{Jansen}{1995}]{jan95} Jansen, D.~J., 1995, PhD thesis, Leiden Univ. 
\bibitem[\protect\citeauthoryear{Koning, Kwok, \& Steffen}{2011}]{kon11} Koning N., Kwok S., Steffen W., 2011, ApJ, 740, 27 
\bibitem[\protect\citeauthoryear{Kwok et al.}{2008}]{kwo08} Kwok S., Chong S.-N., Koning N., Hua T., Yan C.-H., 2008, ApJ, 689, 219-224 
\bibitem[\protect\citeauthoryear{Lagadec et al.}{2011}]{lag11} Lagadec E., et al., 2011, MNRAS, 417, 32 
\bibitem[\protect\citeauthoryear{Lewis}{1989}]{lew89} Lewis B.~M., 1989, ApJ, 338, 234 
\bibitem[\protect\citeauthoryear{Miranda et al.}{2001}]{mir01} Miranda L.~F., G{\'o}mez Y., Anglada G., Torrelles J.~M., 2001, Natur, 414, 284 
\bibitem[\protect\citeauthoryear{Morris \& Podsiadlowski}{2006}]{mor06} Morris T., Podsiadlowski P., 2006, MNRAS, 365, 2 
\bibitem[\protect\citeauthoryear{Nordhaus \& Blackman}{2006}]{nor06} Nordhaus J., Blackman E.~G., 2006, MNRAS, 370, 2004 
\bibitem[\protect\citeauthoryear{Olnon}{1975}]{oln75} Olnon F.~M., 1975, A\&A, 39, 217 
\bibitem[\protect\citeauthoryear{Panagia \& Felli}{1975}]{pan75} Panagia N., Felli M., 1975, A\&A, 39, 1 
\bibitem[\protect\citeauthoryear{Pickett et al.}{1998}]{pic98} Pickett H.~M., Poynter R.~L., Cohen E.~A., Delitsky M.~L., Pearson J.~C., M{\"u}ller H.~S.~P., 1998, JQSRT, 60, 883 
\bibitem[\protect\citeauthoryear{Ramos-Larios et al.}{2012}]{ram12} Ramos-Larios G., Guerrero M.~A., Su{\'a}rez O., Miranda L.~F., G{\'o}mez J.~F., 2012, A\&A, 545, A20 
\bibitem[\protect\citeauthoryear{Renedo et al.}{2010}]{ren10} Renedo I., Althaus L.~G., Miller Bertolami M.~M., Romero A.~D., C{\'o}rsico A.~H., Rohrmann R.~D., Garc{\'{\i}}a-Berro E., 2010, ApJ, 717, 183 
\bibitem[\protect\citeauthoryear{Rizzo et al.}{2013}]{riz13} Rizzo J.~R., G{\'o}mez J.~F., Miranda L.~F., Osorio M., Su{\'a}rez O., Dur{\'a}n-Rojas M.~C., 2013, A\&A, 560, A82 
\bibitem[\protect\citeauthoryear{Sabbadin et al.}{2008}]{sab08} Sabbadin F., Turatto M., Benetti S., Ragazzoni R., Cappellaro E., 2008, A\&A, 488, 225 
\bibitem[\protect\citeauthoryear{Sahai \& Trauger}{1998}]{sah98} Sahai R., Trauger J.~T., 1998, AJ, 116, 1357 
\bibitem[\protect\citeauthoryear{Sahai et al.}{1999}]{sah99} Sahai R., et al., 1999, AJ, 118, 468 
\bibitem[\protect\citeauthoryear{Sahai et al.}{2017}]{sah17} Sahai R., Vlemmings W.~H.~T., Gledhill T., S{\'a}nchez Contreras C., Lagadec E., Nyman L.-{\AA}, Quintana-Lacaci G., 2017, ApJ, 835, L13 
\bibitem[\protect\citeauthoryear{S{\'a}nchez Contreras \& Sahai}{2001}]{san01} S{\'a}nchez Contreras C., Sahai R., 2001, ApJ, 553, L173 
\bibitem[\protect\citeauthoryear{S{\'a}nchez Contreras \& Sahai}{2012}]{san12} S{\'a}nchez Contreras C., Sahai R., 2012, ApJS, 203, 16 
\bibitem[\protect\citeauthoryear{Sandquist et al.}{1998}]{san98} Sandquist E.~L., Taam R.~E., Chen X., Bodenheimer P., Burkert A., 1998, ApJ, 500, 909 
\bibitem[\protect\citeauthoryear{Santander-Garc{\'{\i}}a et al.}{2017}]{san17} Santander-Garc{\'{\i}}a M., Bujarrabal V., Alcolea J., Castro-Carrizo A., S{\'a}nchez Contreras C., Quintana-Lacaci G., Corradi R.~L.~M., Neri R., 2017, A\&A, 597, A27 
\bibitem[\protect\citeauthoryear{Schmidt \& Ziurys}{2016}]{sch16} Schmidt D.~R., Ziurys L.~M., 2016, ApJ, 817, 175 
\bibitem[\protect\citeauthoryear{Soker}{2016}]{sok16} Soker N., 2016, MNRAS, 455, 1584 
\bibitem[\protect\citeauthoryear{Su{\'a}rez et al.}{2015}]{sua15} Su{\'a}rez O., et al., 2015, ApJ, 806, 105 
\bibitem[\protect\citeauthoryear{Teyssier et al.}{2006}]{tey06} Teyssier D., Hernandez R., Bujarrabal V., Yoshida H., Phillips T.~G., 2006, A\&A, 450, 167 
\bibitem[\protect\citeauthoryear{Uscanga et al.}{2014}]{usc14} Uscanga L., G{\'o}mez J.~F., Miranda L.~F., Boumis P., Su{\'a}rez O., Torrelles J.~M., Anglada G., Tafoya D., 2014, MNRAS, 444, 217 
\bibitem[\protect\citeauthoryear{Vickers et al.}{2015}]{vic15} Vickers S.~B., Frew D.~J., Parker Q.~A., Boji{\v c}i{\'c} I.~S., 2015, MNRAS, 447, 1673 
\bibitem[\protect\citeauthoryear{Wilson}{1999}]{wil99} Wilson T.~L., 1999, RPPh, 62, 143 
\end{thebibliography}


\clearpage



\appendix

\section{Model of HCO$^+$ emission from an expanding torus}
\label{app:toroid_model}
In this appendix we describe our modelling of emission of the
HCO$^{+}$(4-3) line from an expanding torus. This includes the
formulation in which we based our model, the calculation procedure,
and our assumptions. Fig.~\ref{fig:toroidGeometry} describes the geometrical configuration of the torus, its related coordinate system ($x$, $y$, $z$, hereafter referred to as the model coordinates), and its relationship to the image plane coordinates $X$, $Y$.

\begin{figure}
  \centering
  \psfrag{Rt}{\small $\Rt$}
  \psfrag{rt}{\small $2\,\rt$}
  \psfrag{t}{\small $\theta$}
  \psfrag{x}{\small $x$}
  \psfrag{y}{\small $y$}
  \psfrag{z}{\small $z$}
  \psfrag{X}{\small $\hat{X}$}
  \psfrag{Y}{\small $\hat{Y}$}
  \psfrag{Z}{\small $Z$}
  \psfrag{n}{\small $\nhat$}
  \psfrag{line}{\tiny line of sight}
  \resizebox{8.5cm}{!}{\includegraphics{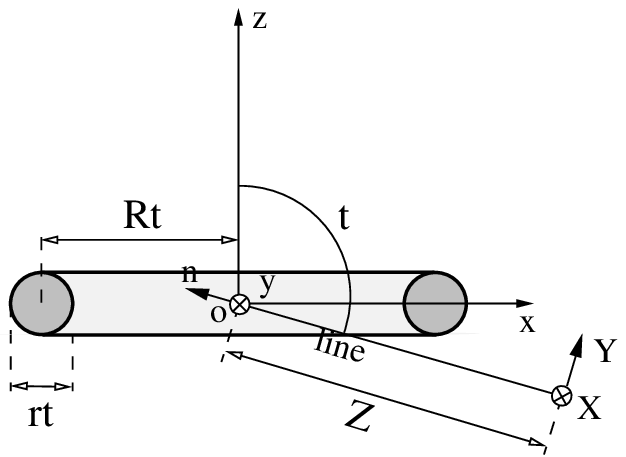}}
  \caption{Model geometry for the expanding torus. A $\otimes$ symbol
    represents a vector pointing into the page. $x, y, z$ are the
    model coordinates. The symmetry axis of the torus, whose centre
    is at the origin $O$, is aligned with the $Oz$ axis. The
    coordinates in the image plane are $X$ and $Y$. The image plane is
    the span of $\hat{X}$ and $\hat{Y}$ and is located at a distance
    $Z$ from the origin of the coordinates system $O$. $\nhat$ is the
    direction vector along the line of sight which is inclined by
    angle $\theta$ with respect to the symmetry axis of the torus}
  \label{fig:toroidGeometry}
\end{figure}

\subsection{Line emission}

A spectral line is characterized by its absorption coefficient ($\kappa_\nu$) and
emissivity ($\eta_{\nu}$), given by
\begin{eqnarray}
  \label{eq:extcoeff}
  \kappa_\nu &=& n_l\,\sigma_{ul}\,\Phi_{ul}(\nu) \\
  \label{eq:emissivity}
  \eta_{\nu} &=& \kappa_\nu\,B_\nu(\Tex) \ ,
\end{eqnarray}
where the indices $l$ and $u$ correspond to the lower and
upper energy levels, respectively. Here, $\sigma_{ul}$ is the absorption cross-section corrected
for stimulated emission, $\Phi_{ul}(\nu)$ the line profile, $\Tex$ is
the line excitation temperature, and $B_\nu$ is the Planck function. 

The condition of LTE allows us to
define a unique equilibrium temperature $T$ in which case we have
$\Tex = T$. Under LTE, the number density of molecules in the
lower state ($n_l$) can be written in the following manner
\begin{equation}
  \label{eq:nl}
  n_l = \Xmol\,\frac{\rho}{m_0}\,g_l\,\frac{\e^{-\frac{E_l}{k\,T}}}{Z(T)} \ ,
\end{equation} 
where $g_l$ is the statistical weight of the lower level, $E_l$ is its
energy, $\Xmol$ is the particle abundance of the considered molecule, $m_0$ is
the mean particle mass taken to be $2.4$ the hydrogen atom
mass\footnote{This corresponds roughly to $\mathrm{H}_2$,
  $\mathrm{He}$ and heavier elements mass fractions of $0.7$, $0.28$ and
  $0.02$ respectively.}, $\rho$ is the density of the medium and $Z(T)$ is
the partition function of the molecule at temperature $T$.

The absorption cross-section is given by
\begin{equation}
  \label{eq:crossSection}
  \sigma_{ul} = \frac{c^2\,A_{ul}}{8\pi \nu^2_{ul}}\,\frac{g_u}{g_l}\,\left(1-\e^{-\frac{h\nu_{ul}}{k\Tex}}\right) \ ,
\end{equation}
where $c$ is the speed of light, $A_{ul}$ the Einstein coefficient for
spontaneous emission, $\nu_{ul}$ is the frequency of the transition, and $g_u$ the statistical weight of the upper level.

For a linear molecule, allowed rotational transitions are between states with angular momentum
quantum numbers $J+1$ and $J$. In this case, the expression
for $\sigma_{ul}$ is given by \citep*[c.f.][]{landau}
\begin{equation}
  \label{eq:crossSectionRot}
  \sigma_{ul} = \frac{8\pi^3}{3 \,c h^2}\,kT_{1} \,\frac{(J+1)^2}{2J+1}\,\mu^2\left(1-\e^{-\frac{h\nu_{ul}}{k\Tex}}\right) \ ,
\end{equation}
where $\mu$ is the dipole moment, $T_{1}$ defines the energy of the
first rotational level, with $E_J = \frac{1}{2}J(J+1)\,kT_{1}$. For the
transition $J = 4 \rightarrow 3$ of the $\HCOp$ molecule, we used
$T_1 = 4.280129\,\K$ and $\mu = 3.888\,\mathrm{debye}$
\citep{pic98}.

Using the relationship between velocity ($V$) and frequency, $V = \frac{c}{\nu_{ul}}\,(\nu_{ul}-\nu)$, 
the line profile $\Phi_{ul}(\nu)$ describing Doppler broadening due to
thermal motion and microturbulence can be expressed in terms of the velocity $V$, as
\begin{equation}
  \label{eq:lineProfile}
  \Phi_{ul}(V) = \frac{c}{\sqrt{\pi}\,b\,\nu_{ul}}\e^{-\frac{1}{b^2}\,\left(V - V_s\right)^2} \ ,
\end{equation}
where $V_s$ is the projection
  of the local macroscopic velocity of the gas on to the line of
  sight.  The linewidth $b$ is given by
\begin{equation}
  \label{eq:lineWidth}
  b^2 = \frac{2k\Tkin}{\mmol} + V_{\rm turb}^2 \ , 
\end{equation}
where $\Tkin$ is the kinetic temperature,  $V_{\rm turb}$  is the
  microturbulence velocity and $\mmol$ is the molecular mass.

\subsection{Intensity maps}

The model and image plane coordinates are defined as depicted in
Fig.~\ref{fig:toroidGeometry}. The line emission at an observation
frequency $\nu$ is obtained by integrating the radiative transfer
equation along rays originating from a particular point $(X, Y)$ in
the image plane and going into the medium. The equation of such a ray
is given by
\begin{equation}
  \label{eq:ray}
  \rr_s = \rr_0 + s\,\nhat \ ,
\end{equation}
where $s$ is the distance along the ray from the image plane and $\nhat$ is the unit vector along the line of sight.  

The point $\rr_0$, located on the image plane at a location $(X,Y)$, is given by
\begin{equation}
  \label{eq:ray0}
  \rr_0 = \left(
    \begin{array}{c}
      -Y\cos{\theta} + Z\sin{\theta} \\
      X \\
      Y\sin{\theta} + Z\cos{\theta}
    \end{array}\right) \ ,
\end{equation}
where $Z$ is the distance from the origin of the model coordinates
system to this plane and $\theta$ is the inclination of the line of
sight with respect to the $Oz$ axis (see
  Fig.~\ref{fig:toroidGeometry}). The unit vector along the line of
sight is
\begin{equation}
  \label{eq:nhat}
  \nhat = \left(
    \begin{array}{c}
      -\sin{\theta} \\
      0 \\ 
      -\cos{\theta}
    \end{array}\right)
\end{equation}

The specific intensity at coordinates $(X,Y)$ in the image plane is
then obtained from the integral form of the radiative transfer
equation
\begin{eqnarray}
  \label{eq:intensity}
  I_\nu(X, Y) &=& \int\limits_{0}^{\infty}\eta_\nu(\rr_s)\,\e^{-\tau_\nu(s)}\,ds  \ , \\
  \label{eq:tau}
 \tau_\nu(s) &=& \int\limits_{0}^{s}\kappa_\nu(\rr_{s'})\,ds' \ ,
\end{eqnarray}
where $\tau_\nu$ is the optical depth. 

If we introduce a new variable $\zeta_\nu$ defined as
\begin{equation}
  \label{eq:zeta}
  \zeta_\nu(s) = \int\limits_{0}^{s}\eta_\nu(\rr_{s'})\,\e^{-\tau_\nu(s')}\,ds' \ ,
\end{equation}
then the integral Eq.~(\ref{eq:intensity}) amounts to seeking for
$\zeta_\nu(\infty)$ while solving the following ordinary differential
equations
\begin{eqnarray}
  \label{eq:eq1}
  \frac{d\zeta_\nu}{ds} &=& \eta_\nu(\rr_s)\,\e^{-\tau_\nu(s)} \ , \\ 
  \label{eq:eq2}
  \frac{d\tau_\nu}{ds} &=& \kappa_\nu(\rr_s) \ ,
\end{eqnarray}
with the following initial conditions: $\zeta_\nu(0) = 0$ and
$\tau_\nu(0) = 0$.

If $\Delta X$ and $\Delta Y$ are the pixel $X$ and $Y$ sizes in the
image plane and $d$ is the distance to the object, an intensity map
expressed in Jy pixel$^{-1}$ is obtained by multiplying
$I_\nu(X, Y)$ by the factor $\frac{\Delta X \, \Delta Y}{d^2}$. Then,
the convolution with the reconstructed beam results in intensity maps
expressed in Jy beam$^{-1}$.

\subsection{Expanding torus}

The material is confined in a torus of mean radius $\Rt$ and transversal radius $\rt$ (Fig.~\ref{fig:toroidGeometry}), so the inner and outer radii of the torus are $R_t-r_t$ and $R_t+r_t$, respectively. The particle density in the torus
is assumed to be constant. The temperature varies as a power law
\begin{equation}
  \label{eq:torusTemperature}
  T(r) = \Tt \,\left(\frac{\Rt}{\rho}\right)^{\gammat} \ ,
\end{equation}
where $\rho = \sqrt{x^2 + y^2}$ is the radial distance in cylindrical
coordinates,  $\Tt$ is the temperature of the torus at 
$\rho = \Rt$, and $\gammat$ is the temperature gradient.

The velocity field, assumed to be a constant radial expansion throughout the
torus, is described by
\begin{equation}
  \label{eq:torusVelocity}
  \vv(r) = \vexp\,\rhohat \ .
\end{equation}

In order to integrate the radiative transfer
equation~(\ref{eq:intensity}), we compute the intersection of a ray
described by Eq.~(\ref{eq:ray}) with the torus by solving for $s$ in
the following quartic equation
\begin{equation}
  \label{eq:quartic}
  a_0 + a_1\,s + a_2\,s^2 + a_3 \, s^3 + s^4 = 0 \ ,
\end{equation}
with 
\begin{eqnarray}
  \label{eq:coeff1}
  a_0 &=& \left(r_0^2-\Rt^2-\rt^2\right)^2 + 4\,\Rt^2\,\left(r_0^2 - \rt^2\right)\\
  \label{eq:coeff2}
  a_1 &=& 4\,\vec{r}_0  . \nhat \, \left(r_0^2 - \Rt^2 - \rt^2\right) + 8\,z_0 \,n_z\,\Rt^2  \\
  \label{eq:coeff3}
  a_2 &=& 4\,\left(\vec{r}_0  . \nhat \right)^2 + 2\,\left(r_0^2 -\Rt^2 -\rt^2\right) + 4\,\Rt^2\,n_z^2 \\
  \label{eq:coeff4}
  a_3 &=& 4\,\vec{r}_0  . \nhat \ .
\end{eqnarray}


\clearpage

\section{Maps of high-velocity CO and HCO$^+$ emission at different velocity ranges}

\label{app:range_maps}

\begin{figure*}
	\includegraphics*[width=0.3\textwidth]{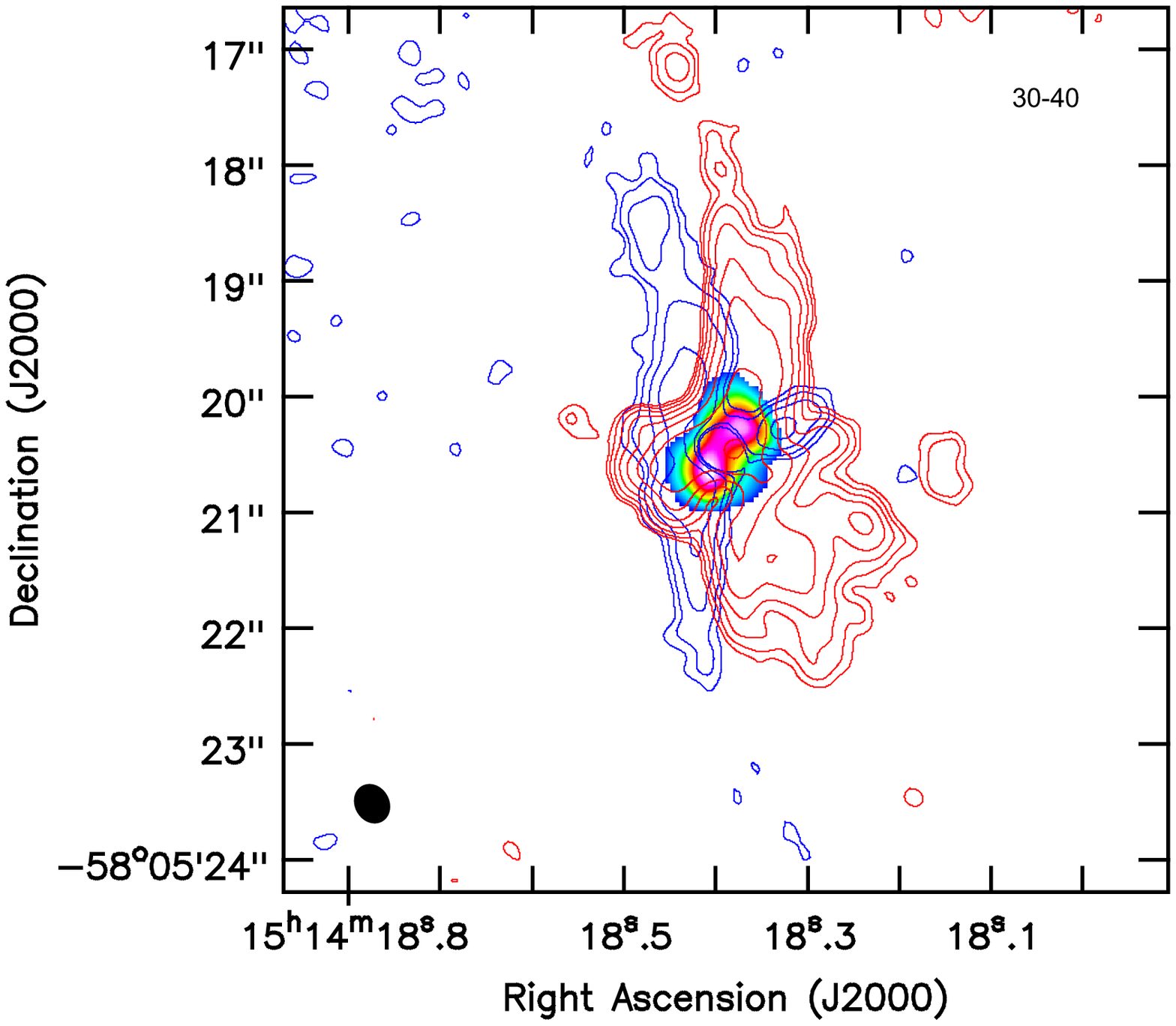}
	\includegraphics*[width=0.3\textwidth]{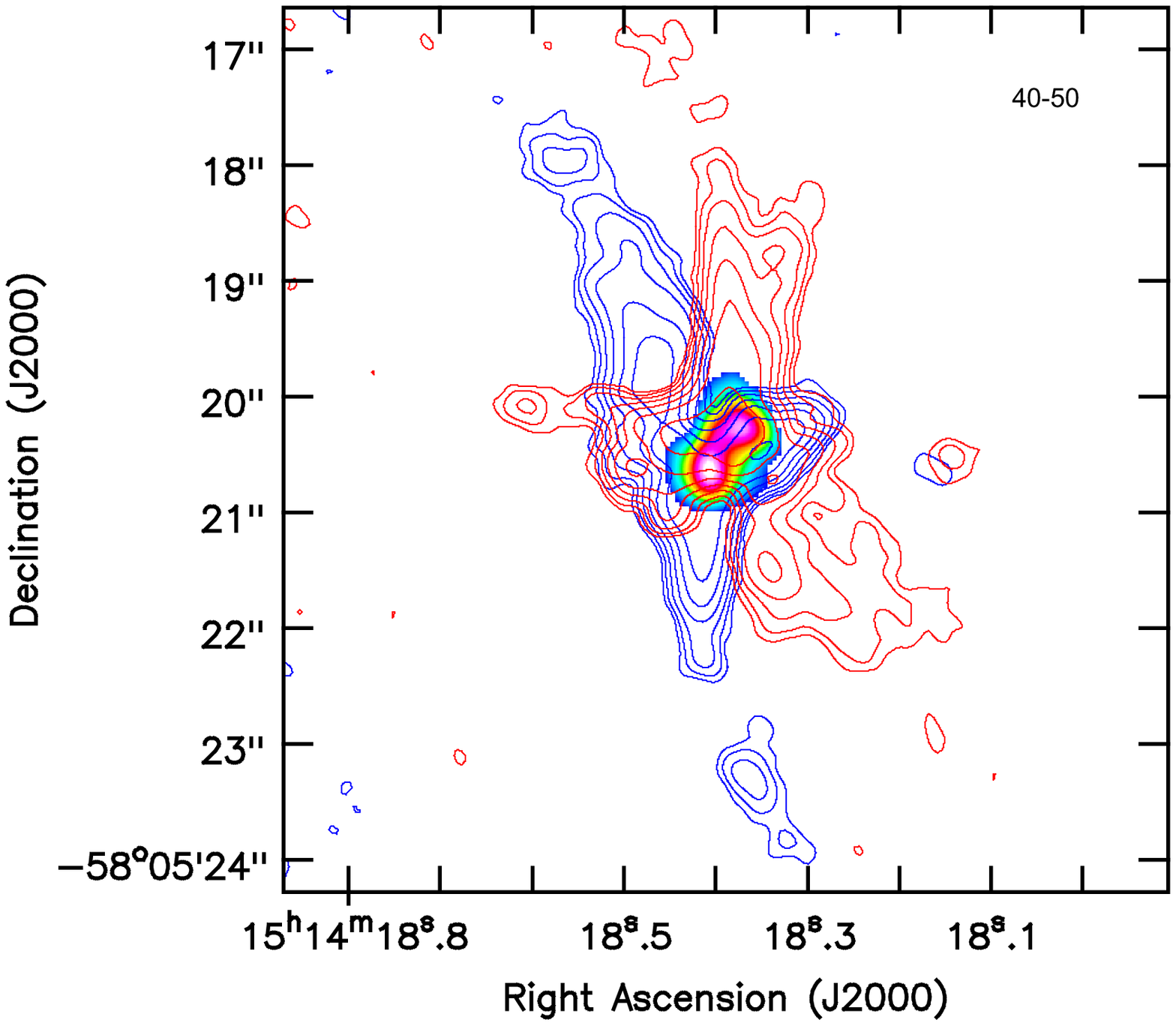}
	\includegraphics*[width=0.3\textwidth]{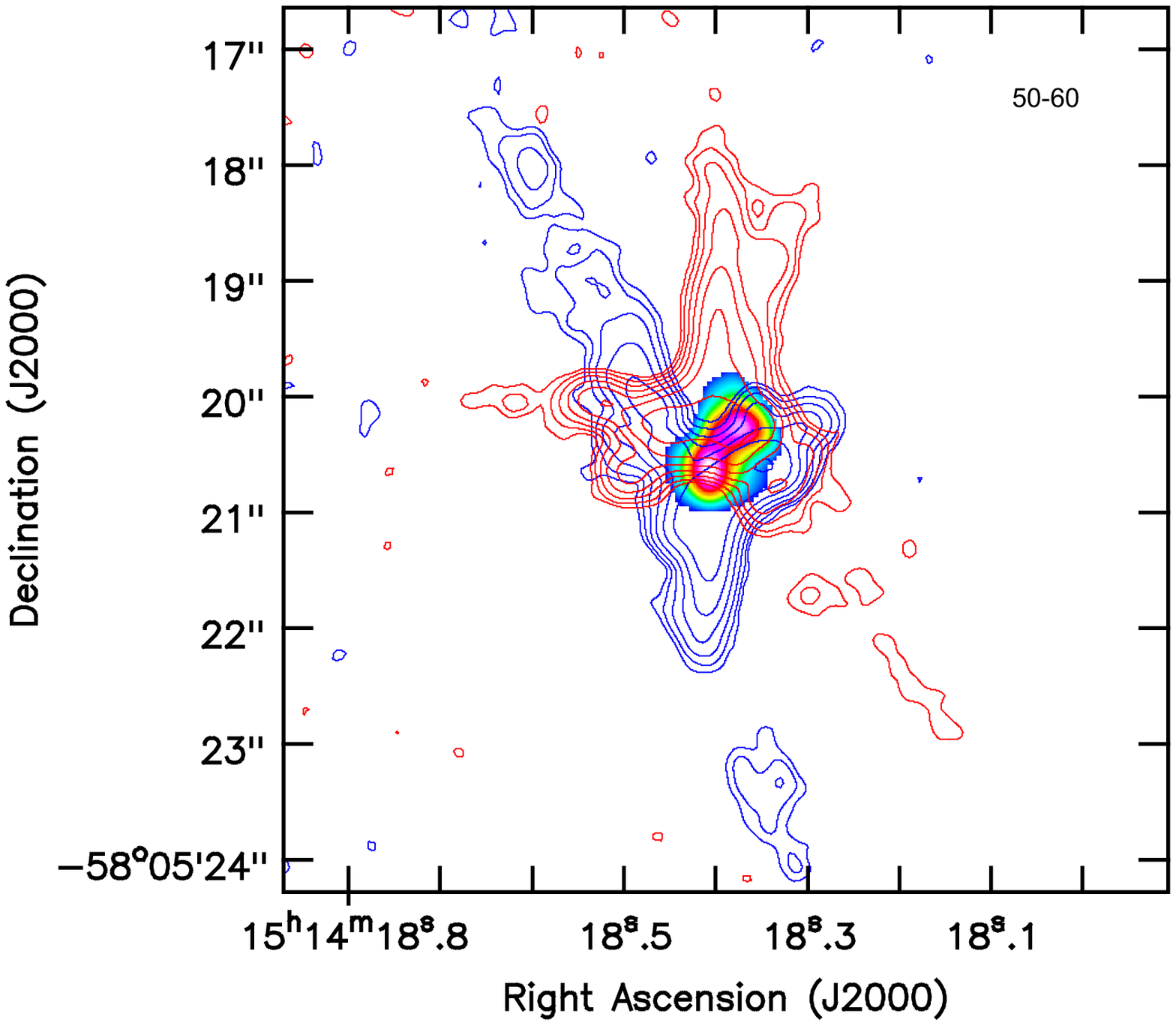}
	\includegraphics*[width=0.3\textwidth]{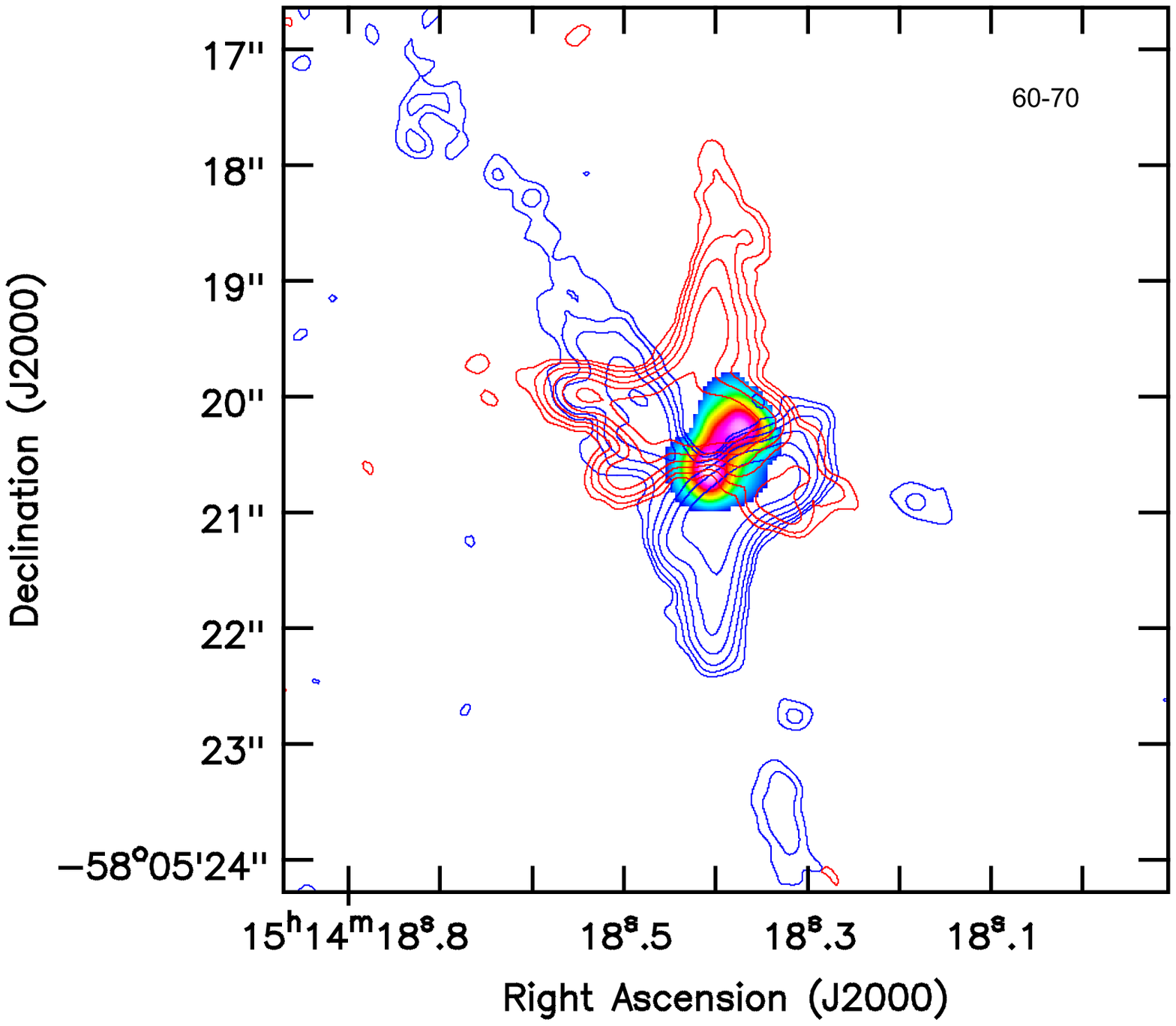}
	\includegraphics*[width=0.3\textwidth]{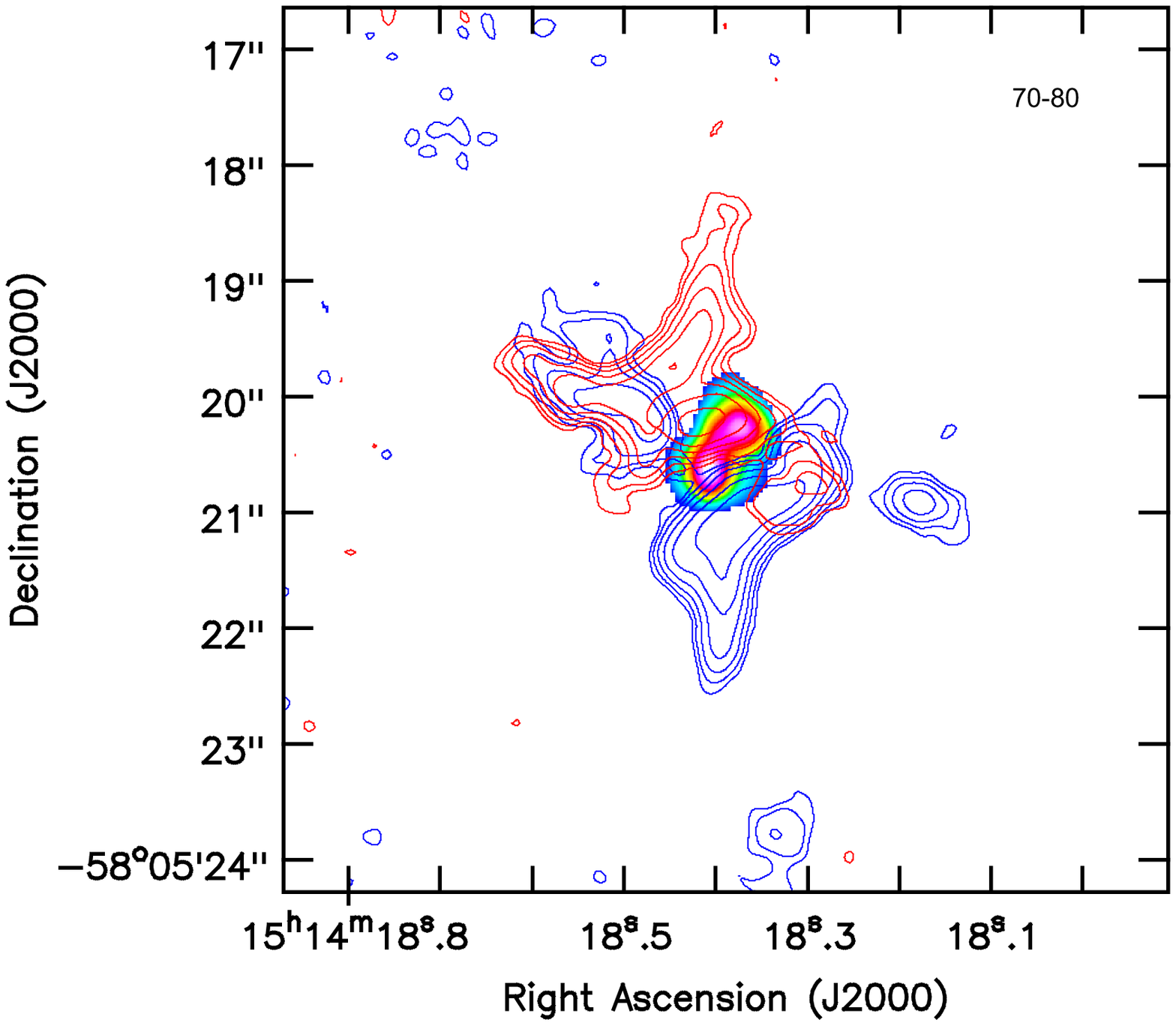}
	\includegraphics*[width=0.3\textwidth]{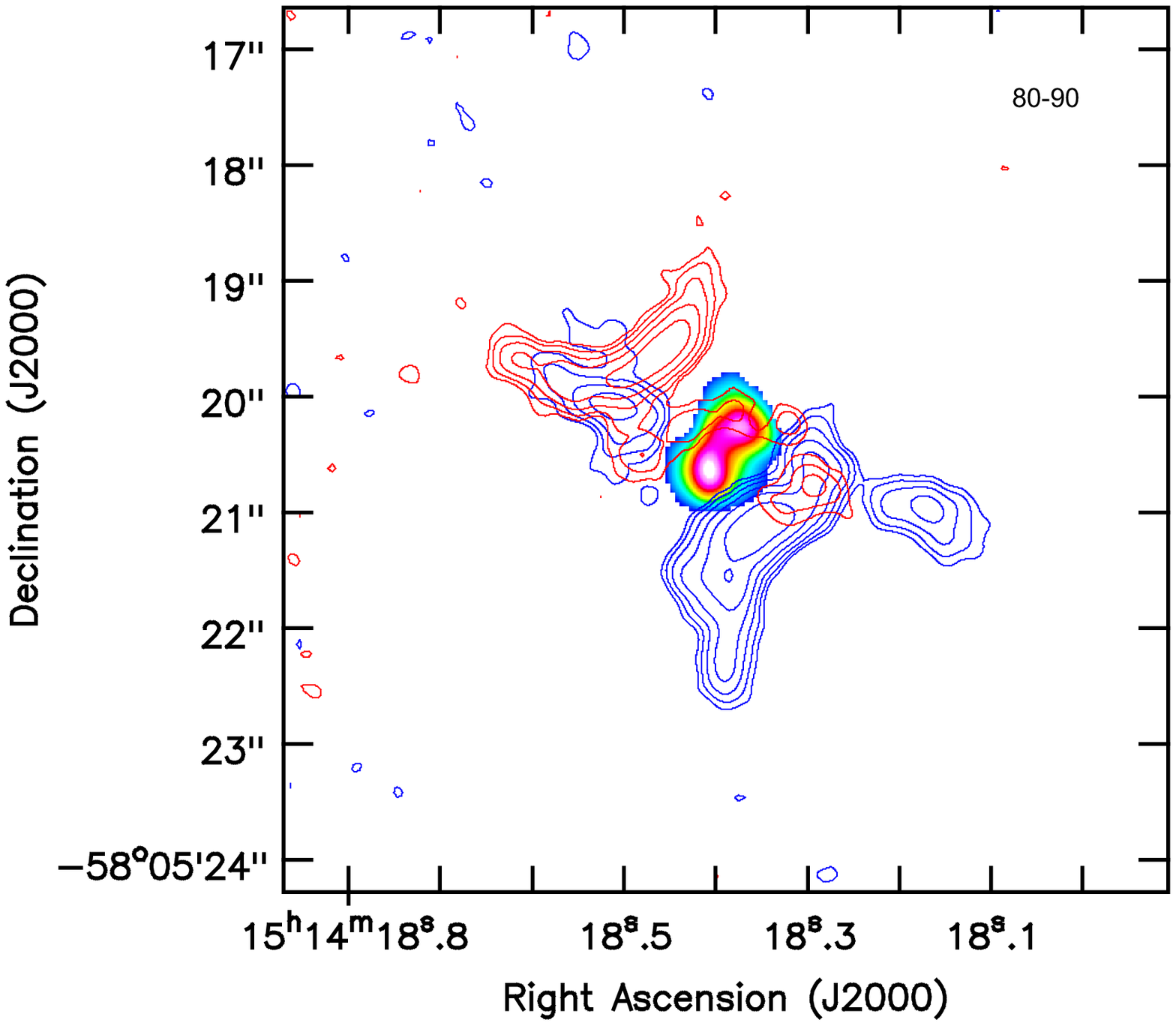}
	\includegraphics*[width=0.3\textwidth]{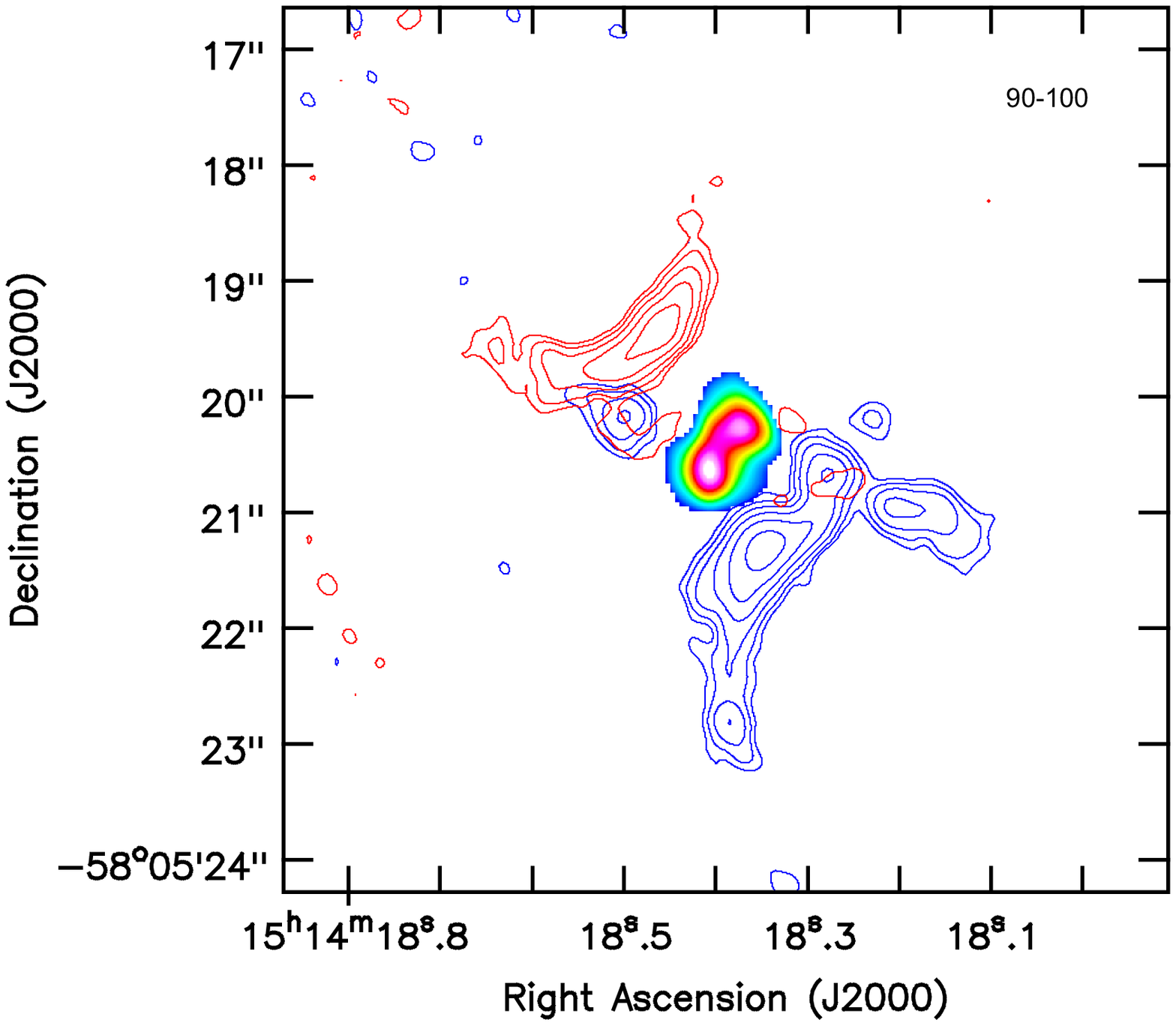}
	\includegraphics*[width=0.3\textwidth]{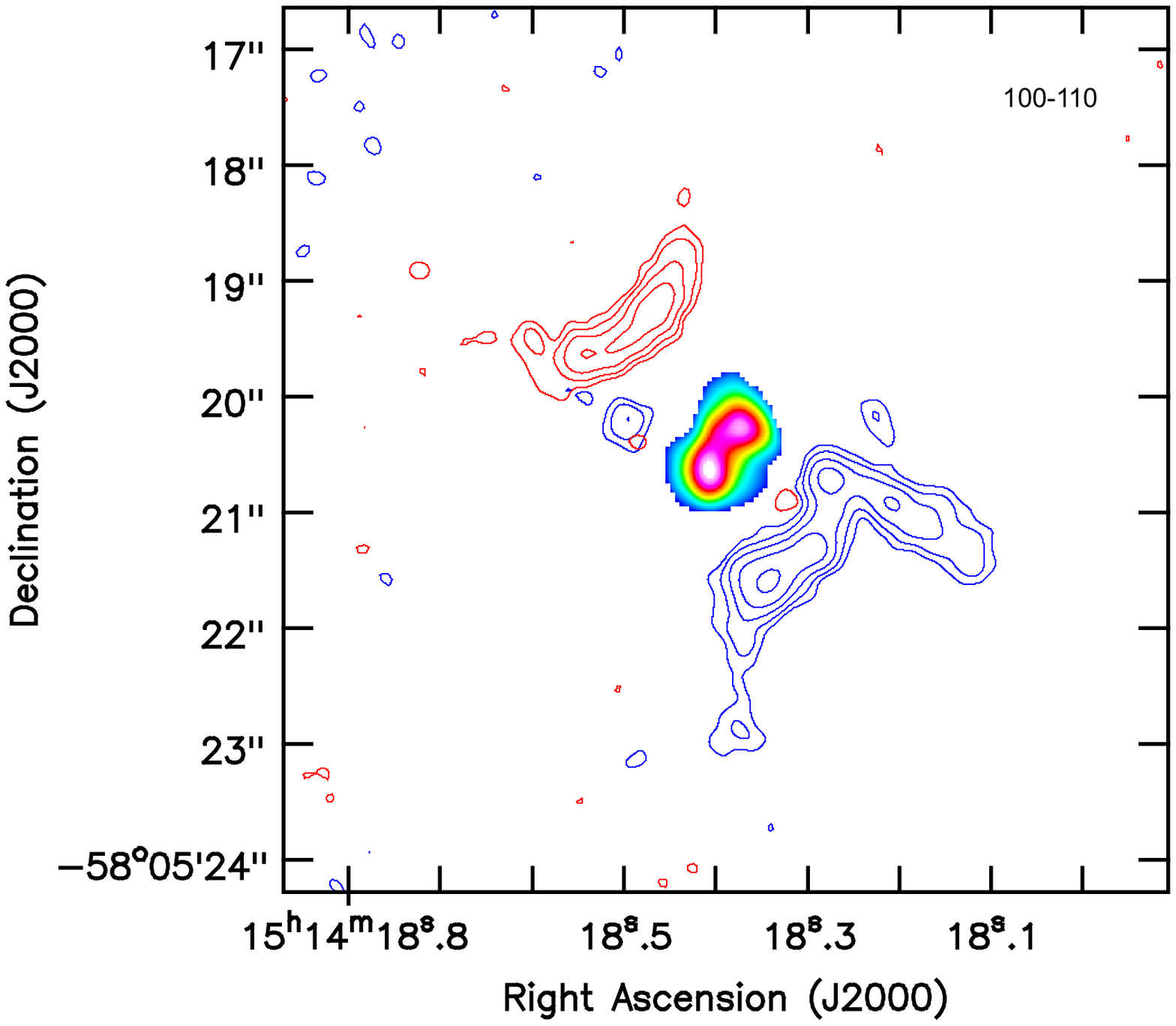}
	\includegraphics*[width=0.3\textwidth]{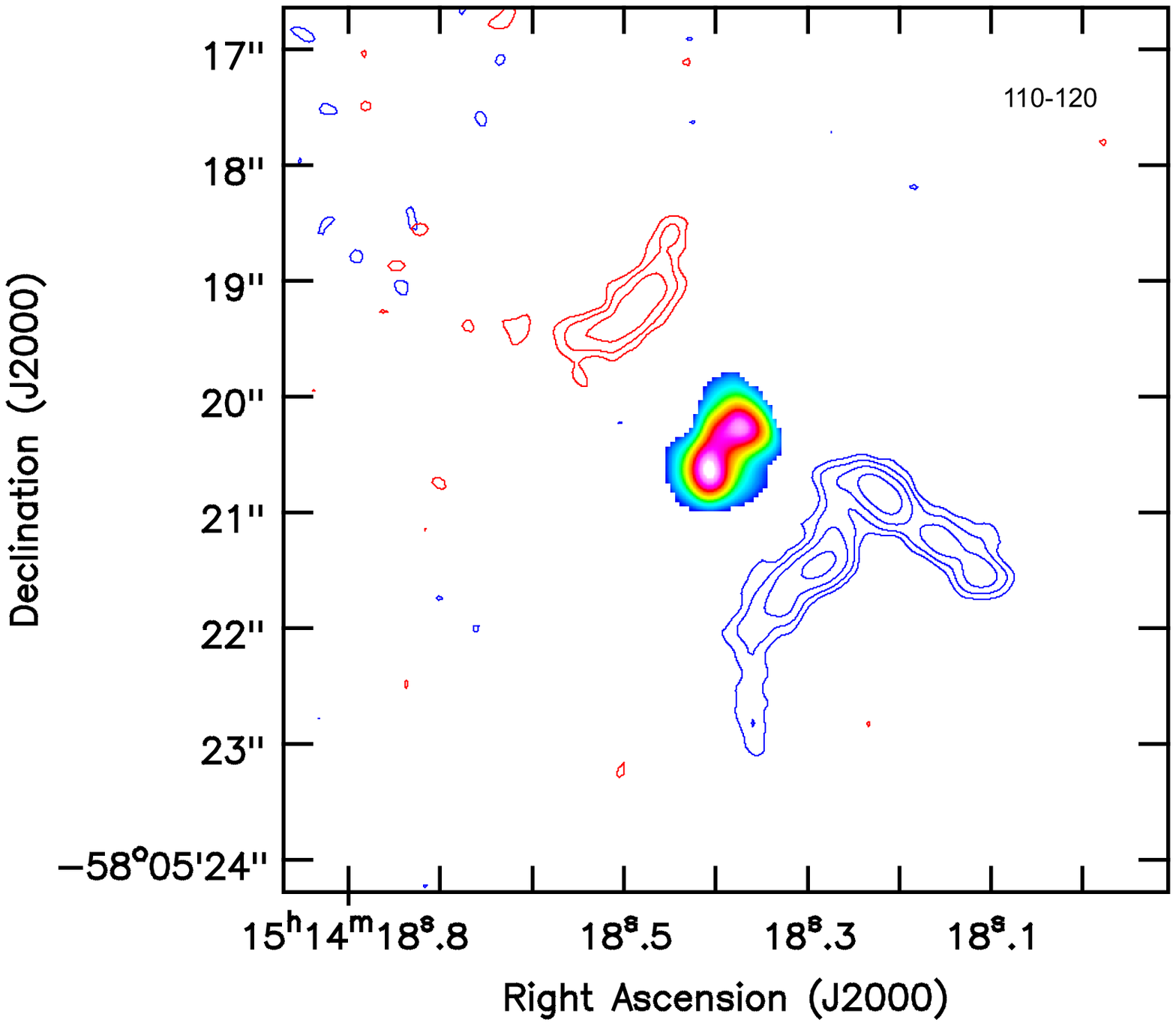}
	\includegraphics*[width=0.3\textwidth]{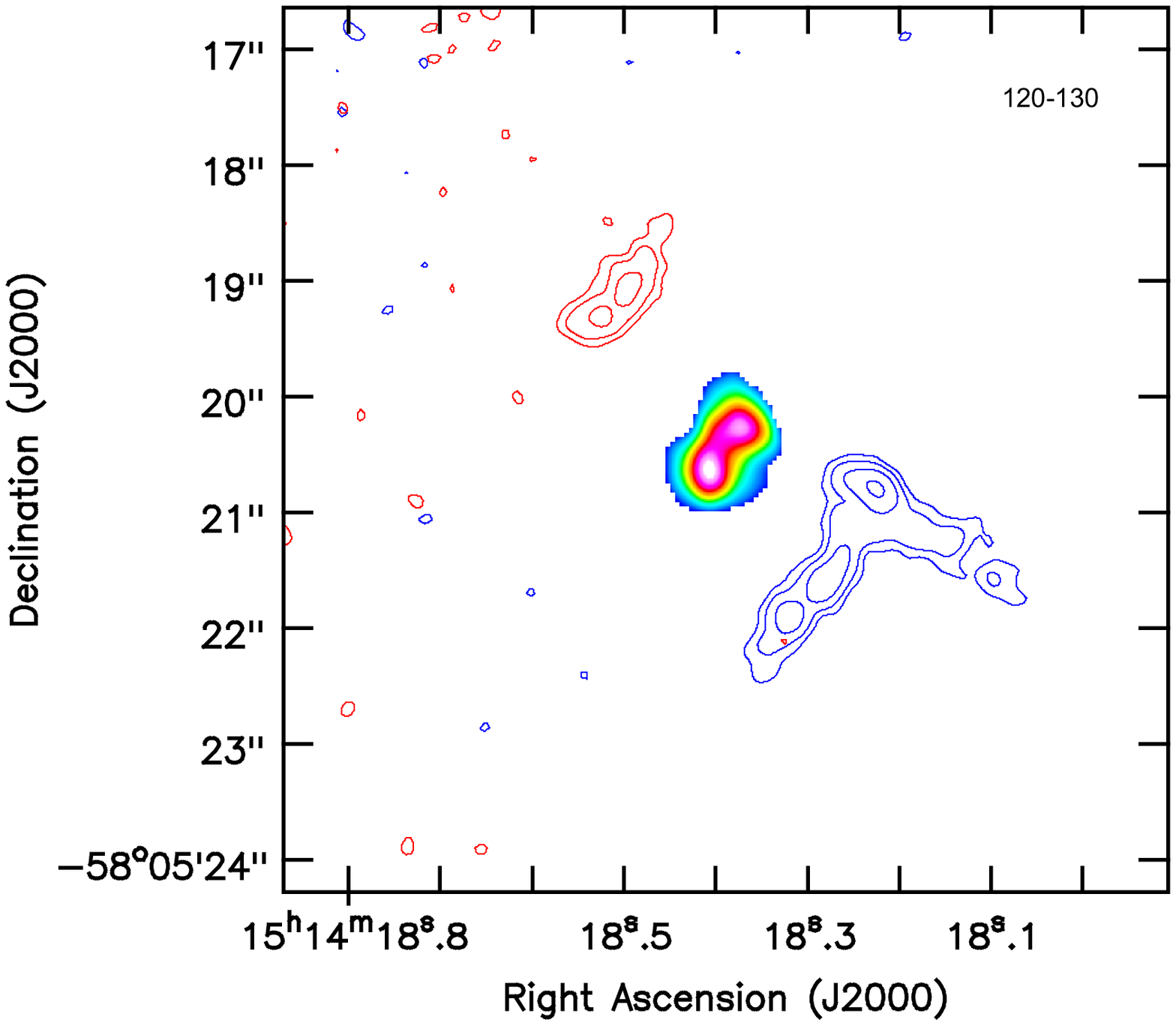}
	\includegraphics*[width=0.3\textwidth]{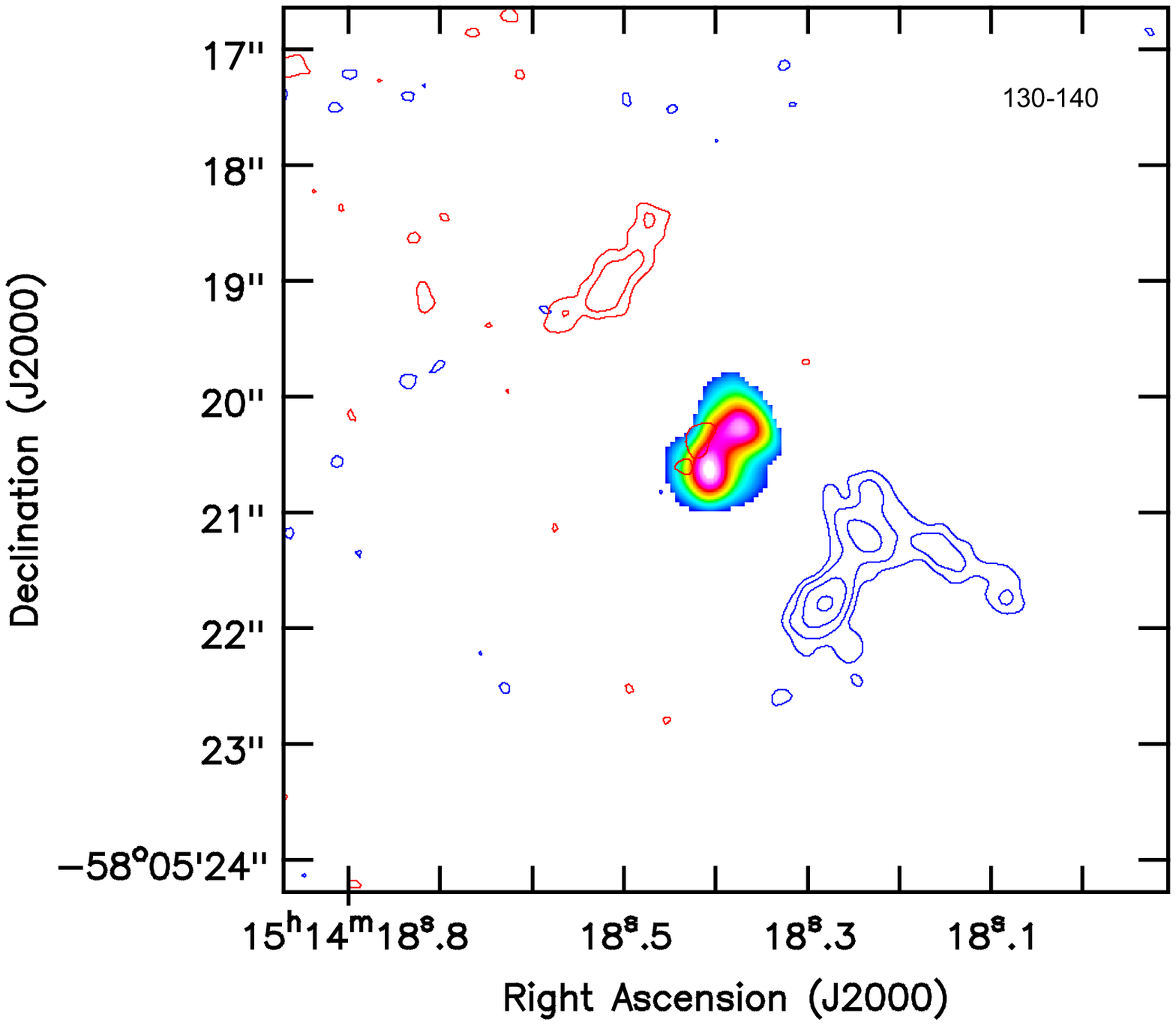}
	\includegraphics*[width=0.3\textwidth]{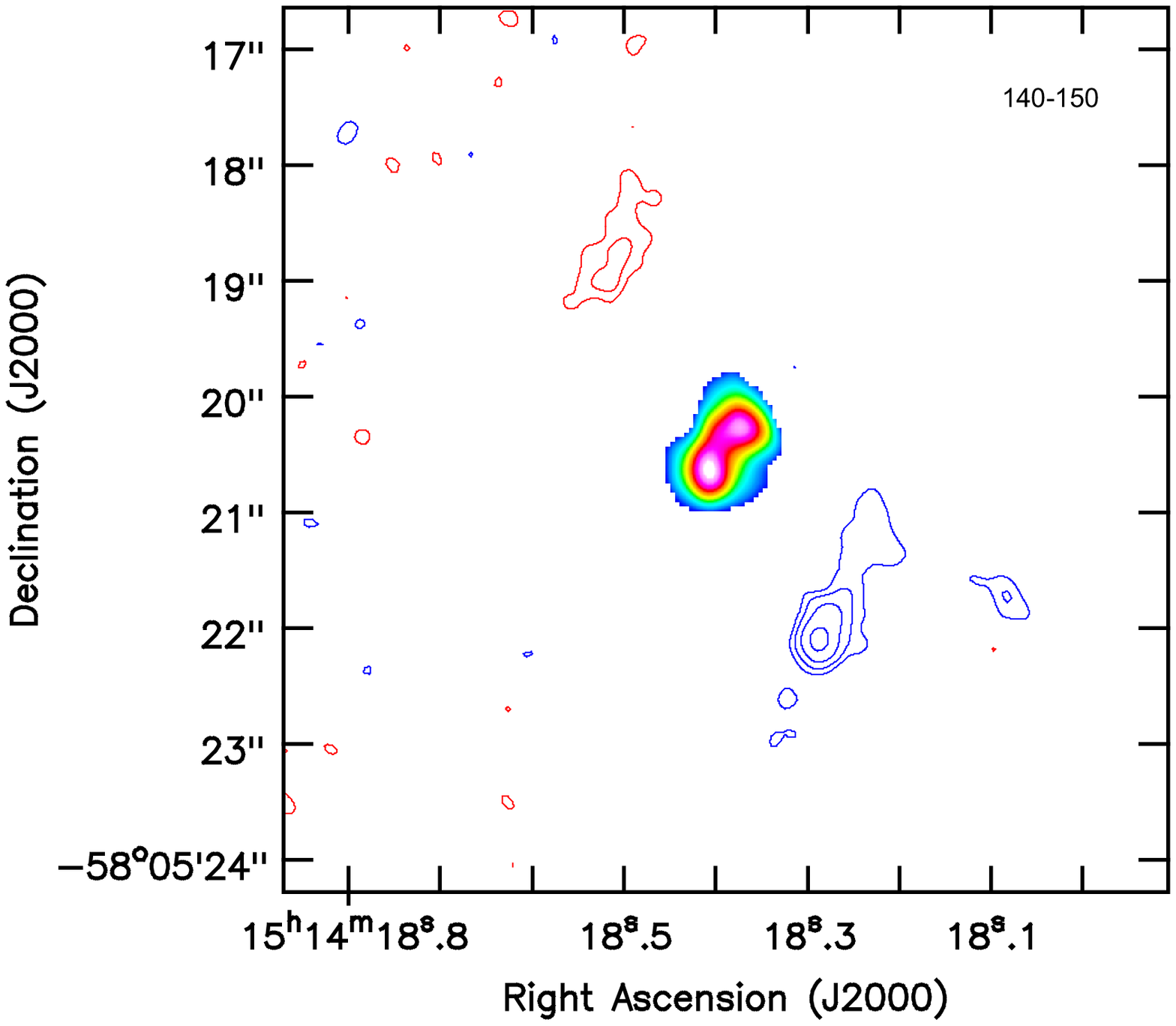}
	\caption{Integrated CO(3-2) emission at different velocity ranges. The range of velocities with respect to the central velocity is shown (in km s$^{-1}$) at the top right corner of each panel. Contour levels are $2^n\sigma$, starting with $n=1$ and increment step $n=1$. The $1\sigma$ rms is 14 mJy beam$^{-1}$ km s$^{-1}$. All maps have the same spatial scale, shown in the panel with velocity range 180-190 km s$^{-1}$.}
	\label{fig:CO_ranges}
\end{figure*}

\begin{figure*}
	\includegraphics*[width=0.3\textwidth]{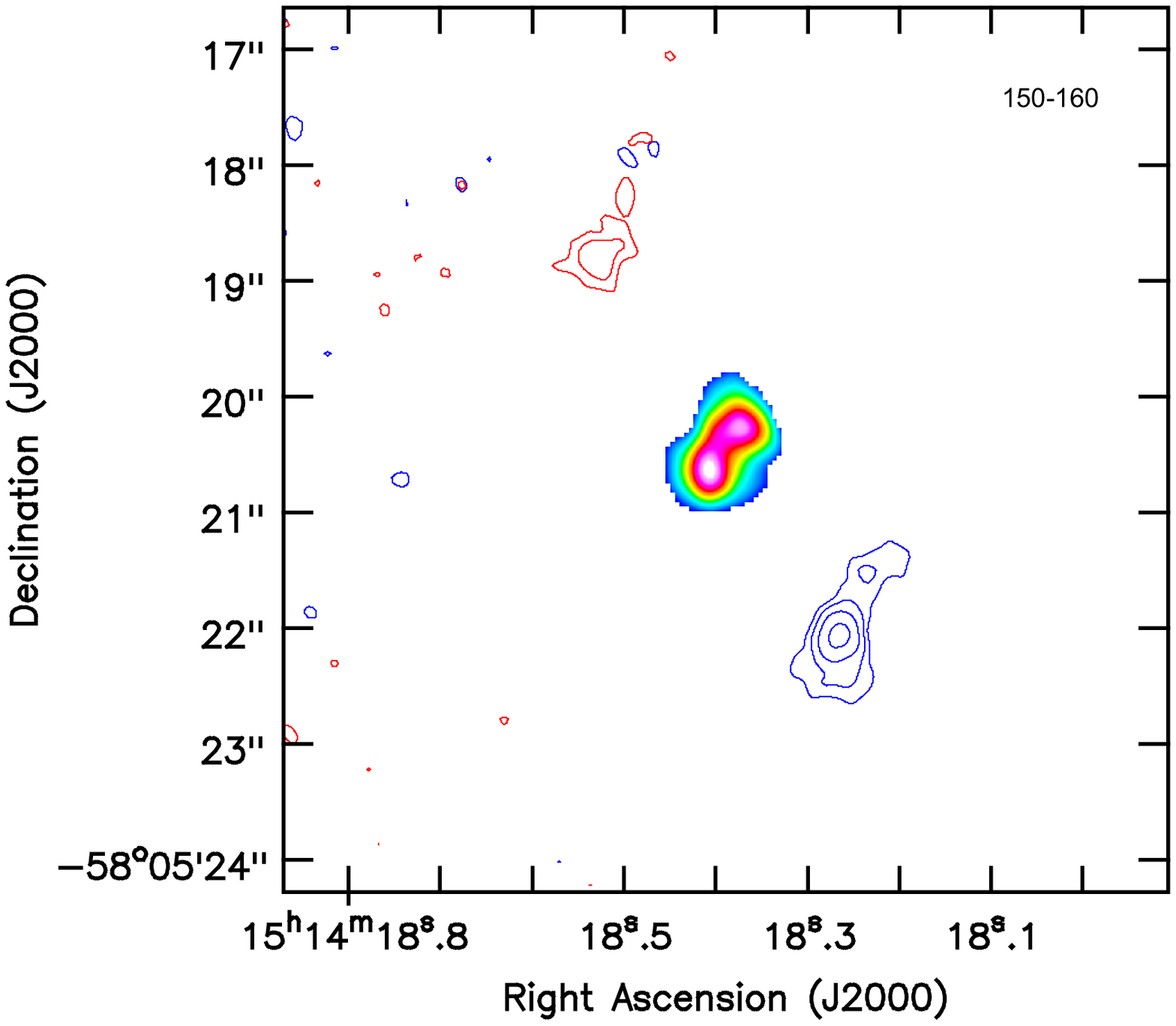}
	\includegraphics*[width=0.3\textwidth]{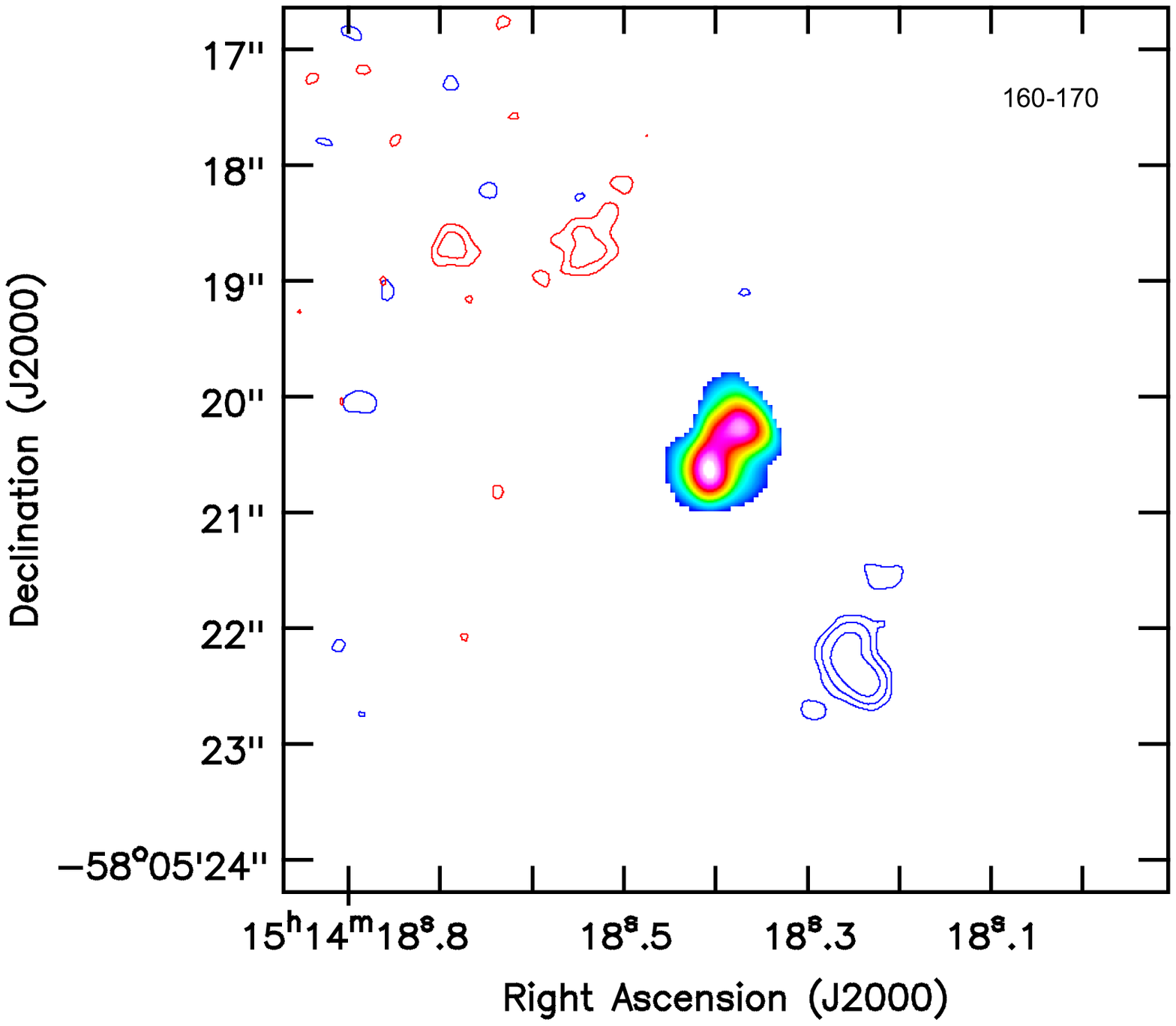}
	\includegraphics*[width=0.3\textwidth]{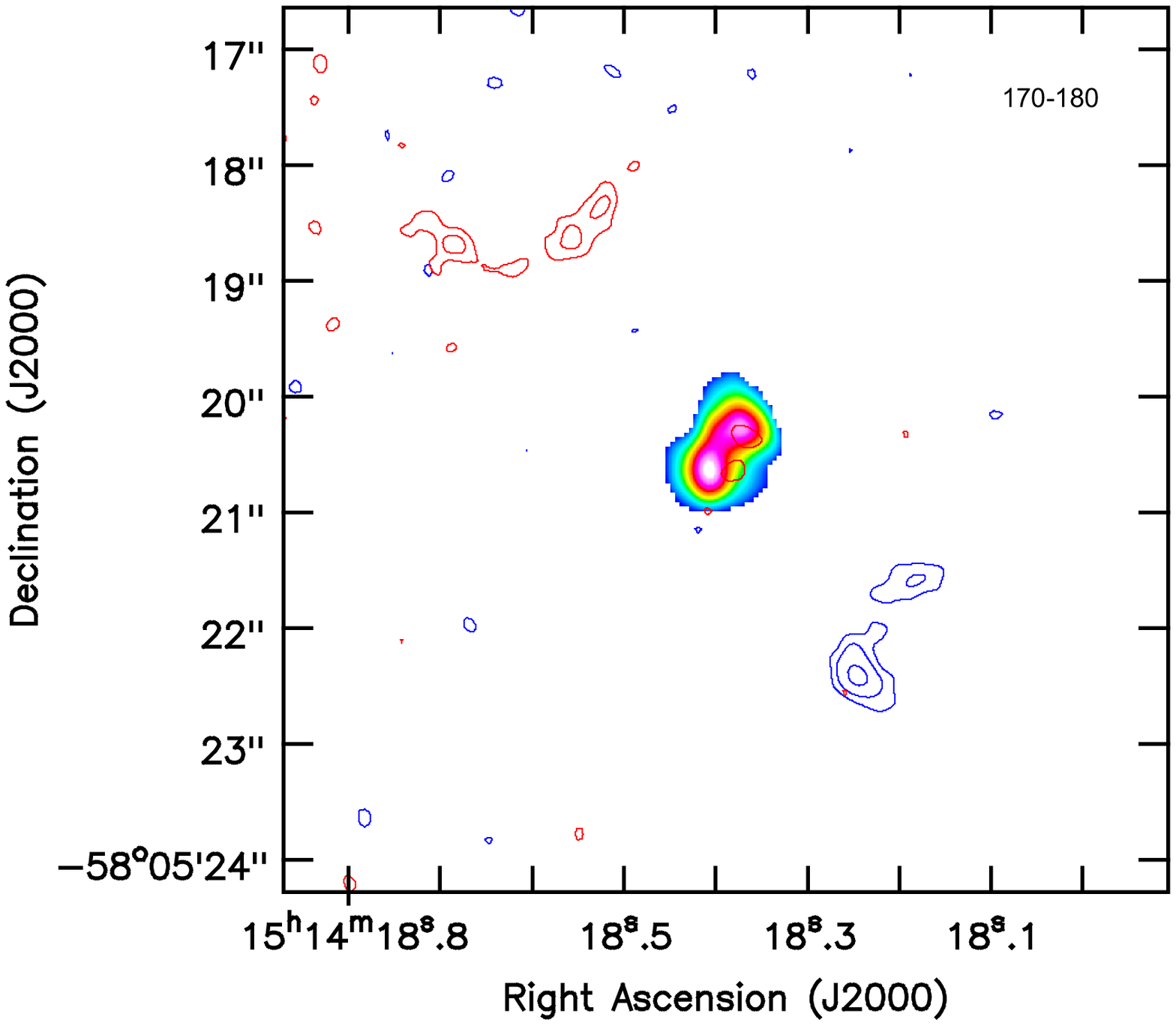}\\\begin{center}
	
	\end{center}
	\includegraphics*[width=0.377\textwidth,valign=t]{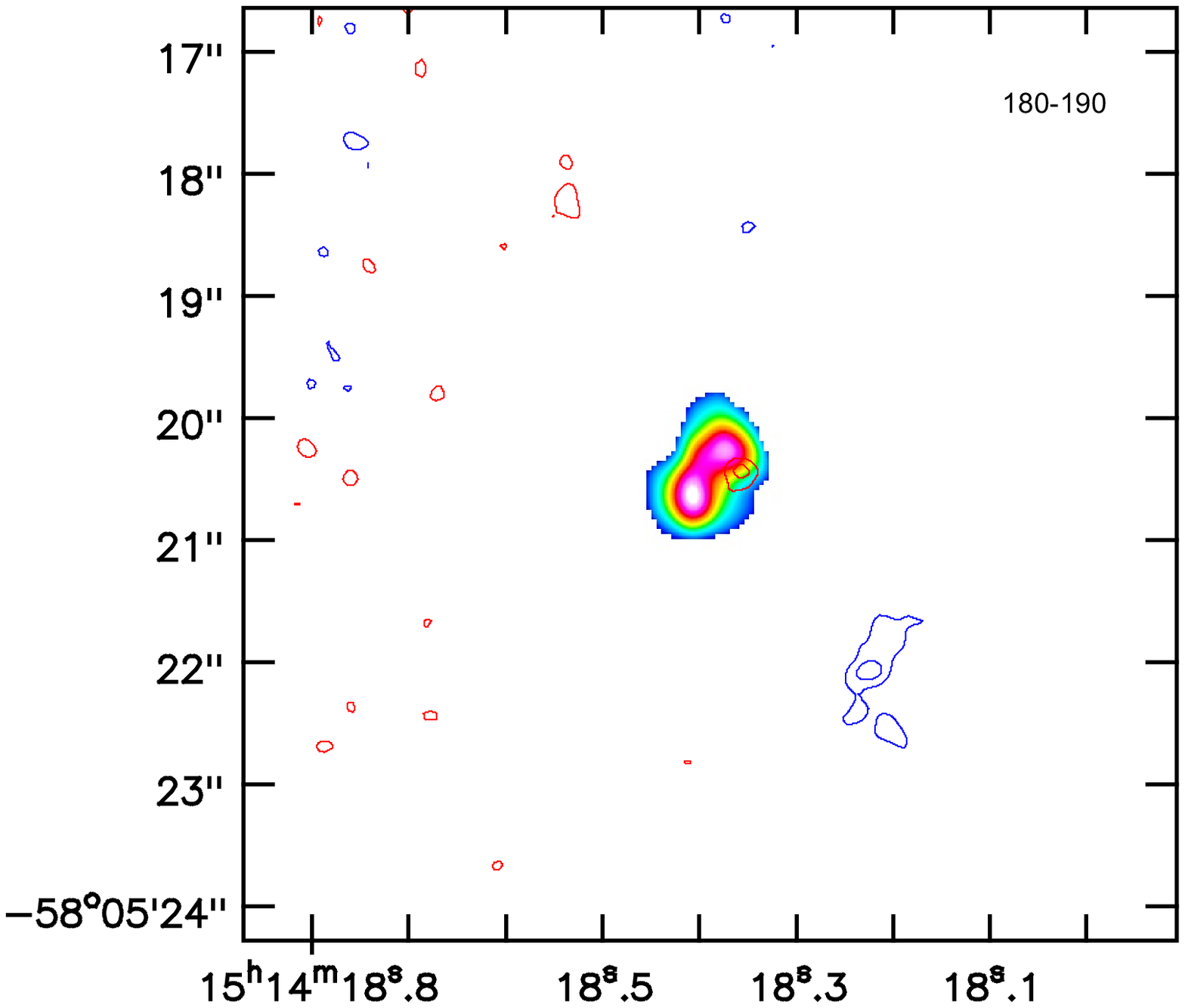}
	\includegraphics*[width=0.3\textwidth,valign=t]{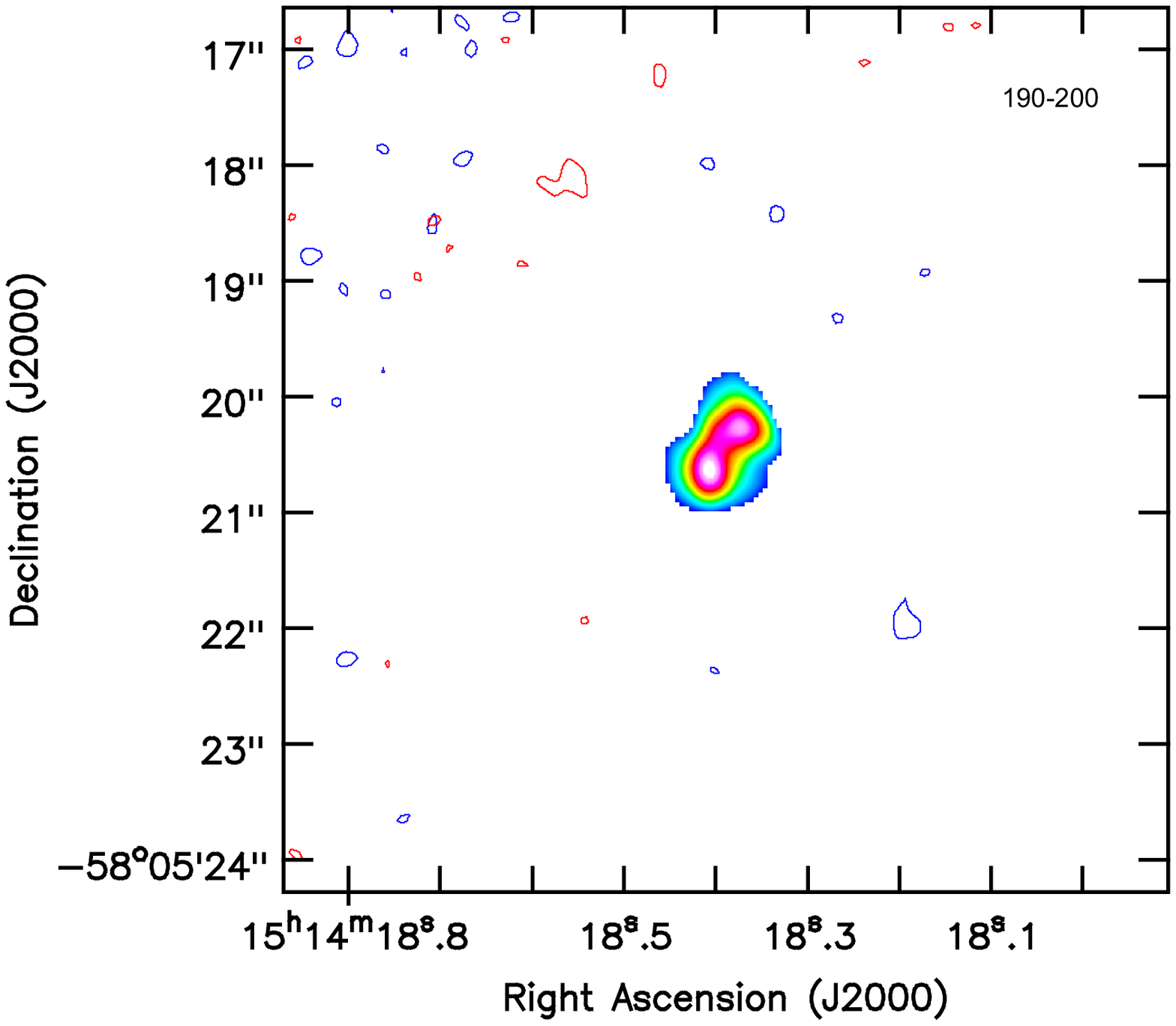}
	\contcaption{}
\end{figure*}

\begin{figure*}
	\includegraphics*[width=0.3\textwidth]{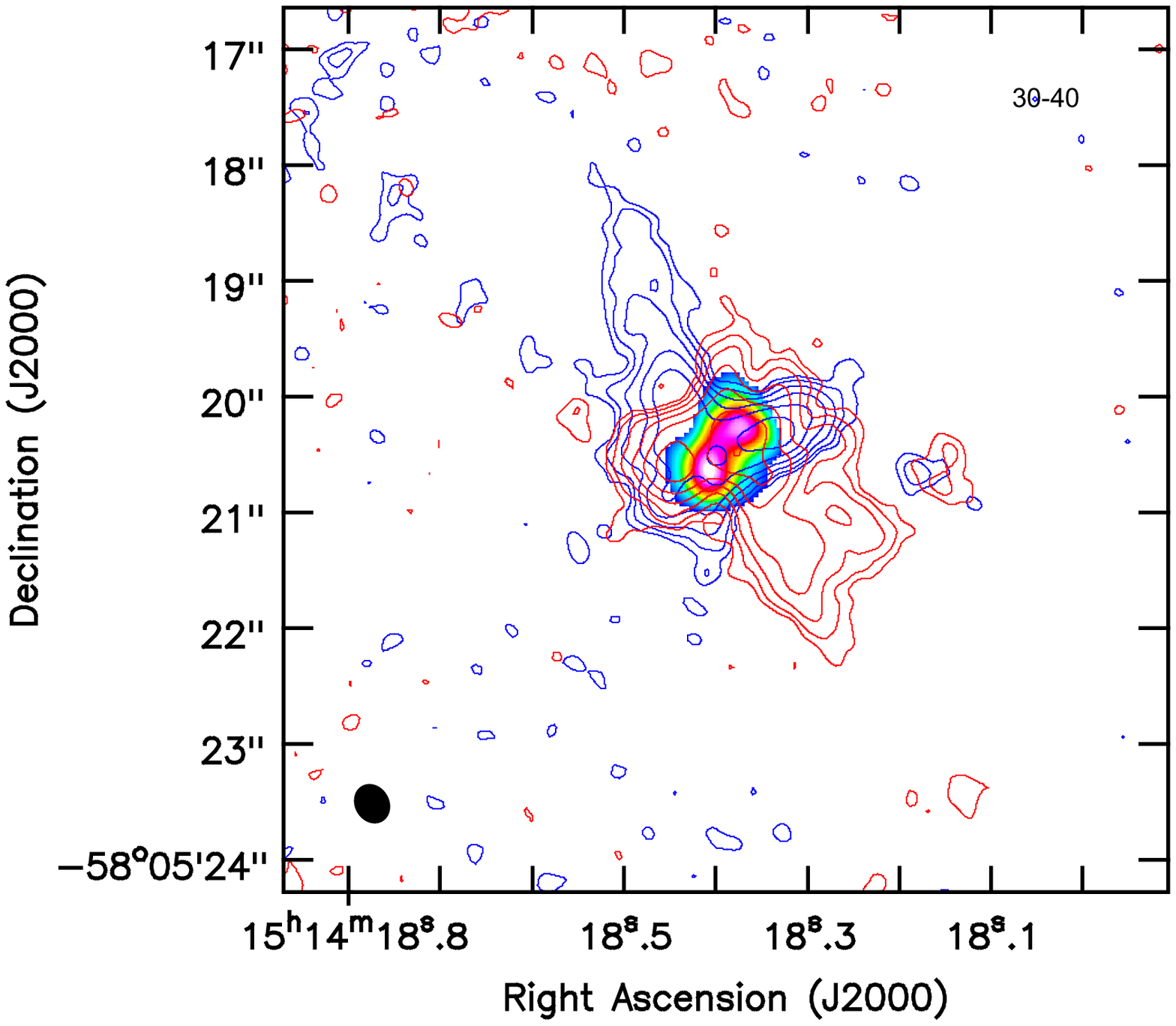}
	\includegraphics*[width=0.3\textwidth]{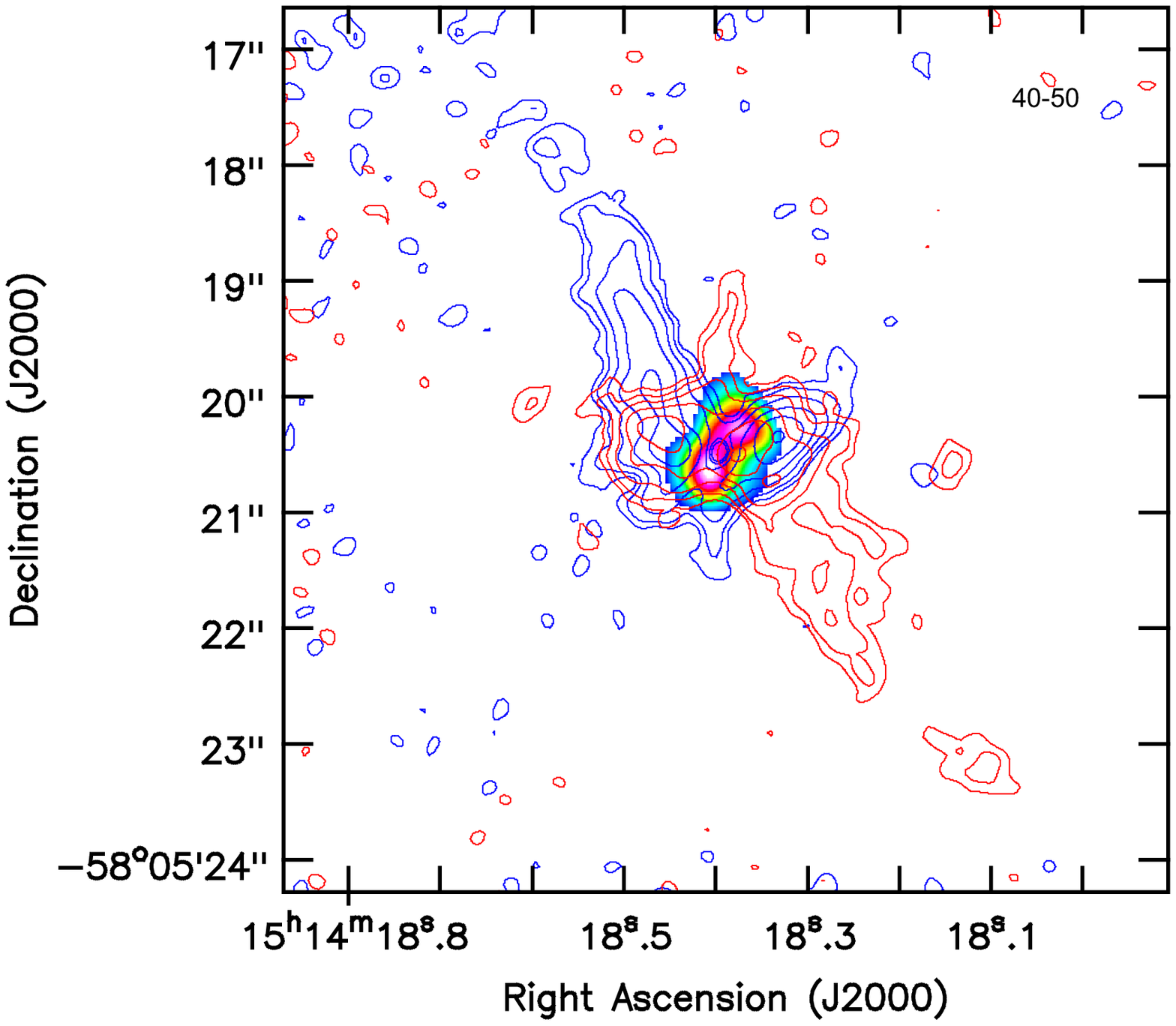}
	\includegraphics*[width=0.3\textwidth]{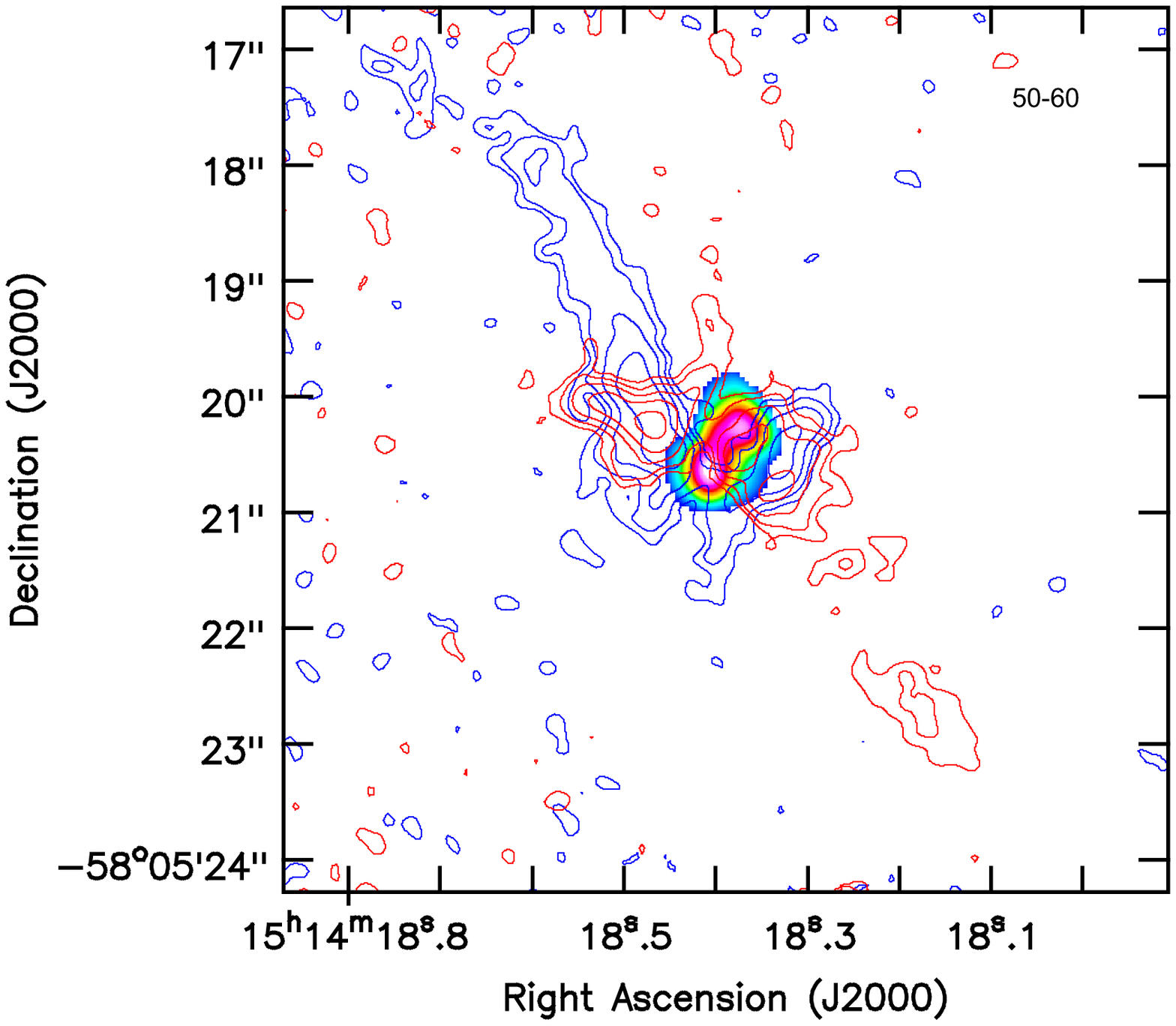}
	\includegraphics*[width=0.3\textwidth]{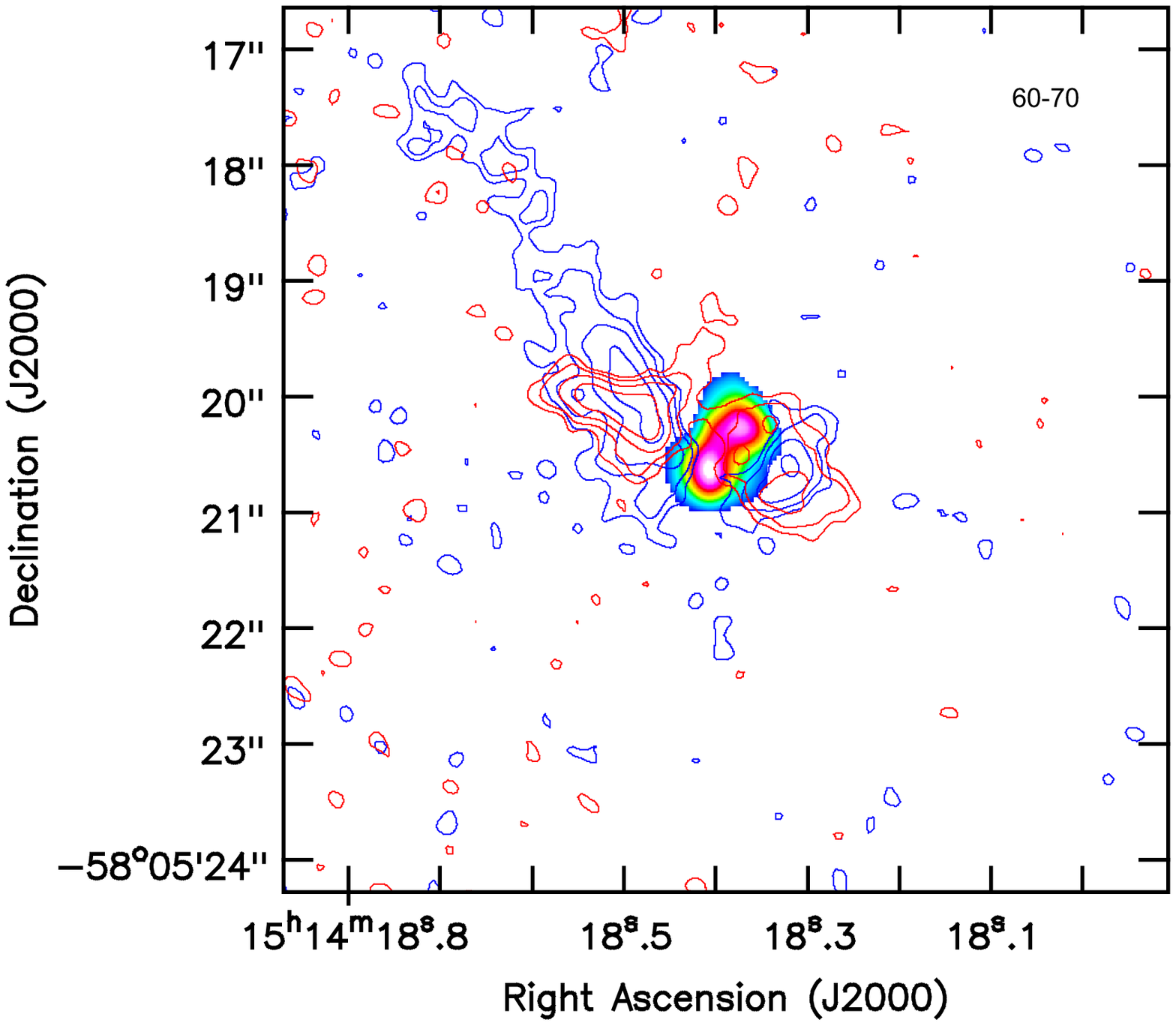}
	\includegraphics*[width=0.3\textwidth]{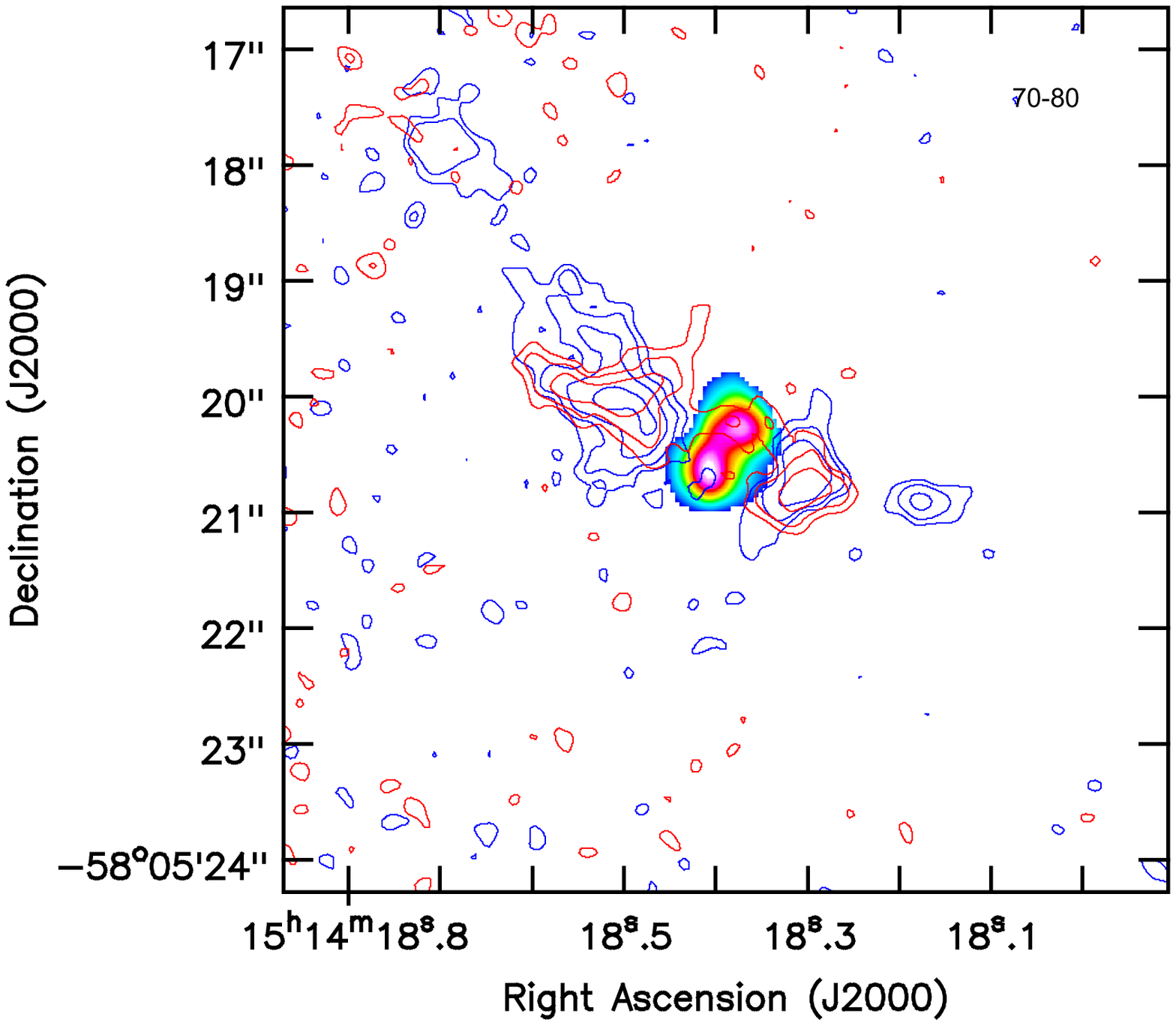}
	\includegraphics*[width=0.3\textwidth]{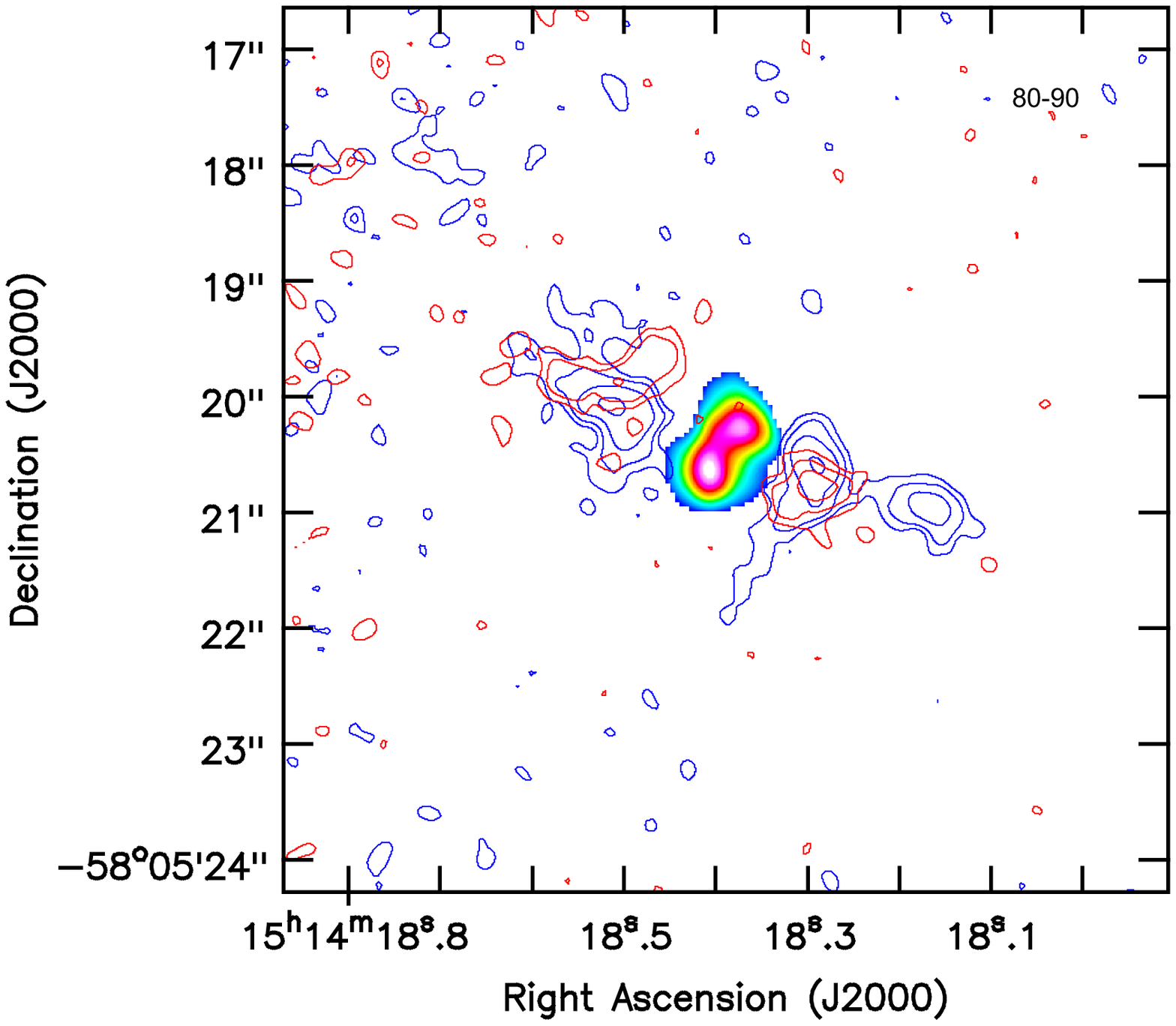}
	\includegraphics*[width=0.3\textwidth]{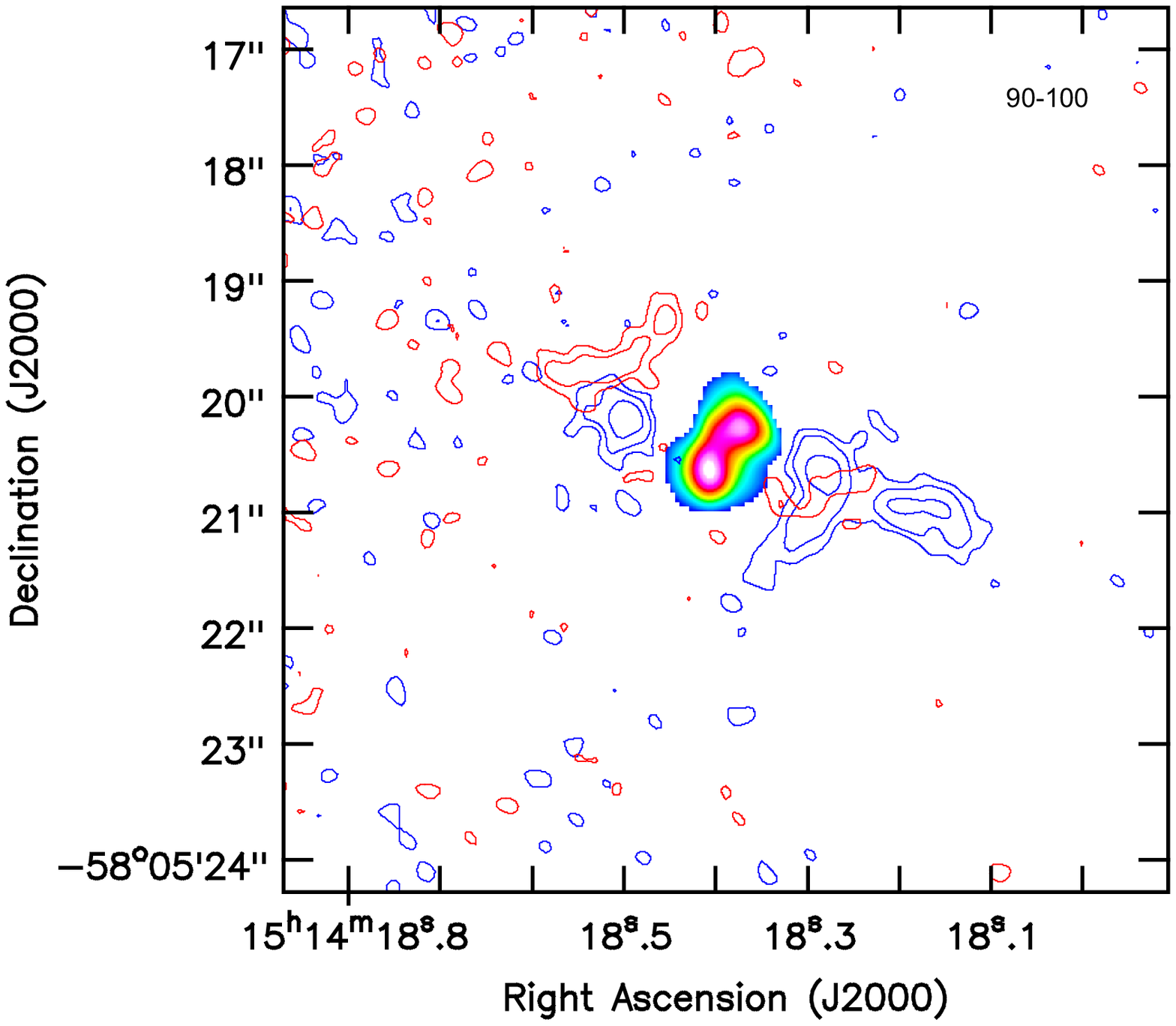}
	\includegraphics*[width=0.3\textwidth]{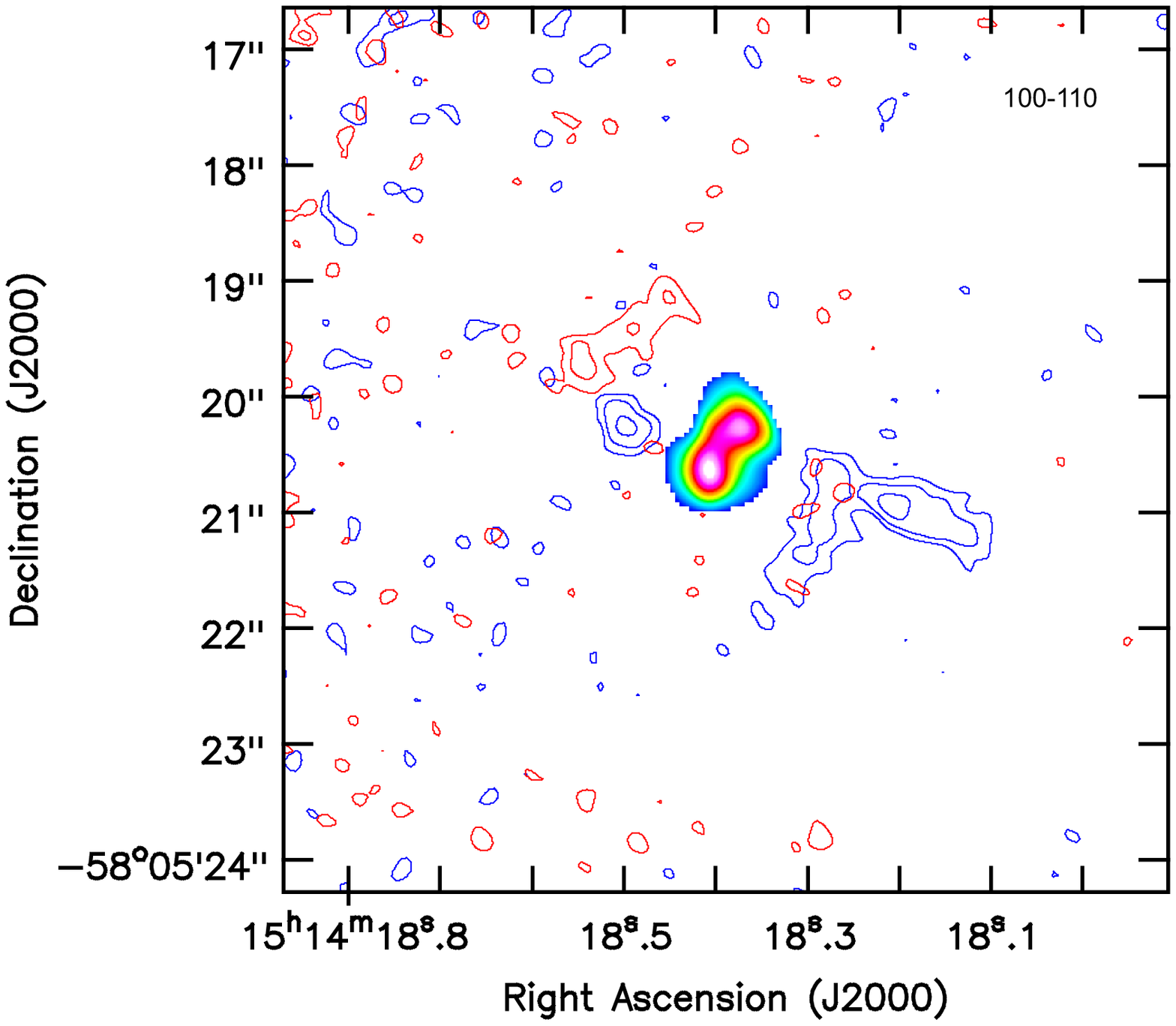}
	\includegraphics*[width=0.3\textwidth]{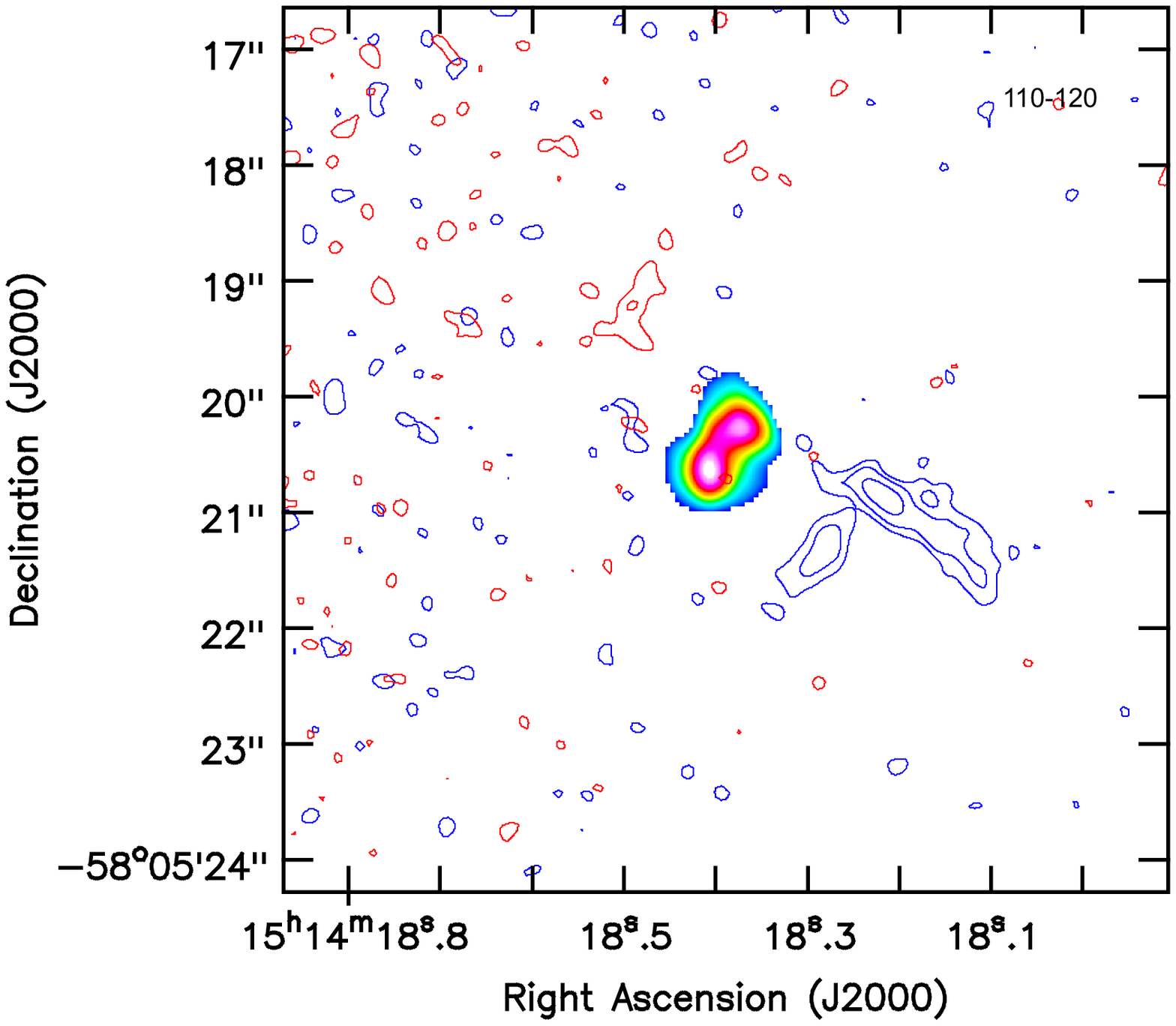}
	\includegraphics*[width=0.3\textwidth]{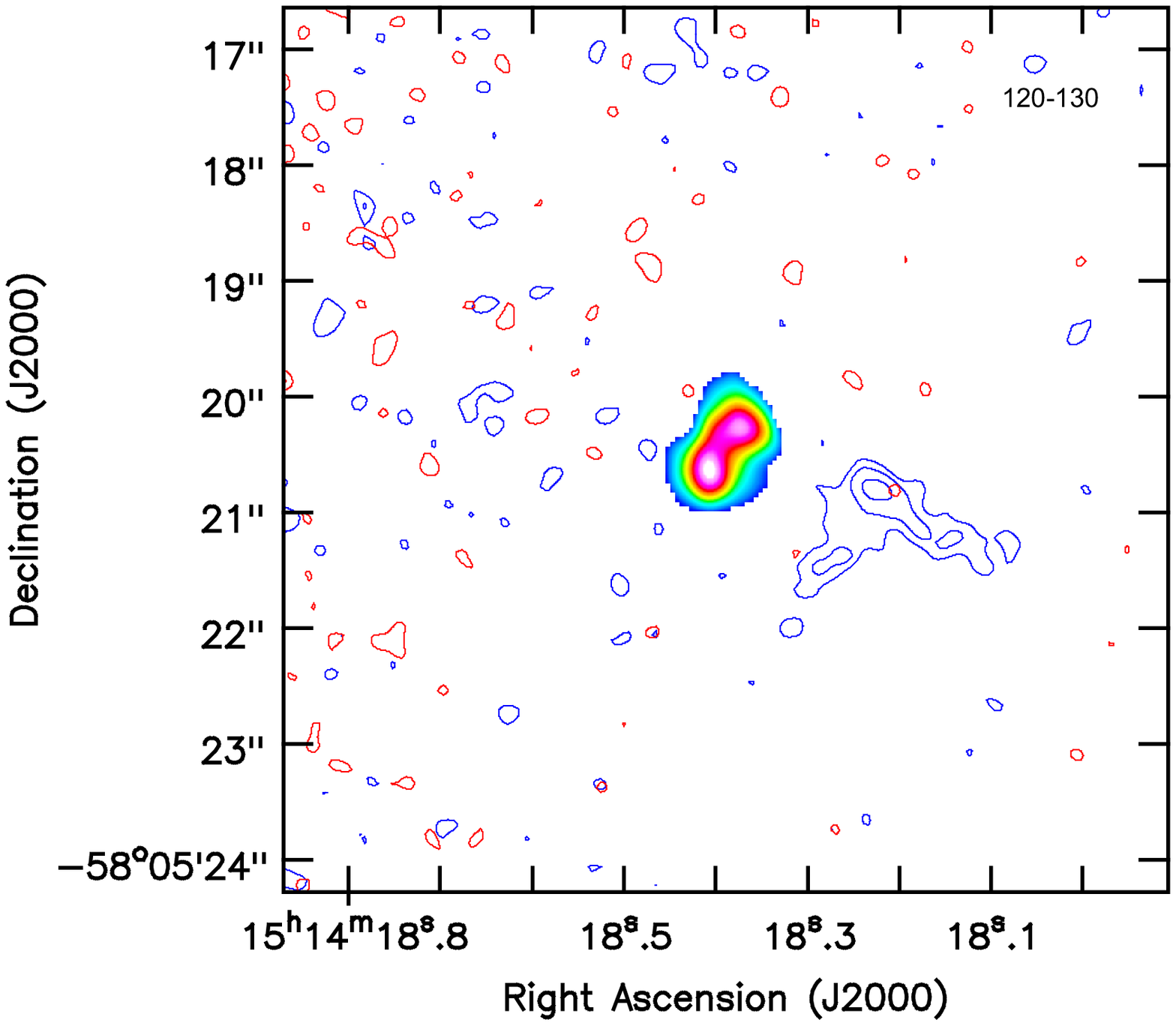}
	\includegraphics*[width=0.3\textwidth]{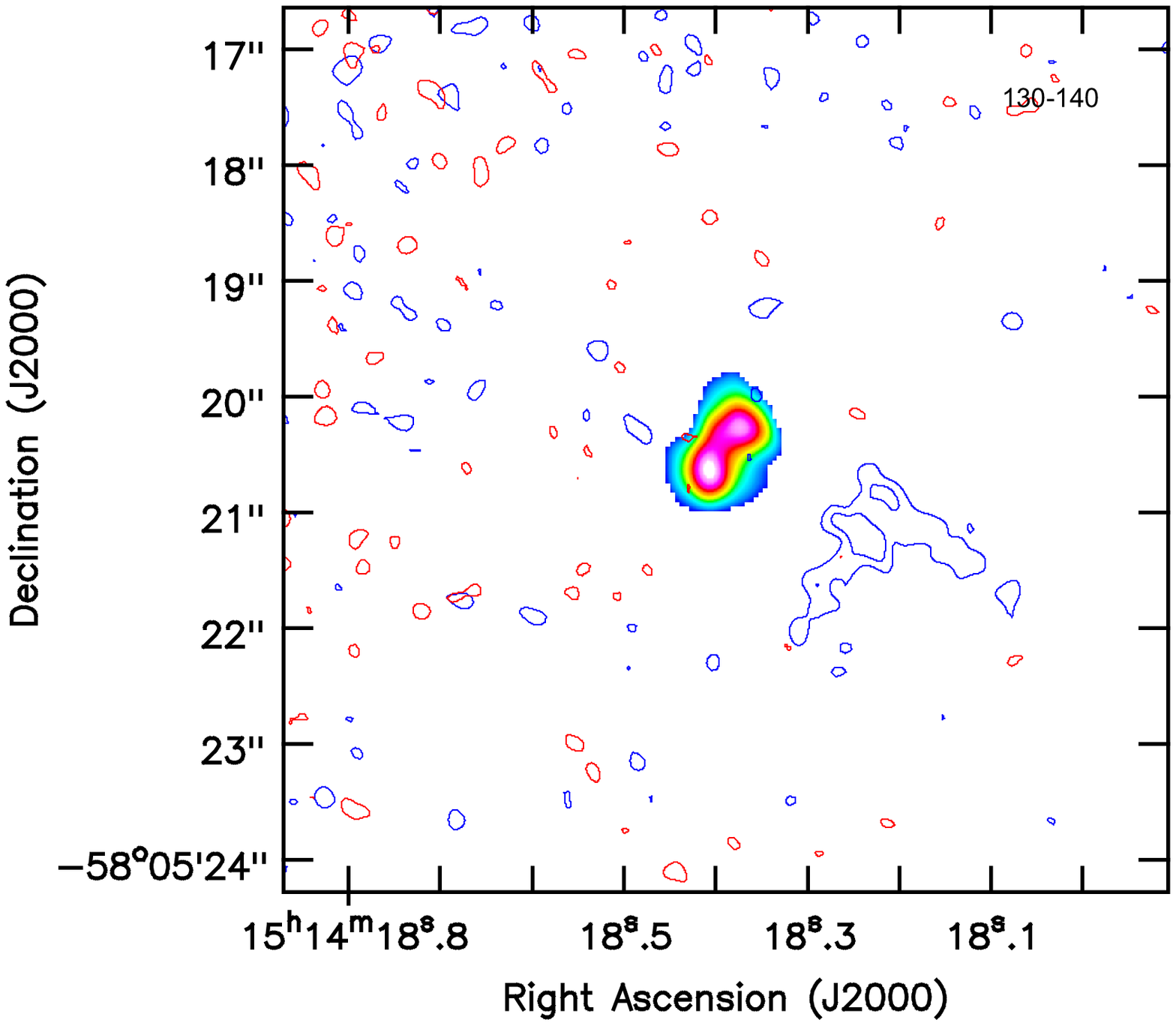}
	\includegraphics*[width=0.3\textwidth]{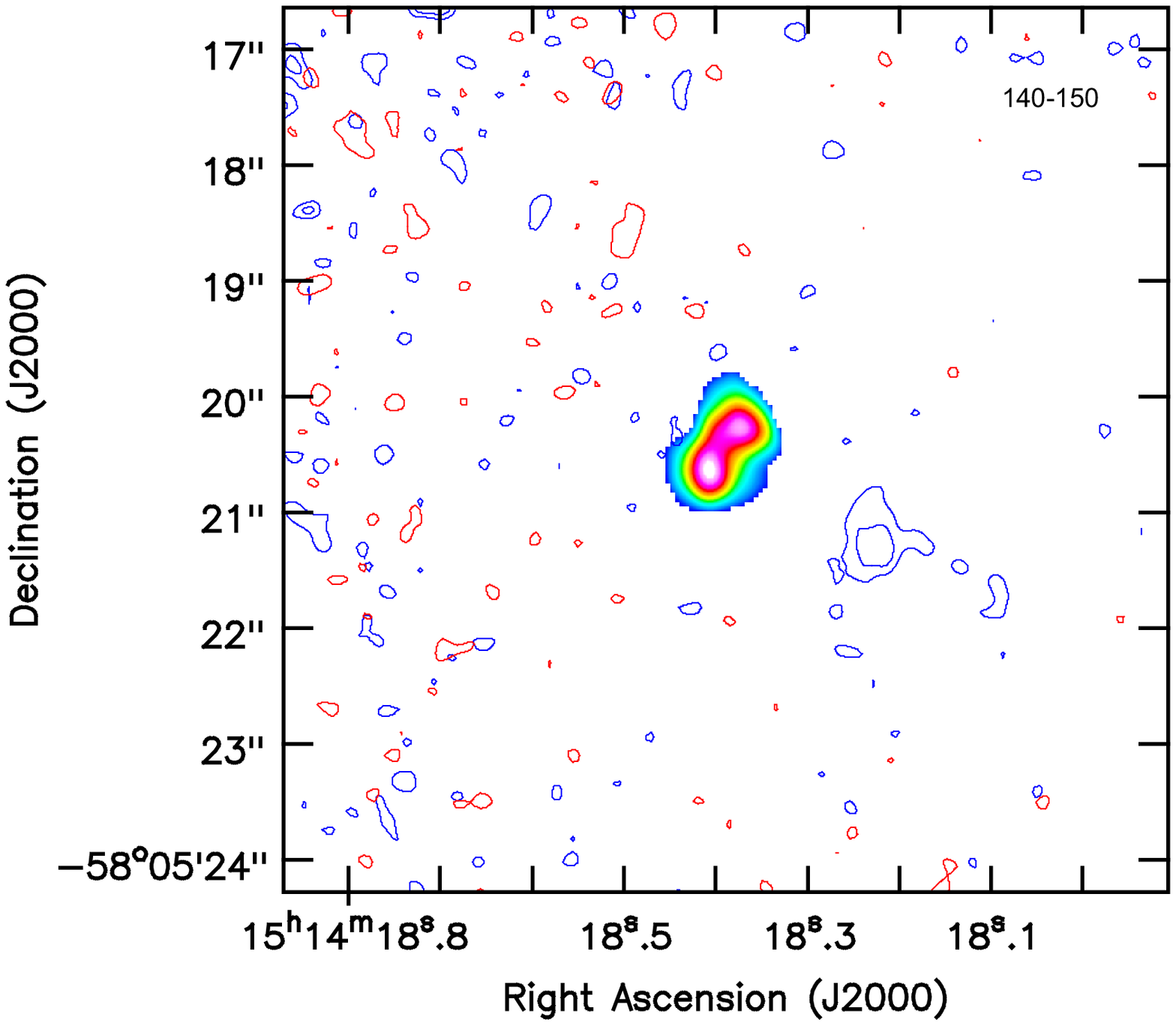}
	\caption{Integrated HCO$^+$(4-3) emission at different velocity ranges. The range of velocities with respect to the central velocity is shown (in km s$^{-1}$) at the top right corner of each panel. Contour levels are $2^n\sigma$, starting with $n=1$ and increment step $n=1$. The $1\sigma$ rms is 14 mJy beam$^{-1}$ km s$^{-1}$. The spatial scale is the same as in Fig. \ref{fig:CO_ranges}.}
	\label{fig:HCO_ranges}
\end{figure*}

\clearpage


\section{Kinematic model of CO emission from a biconical flow}

\label{app:cone_model}

We present our model for the component of the outflow showing some clear
conical features. The model is based on the PV diagram of CO along the
main axis of the CO outflow (see Fig.~\ref{fig:PV}), which shows an X-like shape, and the integrated CO
emission at different velocity ranges (see Fig.~\ref{fig:CO_ranges}), displaying an hyperbolic shape of the
isovelocity regions projected on to the plane of the sky. These two figures suggest a 
configuration presented in Fig.~\ref{fig:coneGeometry}. The initial outflow model is biconical, consisting of two cones with a common apex at the origin $O$ of the
system of Cartesian coordinates. { We discuss the differences with the case of truncated cones, joining at the minor circular surface of their frustum, at the end of this appendix.}


For the sake of simplicity, we use a coordinate system ($X,Y$) on the plane of the sky that is rotated with respect to the equatorial system. The $Y$ axis is parallel with the projection of the cone axes on the plane of the sky. Any rotation in the image plane to account for the actual position angle of the cone will not affect our calculations.

\begin{figure}
  \centering
  \psfrag{t0}{\small $\theta_0$}
  \psfrag{t}{\small $\theta$}
  \psfrag{x}{\small $x$}
  \psfrag{y}{\small $y$}
  \psfrag{z}{\small $z$}
  \psfrag{X}{\small $\hat{X}$}
  \psfrag{Y}{\small $\hat{Y}$}
  \psfrag{1}{\small \circleBox{$1$}}
  \psfrag{2}{\small \circleBox{$2$}}
  \psfrag{3}{\small \circleBox{$3$}}
  \psfrag{4}{\small \circleBox{$4$}}
  \psfrag{n}{\small $\nhat$}
  \psfrag{v}{\small $\vhat$}
  \psfrag{line}{\tiny line of sight}
  \resizebox{8cm}{!}{\includegraphics{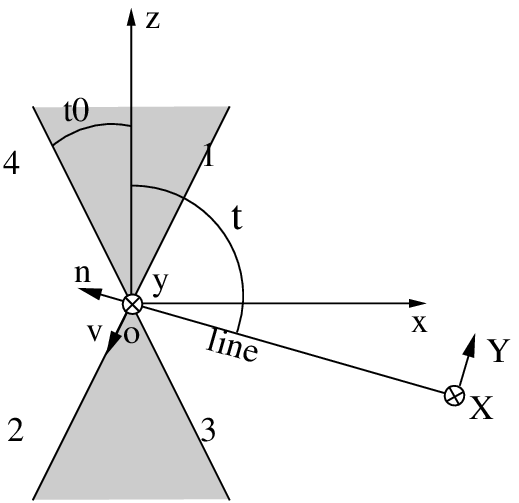}}
  \caption{Description of the geometry of the conical outflow. A
    $\otimes$ symbol represents a vector pointing into the page. The
    coordinates in the image plane are $X$ and $Y$. $\nhat$ is the
    direction vector along the line of sight. $\vhat$ is the direction
    vector along the cone surface.}
  \label{fig:coneGeometry}
\end{figure}


From the biconical configuration shown in
Fig.~\ref{fig:coneGeometry} we can derive the expected PV diagram by
projecting the velocity on to the line of sight (i.e., computing
$\vv . \nhat$, where $\vv$ is the velocity at the surface of the
cones). The PV diagram is derived along the axis of symmetry of the
double cone, $Oz$ (corresponding to $X=0$ in the image plane), and the
position offset of the PV diagram is the coordinate along the $Y$ axis. The conical outflow model
assumes an explosive event with particles having ballistic
trajectories. For a particle ejected well above the escape velocity,
its velocity reaches a constant value. Consequently, $\vv$ is pointing outward from the apex
$O$ and increases linearly with the distance from $O$, the fastest
particles being farther away from the origin.

If $\rr$ is the position vector of a point at the surface of the
double cone, $\vv$ is given by
\begin{equation}
  \label{eq:coneVelocity}
  \vv = \gradv\,\rr \ ,
\end{equation}
where $\gradv$ is the velocity gradient assumed to be
constant. $\gradv$ is also the inverse of the dynamical time-scale
$\tdyn$, elapsed since the explosive event.

In Fig.\ref{fig:coneGeometry} we define four ``branches'' identified
as circled numbers. Note that in the resulting PV diagram (the dashed lines in Fig. \ref{fig:PVCOoutflow}), the
branches $1$ and $2$ on one hand, and $3$ and $4$ on the other hand,
form two straight lines with different slopes ($\vv . \nhat$):

\begin{equation}
  \label{eq:vdotn}
  \vv . \nhat = 
  \begin{cases}
    -\cot{(\theta-\theta_0)}\,\gradv\,Y\, ,& \text{branch $1$  and $2$},  \\
    -\cot{(\theta+\theta_0)}\,\gradv\,Y\, ,& \text{branch $3$ and $4$} \ ,
  \end{cases}
\end{equation}
{ where branches 1 and 4 correspond to $Y \ge 0$, and branches 2 and 3 correspond to $Y \le 0$.}
 These two straight
lines, intersecting at the origin would lead to an X~shaped PV
diagram.

This simple model needs to be modified in order to account for some
supplementary features of the diagram. First, note that the
observed diagram (Fig. \ref{fig:PVCOoutflow}) does not show two intersecting straight lines, but 
branches $2$ and $4$ are shifted upward, while
branches $1$ and $3$ are shifted downward by a similar amount. Moreover, the intersection of branches is not directly observed in the data, due to the presence of the torus. However, if we extrapolate the branches toward the centre, 
branch pair $2$ and $4$ on one hand and the pair $3$ and $1$ on the other hand
intersect near spatial offset zero, as illustrated in the dashed lines of Fig. \ref{fig:PVCOoutflow}. In order to
reproduce these features in the position-velocity diagram, we studied more complex models by spatially moving
the cone vertices along the axis of symmetry by a positive or negative
shift, { i.e., forming separated or truncated cones, respectively}. None of these spatial changes led to a PV diagram in agreement
with the data.

However, we can account for all features of the observed PV diagram
by adding a constant axisymmetrical velocity $V_0$ in the equatorial direction (i.e., perpendicular to the cone axis) to all points on the cone surface, leading to the following modified
position-velocity relation.

\begin{equation}
  \label{eq:vdotnexp}
    \vv . \nhat = 
  \begin{cases}
    -\cot{(\theta-\theta_0)}\,\gradv\,Y -V_0\,\sign{(Y)}\,\sin{(\theta)}\, , & \text{branch $1$ and $2$},  \\
    -\cot{(\theta+\theta_0)}\,\gradv\,Y + V_0\,\sign{(Y)}\,\sin{(\theta)}\, , & \text{branch $3$ and $4$ } \ ,
  \end{cases}
\end{equation}
where $\sign$ denotes the sign function.

This assumption is supported by the fact that the velocity field that
must be added to the linearly varying velocity at the surface of the
cone must have the following properties.
\begin{itemize}
\item The resulting shift in the PV diagram must be independent of the
  offset.
\item Branches $1$ and $3$ are shifted downward toward negative
  velocities in the PV diagram by the same amount. Branches $2$ and
  $4$ are shifted together in the same manner but in the opposite
  direction (upward).
\end{itemize}

The PV diagram described by Eq.~(\ref{eq:vdotnexp}) is shown as solid black lines in the bottom panel of
Fig.~\ref{fig:PVCOoutflow} and is in good agreement with the observed
diagram with an inclination of the symmetry
axis of the cones with respect to the line of sight 
$\theta \simeq 106 \degr$, a half opening angle
$\theta_0 \simeq 28 \degr$, a velocity gradient 
$\gradv = 73\,\kms\,\mathrm{arcsec}^{-1}$ and $v_0 \simeq 23$ km s$^{-1}$.
Also note that the central LSRK velocity of the double cone is $-41$ km s$^{-1}$, which does not coincide with 
the central velocity of the expanding torus 
($-33$ km s$^{-1}$).

{ The presence of a velocity $V_0$ implies that the present geometry of the conical outflow cannot exactly be that shown in Fig. \ref{fig:coneGeometry}, with full conical surfaces, but it must consist of two truncated cones, as illustrated in Fig. \ref{fig:toy}. We have thus considered this more general case of truncated cones joining, on the $xy$ plane, at the minor circular surface of their frustum, with $L$ being the radius of this surface (Fig.\ref{fig:truncated}).
In this case, the velocity gradient in the PV diagram does not change, as the projected velocities are the same as in the case with two cones joining at the vertex. There is, however, a small shift in the spatial offsets, since the surface points at the bases of the truncated cones are at offset $Y\ne 0$. The result is that branches 1 and 3 in the PV diagram (Fig. \ref{fig:PVCOoutflow}) will be shifted toward the right, and branches 2 and 4 toward the left, by an amount $Y_0=|L\cos\theta |$. In quantitative terms, assuming $L=0.3$ arcsec (the mean radius of the torus), this shift would be $\simeq 0.08$ arcsec, which is a factor of 3 smaller than the beam size. Thus, the PV diagrams do not significantly change in the case of truncated cones, for the particular geometry and orientation of IRAS 15103-5754.}

\begin{figure}
	\includegraphics*[width=0.48\textwidth]{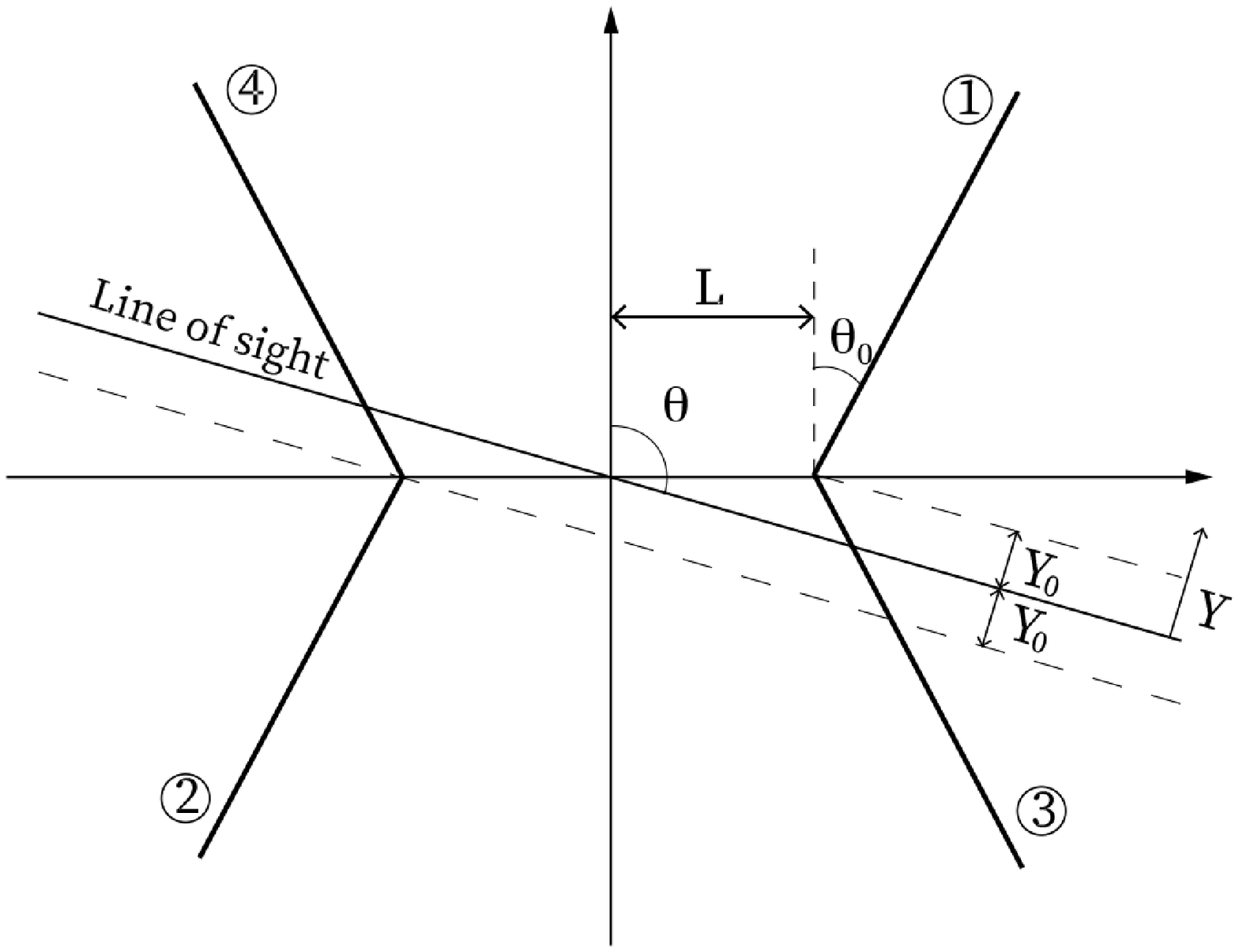}\\
	\caption{ Geometry in the case of truncated cones, showing the shift of projected spatial offsets at their bases, $Y_0$. }
	\label{fig:truncated}
\end{figure}

\bsp	
\label{lastpage}
\end{document}